\documentclass [11pt,a4paper]{article}
\usepackage{jcappub}

\usepackage{amsmath,amssymb,amsfonts,amsthm}
\usepackage{bm,bbm}
\usepackage{graphicx}
\usepackage[all]{xy}
\usepackage{pifont}

\newcommand{\be}{\begin{equation}}
\newcommand{\ee}{\end{equation}}
\newcommand{\ba}{\begin{eqnarray}}
\newcommand{\ea}{\end{eqnarray}}
\def\bs{\begin{subequations}}
\def\es{\end{subequations}}
\def\a{\alpha}
\def\b{\beta}
\def\de{\delta}
\def\g{\gamma}
\def\la{\lambda}
\def\k{\kappa}
\def\e{\epsilon}

\def\Om{\Omega}
\def\om{\omega}
\def\G{\Gamma}

\def\s{\sigma}
\def\vr{\varrho}
\def\vp{\varphi}
\def\N{\nabla}
\def\B{\Box}
\def\cA{\mathcal{A}}

\def\cC{\mathcal{C}}
\def\cD{\mathcal{D}}

\def\cF{\mathcal{F}}
\def\cG{\mathcal{G}}

\def\cJ{\mathcal{J}}
\def\cK{\mathcal{K}}
\def\cL{\mathcal{L}}
\def\cM{\mathcal{M}}
\def\cN{\mathcal{N}}

\def\cP{\mathcal{P}}

\def\cR{\mathcal{R}}
\def\cS{\mathcal{S}}
\def\cT{\mathcal{T}}
\def\cV{\mathcal{V}}

\def\ds{d_{\rm S}}
\def\dh{d_{\rm H}}
\def\dw{d_{\rm W}}

\def\p{\partial}
\def\bp{\bar{\partial}}
\def\bg{\bar g}
\def\bN{\bar\nabla}
\def\bB{\bar\B}
\def\bR{\bar R}

\def\tg{\tilde g}

\newcommand{\Eq}[1]{(\ref{#1})}
\def\com{\color{magenta}}
\def\cob{\color{blue}}
\newcommand{\book}[5]{\emph{#1}, #2, #3 #4 (#5)}
\newcommand{\oarX}[1]{\href{http://arxiv.org/abs/#1}{{\ttfamily\com #1}}}
\newcommand{\arX}[1]{\href{http://arxiv.org/abs/#1}{{\ttfamily\com arXiv:#1}}}
\newcommand{\doij}[5]{\href{http://dx.doi.org/#1}{\cob {\it #2} {\bf #3} (#5) #4}}
\newcommand{\doin}[6]{\href{http://dx.doi.org/#1}{\cob {\it #2} {\bf #3 #4} (#6) #5}}
\newcommand{\doinn}[5]{\href{http://dx.doi.org/#1}{\cob {\it #2} {\bf #3} (#5) #4}}
\newcommand{\ndoin}[6]{\href{#1}{\cob {\it #2} {\bf #3 #4} (#6) #5}}
\newcommand{\ndoinn}[5]{\href{#1}{\cob {\it #2} {\bf #3} (#5) #4}}
\newcommand{\tia}[1]{\textit{#1},}

\def\lp{\ell_{\rm Pl}}

\def\tp{t_{\rm Pl}}
\def\rme{e}
\def\rmd{d}
\def\rmi{i}

\begin{document}

\title{Multi-scale gravity and cosmology}

\author{Gianluca Calcagni}

\affiliation{Instituto de Estructura de la Materia, CSIC,\\ Serrano 121, Madrid, 28006 Spain}

\emailAdd{calcagni@iem.cfmac.csic.es}

\abstract{The gravitational dynamics and cosmological implications of three classes of recently introduced multi-scale spacetimes (with, respectively, ordinary, weighted and $q$-derivatives) are discussed. These spacetimes are non-Riemannian: the metric structure is accompanied by an independent measure-differential structure with the characteristics of a multi-fractal, namely, different dimensionality at different scales and, at ultra-short distances, a discrete symmetry known as discrete scale invariance. Under this minimal paradigm, five general features arise: (a) the big-bang singularity can be replaced by a finite bounce, (b) the cosmological constant problem is reinterpreted, since accelerating phases can be mimicked by the change of geometry with the time scale, without invoking a slowly rolling scalar field, (c) the discreteness of geometry at Planckian scales can leave an observable imprint of logarithmic oscillations in cosmological spectra and (d) give rise to an alternative mechanism to inflation or (e) to a fully analytic model of cyclic mild inflation, where near scale invariance of the perturbation spectrum can be produced without strong acceleration. Various properties of the models and exact dynamical solutions are discussed. In particular, the multi-scale geometry with weighted derivatives is shown to be a Weyl integrable spacetime.}



\maketitle


\section{Introduction}

Nature is irregular. The relative simplicity of quantum and atomic physics does not prevent the emergence of complex dynamics at scales where the fundamental degrees of freedom interact with the environment and with themselves to the point where a fundamental description of the system must be superseded by an effective one. For instance, the kinematics of ideal particles or elementary molecules is not sufficient, by itself, to provide an adequate description of the diffusion of a particle in a complex medium. Even ordinary Brownian motion, which is the simplest transport model going beyond kinetic theory, is inapplicable to the majority of diffusing systems, which show anomalous properties requiring a still higher level of sophistication \cite{bH,MeK00,Zas02,Sok12}. A whole toolbox of techniques borrowed from multi-fractal geometry \cite{Har01,Fal03},  transport, probability and chaos theories, statistical systems and anomalous diffusion, becomes then necessary. 

In recent years, many quantum-geometry models showed properties calling for this alternative but well-honed set of tools. The chief hint that quantum gravity has a complexity far beyond its first principles is \emph{dimensional flow}, namely the change of the dimensionality of effective spacetime with the probed scale \cite{tH93,Car09,fra1,Car10}. A canonical list of examples includes causal dynamical triangulations \cite{AJL4,BeH,SVW1}, asymptotically-safe quantum Einstein gravity \cite{LaR5,CES}, spin foams \cite{Mod08,CaM,MPM}, Ho\v{r}ava--Lifshitz gravity \cite{SVW1,CES,Hor3}, non-commutative geometry \cite{Con06,CCM,AA} and $\kappa$-Minkowski spacetime \cite{Ben08,ACOS}, non-local super-renormalizable quantum gravity \cite{Mod11}, spacetimes with black-hole \cite{CaG,Mur12,AC1}, fuzzy spacetimes \cite{MoN}, multi-fractal spacetimes \cite{fra1}, random combs \cite{DJW1,AGW} and random multi-graphs \cite{GWZ1,GWZ2}.

Dimensional flow is typically encoded in the running of the spectral dimension $\ds$. When it decreases to values lower than 4 at small scales, the ultraviolet (UV) properties of quantum gravity seem to improve. In most of the quantum-gravity approaches, the UV limit of the spectral dimension of spacetime is 2. To better understand the link between renormalizability, dimensional flow and the almost universal value $\ds\sim 2$, it is important to notice that the same value of the spectral dimension may appear in widely different theories without implying a physical duality among them. This is due to the fact that the spectral dimension is obtained from a diffusion equation, but different diffusion equations can give rise to the same correlation properties \cite{CES,frc4}. To remove this degeneracy, it is necessary not to limit the attention to the value of the spectral dimension and study the full solution of the diffusion equation associated with the given effective quantum geometry. 

This is a first step towards a fuller characterization of quantum geometries with the techniques of the above-mentioned alternative toolbox. To illustrate this and other applications, the framework of \emph{multi-scale spacetimes} has been developed. Geometry is described by a continuum in $D$ topological dimensions (for instance, $D=4$) where points do not contribute with the same weight to the measure. The usual Lebesgue measure $\rmd^D x$ is replaced by one with a non-trivial weight distribution,
\be\label{xv}
\rmd^D x\to\rmd^D x\,v(x)\,.
\ee
This weight $v(x)$ carries in itself a hierarchy of fundamental time-length scales $(t_1,\ell_1)$, $(t_2,\ell_2)$, \dots, which characterize the geometry. The most general measure is a `multi-fractal' distribution, possibly quite irregular and discontinuous. Assuming that $v(x)$ is smooth except at a finite number of singular points preserves most of the features of a multi-scale measure \cite{fra1,fra2,fra3} (see also \cite{HX1,HX2,STW} for applications), but to make practical advances it is convenient to make a further simplification and move to multi-scale factorizable measures \cite{frc6,frc7,frc8,frc10}, i.e., such that $v(x)=v_0(x^0)v_1(x^1)\dots v_{D-1}(x^{D-1})$. In this restricted class, particular importance is attached to multi-fractional \cite{ACOS,frc4,frc7,fra4,frc1,frc2,frc3,fra6,frc5,fra7,frc9} and log-oscillating \cite{ACOS,fra4,frc2} measures, which realize anomalous scaling and symmetry properties of `irregular' geometries in a most immediate manner.

Only one time or length scale is sufficient for the geometry to be multi-scale and, in fact, most quantum-gravity approaches fall into this simplest example. Consider the one-dimensional measure
\be\label{2}
v(x)=1+\left|\frac{x}{\ell_*}\right|^{\a-1}\,,
\ee
where $x$ scales as a length, $\ell_*$ is a characteristic length and $\a$ is some positive number smaller than 1. This measure is called \emph{binomial} since it is made of only two terms. The length $q(R)$ of an interval of size $R$ is not linear with $R$ but acquires an extra term, $q(R)=\int_0^R\rmd x\,v(x)=R+(\ell_*/\a)(R/\ell_*)^\a=\ell_*[(R/\ell_*)+(R/\ell_*)^\a/\a]$. For intervals much larger than $\ell_*$, the length is simply $q(R)\sim R$, while for `microscopic' intervals with $R\ll \ell_*$, $q(R)\sim R^\a$: the geometric properties and measurement units of the system change according with the scale. One can extend the discussion to $D$ Euclidean dimensions and calculate the volume $\cV^{(D)}(R)$ of a $D$-ball with radius $R$. Replicating eq.\ \Eq{2} with same $\a$ for all dimensions, one obtains
\be\label{dibal}
\cV^{(D)}(R)=\int_{D\textrm{-ball}}\rmd^Dx\, v(x)=\ell_*^D\left[\Omega_1 \left(\frac{R}{\ell_*}\right)^D+ \Omega_\a \left(\frac{R}{\ell_*}\right)^{D\a}\right]\,,
\ee
where $\Omega_{1,\a}$ are unit volume prefactors. Again, the volume of large balls scale as $R^D$, while the scaling of small balls is $\a$ times smaller. The $D$-ball scaling gives an operational definition of the Hausdorff dimension $\dh$ of spacetime (time is treated as a Euclidean coordinate). In the binomial example, $\dh\sim D\a$ for scales $R\ll\ell_*$, while at large scales one recovers the ordinary behaviour $\dh\sim D$.

The multi-scale framework can be regarded either as stand-alone or, when applied to some model of quantum gravity, effective. Its advantage is that one can carry an analytic study of classical and quantum mechanics and field theory on these geometries by combining the simplicity of well-known formalisms of theoretical physics with the lore of the alternative toolbox. Multi-fractal geometry and transport theory can characterize anomalous and quantum geometries with an extended set of topological and geometric indicators (Hausdorff, spectral and walk dimension \cite{frc4,frc1,frc2,frc7}; harmonic structures \cite{frc2}); probability theory and statistical-systems techniques label the same geometries even better through stochastic indicators (a natural random walker can be associated with a given geometry \cite{CES,frc4,frc7}); chaos theory and complex systems can provide transition mechanisms from small-scale discrete to large-scale continuous geometries \cite{fra4,frc2}; and so on. The price to pay is the loss of translation and Lorentz invariance at small scales and early times, which is however recovered in the infrared. Loss of Lorentz invariance is not typical of this setting, and there are other quantum-gravity frameworks (fundamental or phenomenological) where an intrinsic hierarchy of characteristics scales demands the abandoning of the principles of special relativity, for instance when the Planck length $\lp$ acts as a minimal distance \cite{AC1,Pad98,Pad99}.

So far, most of the results in the multi-scale framework have been obtained in the absence of gravity. The purpose of this paper is to include gravitation in the picture and study the cosmological consequence of having an anomalous geometry. Depending on the symmetries of the dynamics, we can distinguish various multi-scale theories: we will concentrate on three classes, respectively with ordinary, weighted, or `$q$' derivatives. For each multi-scale class of models under examination, we will define a gravitational action compatible with the non-trivial measure and differential structure. Then, we will specialize to the Friedmann--Lema\^{i}tre--Robertson--Walker (FLRW) metric of maximally symmetric (i.e., homogeneous and isotropic) spacetimes. Cosmology with a no-scale fractional measure was considered already in \cite{fra2,STW}. However, the presence of a hierarchy of scales in the measure gives rise to far richer scenarios. In \cite[section 6.2.5]{frc2}, it was briefly commented that this framework can solve the cosmological problems of the hot big bang model in a characteristic way. Here we present the details of that claim:
\begin{itemize}
\item \emph{Acceleration and cosmological constant.} First, one can get an effective dynamical cosmological constant (hence, acceleration) without invoking a slow-rolling scalar field or any exotic type of matter. The reason, explored in the theory with weighted derivatives (section \ref{graco4}), is that dynamics is reinterpreted under the multi-scale paradigm. In the Einstein equations, the profile $v(x)$ is \emph{fixed} by the requirement that spacetime possesses some of the characteristics of a multi-fractal (multiple scaling and, eventually, discrete scale invariance). Then, the measure profile determines the metric $g_{\mu\nu}(x)$. To make the equations of motion self-consistent, a particular potential $U(v)$ must be tailored. This is a conceptual leap from what done in standard field theory or cosmology, where the profile of a scalar field $\phi(x)$ is determined by the dynamics when an ad-hoc potential $W(\phi)$ is introduced. The rigidity of the profile $v(x)$ will strongly constrain the dynamics as well as the form of $U(v)$. The latter acts as a dynamical cosmological constant $\Lambda(x)\propto U[v(x)]$, which can fuel an accelerated cosmic expansion. The hierarchy of scales present in $\Lambda$ also appear throughout the dynamics and it can be constrained by independent observations, which establish upper bounds for the scales below which effects of anomalous geometry are tolerated. In such a way, a given measure profile in a given multi-scale theory will predict a specific evolution $\Lambda(x)$. The cosmological constant problem is thus reformulated and given a fresh insight. The evolution of the universe follows the rolling of $v$ down its potential $U$, from a point where $v(t)$ is non-trivial down to a minimum $U_{\rm min}\propto \Lambda_{\rm today}$ where $v\sim 1$ (ordinary geometry). Thus, the cosmological evolution reflects the multi-scale properties of the geometry and the details of dimensional flow, by which it is governed.
\item \emph{Alternative to inflation.} A second mechanism which can solve the traditional cosmological puzzles is simply based on the fact that the measure $v$ changes the behaviour of cosmological horizons, to the point where acceleration may no longer be needed. While in standard inflation the universe expands at an exponential rate and the proper Hubble horizon remains almost constant, in these alternative scenarios it is the Hubble horizon which drastically shrinks and expands because of the geometric effects. This may happen also in the absence of a potential $U$. In section \ref{cos}, we explore this possibility in a particular model with $q$-derivatives capitalizing on the following feature.
\item \emph{Early universe with discrete symmetry.} A most graphic illustration of the multi-scale framework, and a novelty in cosmology, is logarithmic oscillations. They reflect a discrete symmetry of the spacetime measure typical of deterministic fractal sets, a rescaling $x\to \la_\om^m x$ of the coordinates by (the integer power of) a fixed ratio $\la_\om$. This \emph{discrete scale invariance}, or DSI \cite{Sor98}, is present whenever the geometric structure of spacetime is postulated to be that of a self-similar fractal. At sufficiently large scales, log-oscillations are coarse-grained and spacetime becomes a continuum with conventional symmetries. However, at early times they deeply affect the evolution of the universe, which undergoes log-cyclic phases of contraction and expansion. Overall, these cycles let the universe expand so much that a traditional inflationary era is not necessary and it is either completely dispensed with or reduced to a very moderate acceleration (which we call `mild cyclic inflation'). This scenario is strikingly similar to emergent cyclic inflation (CI) \cite{Bis08,BiA,BiM,BMSh,BKM1,BKM2,DuB}, as we shall discuss in sections \ref{cos} and \ref{ci}.
\item \emph{Big bang problem.} Exact cosmological solutions can show a big bounce removing the initial singularity. Although the details differ from model to model, this is an exclusive consequence of the multi-scale geometry.
\item \emph{Miscellaneous mathematical results} will also be obtained. For example (section \ref{wist}), it comes as a surprise that the theory with weighted derivatives is, in fact, a case of \emph{Weyl integrable spacetime} (WIST) \cite{Wey52,NOSE,NoB08,PS}. Also, multi-fractional measures with the same anomalous scaling can be written in different forms, in particular by changing the position of their singularities. This corresponds to a change in the presentation of the theory rather than to a different geometric background \cite{frc1}. In the context of cosmology, it only amounts to an overall shift in the time-line of the universe, at least in the examples of section \ref{twdco}.
\end{itemize}

\subsection{Outline}

The plan of the paper is the following. After commenting on the existing literature of fractional cosmology and dynamical cosmological constants (section \ref{compa}), in section \ref{mink} we review multi-scale spacetimes and their different measures in a Minkowski embedding. Section \ref{graco} discusses, in most general terms, the conceptual novelties of the multi-scale paradigm in the presence of gravity. First (section \ref{coco0}), it is compared with scalar-tensor and unimodular theories, with which multi-scale models share some similarities. Technically and physically, all these classes of models differ from one another. Next, the generation of an inflation-like era and the cosmological constant problem are revised in the framework of multi-scale spacetimes (section \ref{coco1}). The double requirement that local frames should reproduce the multi-scale geometry everywhere (section \ref{graco1}) and that some modified notion of diffeomorphism invariance should survive (section \ref{graco2}) make the $q$-theory theoretically more pleasant than the models with ordinary and weighted derivatives, which have anyway many interesting properties. 

The gravitational actions and the equations of motion of three multi-scale theories are discussed in sections \ref{graco3}--\ref{graco5}. Section \ref{graco3} is mainly a review of the theory with ordinary derivatives \cite{fra2}, with some small extension. In section \ref{graco4}, we find that the theory with weighted derivatives is nothing but a Weyl integrable spacetime and we construct solutions where the big bang is replaced by a bounce and where the cosmological constant is reinterpreted as a purely geometry potential. Section \ref{graco5} is devoted to the $q$-theory, where the mechanism of log-oscillations is analyzed in detail. Conclusions are in section \ref{concl}.

The length and uneven technical level of the paper may not be palatable for the cosmologist interested in grasping the main physical features of the theory. We therefore suggest a first-reading pattern. The Minkowski setting of section \ref{mink} assumes no prior knowledge of the motivations and details of multi-scale measures and Lagrangians. This part is meant to provide the unfamiliar reader with a self-contained systematic introduction to the subject and summarize the status of each proposal. The different measures used in multi-scale spacetimes are motivated in section \ref{meas} with some basic arguments of fractal geometry. In cosmological applications, we will employ simplified versions of these measures, the binomial fractional measure given by eqs.\ \Eq{binom} and \Eq{qt} and the log-oscillating measure \Eq{logos}. However, the formalism works with arbitrary measures all the way down to the equations of motion and, in the case of the $q$-theory, also of their solutions. The replacement \Eq{xv} changes the kinetic terms in field-theory Lagrangians: section \ref{must} introduces these modifications for a scalar field. Section \ref{graco} is mainly theoretical and may be skipped, although sections \ref{coco0} and \ref{coco1} should be of interest for the reader acquainted with scalar-tensor theories. A quantitative notion of how non-trivial measures can affect the cosmic evolution is given in section \ref{twdco} for the theory with weighted derivatives (the only prerequisites being the action \Eq{eha11} and the equations of motion \Eq{ee4}) and the whole section \ref{graco5} for the $q$-theory.

\subsection{Comparison with other literature}\label{compa}

Spacetimes with non-integer or scale-dependent dimensions have made their sporadic appearance in the literature since the 1970s but, depending on the model, either progress had been technically difficult from the start or no attempt to a systematic construction of the physics (from classical static spacetimes to field theory and quantum mechanics, to gravity and cosmology) was made. In \cite{frc2}, these independent early works were discussed side by side with the more recent multi-scale spacetimes programme. Here we complete the comparison by looking at other proposals where fractional calculus was applied to cosmology. The reader wishing to catch all the technical details is advised to read this subsection after the rest.

Before the introduction of multi-scale spacetimes, a `fractional action cosmology' was obtained from Friedmann equations stemming from symmetry-reduced geodesic equations, where the measure profile was fixed to be \cite{El07a,El07b,El07c,El08}
\be\label{fac}
v(t)\sim t^{\a-1}\,. 
\ee
This phenomenological model was later developed in parallel with our programme \cite{El10a,El10b,Shc10,JRMRE,DCJ,DJC,El12a,Shc12} from a mini-superspace action. Attempts to extract the dynamics from covariant equations of motion \cite{El08,El10a} still considered a non-trivial measure only along the time direction, eq.\ \Eq{fac}. This measure is of fractional type (eq.\ \Eq{mea1} below) and it is neither associated with a \emph{multi}-scale geometry nor motivated by more than a mathematical curiosity of applying fractional calculus to cosmology. As far as the comparison can go, the dynamics of fractional action cosmology is much simpler than ours, and the multi-scale theory which most resembles it is the one with ordinary derivatives (section \ref{graco3}). The dynamical equations \Eq{00} and \Eq{trm} with $v$ given by eq.\ \Eq{fac}, first considered in \cite{fra2}, are somewhat similar (but not equivalent) to those of fractional action cosmology. Also, a time-dependent cosmological constant was noticed to be necessary to obtain self-consistent dynamics in certain situations \cite{El10a,El10b,Shc10,Shc12}, but this observation was not framed into a theoretical interpretation. Here we do have such a frame (still not clearly formulated in \cite{fra2}), which allows us to go beyond a pedantic treatment of dynamical solutions. Moreover, the measure profile we will work with in the theory with weighted derivatives (eq.\ \Eq{binom}) will be more realistic than eq.\ \Eq{fac}.

Phenomenological Friedmann equations with fractional derivatives were considered in \cite{El05,Rob09,Mun10}. Their derivation from a variational principle is unclear, although some formalization has been attempted \cite{Shc10,Vac10}. This framework is not conceptually unified and we do not know whether it can bear some resemblance with the cosmology of the multi-scale theory with fractional derivatives (see below), which we will not develop here.

A dynamical cosmological constant $\Lambda(t)$ is not a novelty of multi-scale spacetimes or of fractional action cosmology. Running couplings were obtained also in fractal-related cosmological models with variable dimension \cite{MN}. Their equations of motion are quite different from ours and a comparison is difficult. Also, in varying-speed-of-light (VSL) models \cite{Mag03} the speed of light $c\to c(x)$ becomes dynamical and it can provide an alternative mechanism to inflation \cite{Mof92,AlM,BaM1,BaM2,BaM3,Bar99}. Criticism to early VSL scenarios \cite{LSV,ElU} can be overcome in more recent formulations \cite{Mag03,Mag00,MBS,Ma08a,Ma08b,Ma08c,Mag10}. The particle horizon (i.e., the distance traveled by light since the big bang at, say, $t=0$) is modified in the early universe by a non-trivial geometric effect. In VSL models, the comoving particle horizon $r_{\rm p}=\int_0^t \rmd t'\,c(t')/a(t')$ is affected by a non-constant $c$, while in multi-scale spacetimes it is the integration measure to be modified: $r_{\rm p}=c\int_0^t \rmd t'\,v(t')/a(t')$. The main reason why one can dispense with standard inflation is apparently similar in both approaches, which in fact have many features in common \cite{frc8}. Still, there is a crucial difference: While the profile $c(t)$ is determined dynamically together with the scale factor $a(t)$, the measure $v(t)$ is fixed by construction with the tools of multi-fractal geometry and complex systems, and it dynamically determines $a(t)$ via Einstein equations (and, as in section \ref{graco4}, a geometric potential $U(v)$). This is the point where multi-scale theories depart conceptually from all models with a dynamical scalar, let them be varying-$c$, varying-$e$, or other scalar-tensor scenarios \cite{DeS}. 

Another example of an effectively varying $\Lambda$ can be found within the theory of quantum gravitation based on asymptotic safety \cite{LaR5,CES,Wei79,Reu1,Nie06,RSnax}, a powerful principle which strongly constrains, after a cut-off identification, the evolution of the gravitational field at early times and small scales. The renormalization-group-improved cosmological dynamics of this theory \cite{BoR1,ReS3,BoR3} is encoded in two Friedmann equations, which are the usual ones except for the replacements $\Lambda\to \Lambda(t)$ and $G\to G(t)$. A third equation makes it possible to have running couplings, $\dot\Lambda+8\pi\rho \dot G=0$, where $\rho$ is the energy density of matter. The resulting possibility to have an alternative acceleration mechanism without an inflaton was clearly recognized there, but the big bang problem still persisted. The latter, on the other hand, seems to be at least relaxed in multi-scale theories.

Nowhere in the above-mentioned scenarios do log-oscillations appear and play such a role as in multi-scale geometry, where they can provide both a modification of and an alternative mechanism to inflation.\footnote{In a non-perturbative model of asymptotic safety, the effective action is a log-oscillating function of the Ricci scalar, but de Sitter still is the simplest cosmological solution with acceleration \cite{Bon12}.} They also constitute an original realization of cyclic inflation, which will be compared with in section \ref{ci}.


\section{Multi-scale Minkowski spacetimes}\label{mink}

A \emph{multi-scale Minkowski spacetime} is the multiplet $\cM^D = (M^D,\,\vr,\,\p,\,\cK)$ specified by an embedding space (in this case, ordinary $D$-dimensional Minkowski spacetime $M^D$), a Lebesgue--Stieltjes measure $\vr$ for the action, a differential structure defined by some calculus rules with derivative operators $\p$, and a set of symmetries for the Lagrangian determined by how the operators $\p$ combine to give the Laplace--Beltrami operator $\cK$ in kinetic terms.

A generic Lebesgue--Stieltjes measure $\vr$ does not allow one to employ all the conventional tools of continuous geometry in mechanics and field theory. For instance, we would like to have a well-defined $D$-dimensional transform between position and momentum space, extend the multi-index covariant formalism to these geometries, calculate Noether currents, propagators, scattering amplitudes, and so on. For these technical reasons, the action measure is assumed to be of the form $\rmd\vr(x)=\rmd^Dx\,v(x)$, where the weight $v(x)$ is \emph{factorizable} in the coordinates and positive semi-definite:
\be
v(x)=\prod_{\mu=0}^{D-1} v_\mu(x^\mu)\,,\qquad v_\mu(x^\mu)\geq 0\,,
\ee
where the $D$ weights $v_\mu$ may differ from one another (`anisotropic' configuration). Thanks to factorizability, one can formally re-express the measure as a standard Lebesgue measure $\rmd\vr(x)=\rmd^D q(x)$, where
\be
q^\mu(x^\mu):=\int^{x^\mu}\rmd {x'}^\mu\,v_\mu ({x'}^\mu)
\ee
are $D$ distributions of the coordinates $x$. These distributions, which we call `geometric coordinates,' have anomalous scaling $q(\la x)=f(\la,x)$ under a dilation $x\to \la x$, contrary to the standard Lebesgue measure where $\rmd^D(\la x)=\la^D\,\rmd^Dx$.


\subsection{Examples of multi-scale measures}\label{meas}

Before introducing any dynamical field, we discuss the structure of the measure. In the absence of gravity, the differential structure of $\cM^D$ is not dynamical and the coordinate profile $v(x)$ is fixed. The requirement that geometry be multi-scale or, more specifically, multi-fractal, turns out to be rather stringent in determining $v$. In general, the measure $\vr$ possesses most or all the qualities that characterize a \emph{fractal}. The latter is a set of points endowed with some characteristics: 
\begin{enumerate}
\item[(i)] its structure is fine, i.e., one finds details (points) at any scale when zooming in the set; 
\item[(ii)] its structure is irregular, i.e., ordinary differentiability is given up; 
\item[(iii)] the spectral dimension $\ds$ of the set cannot exceed its Hausdorff dimension, $\ds\leq \dh$, and its walk dimension is related to the other two by $\dw=2\dh/\ds$.
\end{enumerate}
Other properties such as (iv) self-similarity or self-affinity and (v) having non-integer spectral or Hausdorff dimension are violated in many cases (for instance, random fractals are not self-similar; space-filling curves, the boundary of the Mandelbrot set, some diamond fractals and other sets have integer dimension; and so on).

A popular example of fractal is the middle-third Cantor set $\cC=\cS_1(\cC)\cup\cS_2(\cC)$, defined by two similarity maps $\cS_{1,2}$ acting on the interval $[0,1]$. For each point $x$ in the interval, $\cS_1(x)= (1/3)x$ and $\cS_2(x)=(1/3)x+2/3$. The coefficients in front of $x$ are called similarity ratios, which are equal to $\la_1=\la_2=1/3$ in this case. The set is \emph{self-similar} because smaller portions of it, of size $1/3$ with respect to the original, have the same structure of the whole.

Let us consider an instance of measure which is not multi-scale but, still, has anomalous scaling. Let $\cF$ be a deterministic fractal embedded in $\mathbb{R}_+\ni x$. Deterministic fractals are sets determined by some recursive mappings; self-similar fractals are a special case. Suppose we want to calculate the integral of a smooth function $f:\,\mathbb{R}\to \mathbb{R}$ over $\cF$. The symmetries of $\cF$ determine the Borel measure $\vr_\cF$ over which one integrates. Recasting the problem in Laplace momentum space and taking the large-momentum limit ${\rm Re}(p)\to +\infty$, one can show that calculus on these fractals is approximated by continuous fractional calculus, and \cite{RYS,YRZ,RYZLM,Yu99,QL,RQLW,RLWQ}
\be\label{appr}
\int_{\cF} \rmd\vr_\cF(x)\,f(x) \approx \int \rmd x\, \frac{x^{\a-1}}{\Gamma(\a)}\,f(x)\,,
\ee
where $\G$ is Euler's function. The integral on the right-hand side is a \emph{fractional integral} of real order $0<\a\leq 1$, corresponding to the Hausdorff dimension $\dh\sim\a$ of the set. Applications of real-order fractional integrals to complex and statistical systems have been mentioned elsewhere \cite{frc1,frc2}. Here we are only interested in the mathematical meaning of eq.\ \Eq{appr}. This approximation is a coarse-graining of the fine structure of the fractal by a randomizing process, where the oscillations typical of deterministic fractal kernels are canceled by mutual interference \cite{LMNN,NLM}. For self-similar fractals defined by some similarity maps $\cS_i$, this is nothing but a `continuum approximation' where one is taking the limit of infinitely refined similarities (i.e., infinitely small similarity ratios $\la_i$) \cite{frc1,frc2}. Thus, a fractional integral of real order represents either the averaging of a smooth function on a deterministic fractal, or a random fractal support.

Taking the Cartesian product of $D$ fractional measures, we obtain the first example of weight $v(x)$, describing spacetimes with a randomized structure and fixed Hausdorff dimension \cite{frc1}:
\be\label{mea1}
\text{fractional measure:}\qquad v_\a(x)= \prod_\mu v_{\a_\mu}(x^\mu)= \prod_\mu \frac{|x^\mu|^{\a_\mu-1}}{\Gamma(\a_\mu)}\,,
\ee
corresponding to $\rmd\vr_\a(x)=\rmd^D\,v_\a(x)$ and to the geometric coordinates
\be
q^\mu_\a(x^\mu):= \frac{{\rm sgn}(x^\mu)|x^\mu|^{\a_\mu}}{\Gamma(\a_\mu+1)}\,.
\ee
The scaling property $\vr_\a(\la x)=\la^{D\a} \vr_\a(x)$ (where $\a:=\sum_\mu\a_\mu/D$) immediately determines the Hausdorff dimension of this spacetime, $\dh =D\a$, which can be also obtained by self-similarity theorems \cite{frc1} or as the volume scaling of eq.\ \Eq{dibal}.

The second example generalizes the previous case to a geometry changing with the probed scale. For that, it is sufficient to sum over a finite number of exponents $\a$ \cite{frc2}, which we do in a way preserving factorizability \cite{frc6}:
\be\label{mea2}
\text{multi-fractional measure:}\qquad v_*(x)=\prod_\mu \left[\sum_{n=1}^N g_n\, v_{\a_n}(x^\mu)\right],
\ee
where the dimensionful coupling constants $g_n=g_n(\{\ell_n\})>0$ depend on a hierarchy of characteristic length scales $\ell_n$. Geometric coordinates are given by
\be
q^\mu_*(x^\mu):= \sum_{n=1}^N g_n\,q^\mu_{\a_n}(x^\mu)\,.
\ee
The Hausdorff dimension is now scale dependent, as one can easily show by computing the $D$-ball volume: depending on whether the radius is smaller or greater than each of the characteristic scales $\ell_n$, $\cV^{(D)}\sim R^{D\a_1},\,R^{D\a_2},\dots$, and so on. The simplest case of dimensional flow is $N=2$ (binomial measures such as eq.\ \Eq{2}), where there is only one characteristic scale $\ell_1=\ell_*$ defining large and small scales and both $\dh$ and $\ds$ vary monotonically \cite{frc2,frc4,frc7}. 

The third and last example stems from a relaxation of the approximation \Eq{appr}. On deterministic fractals, the heat kernel displays ripples with a logarithmic oscillatory pattern in diffusion time $\s$ (e.g., \cite{DIL,KiL,Tep05,LvF,Akk1,ABS,Kaj10}): $\cP(\s)=(4\pi \s)^{-\ds/2}\,F(\s)$, where $F$ is periodic in $\ln\s$ \cite{KiL,Kaj10}. As far as we known, these log-oscillations seem to be originated by the high degree of symmetry of these sets. The real-order approximation corresponds to averaging over a log-period, so that only the zero mode of the harmonic expansion of the measure survives. Including also the other frequencies, one gets a multi-fractional measure with complex exponents \cite{NLM}: for each harmonic $\om>0$, $v_\a\to v_{\a,\om}= c_+ |x|^{\a+\rmi\om-1}+c_- |x|^{\a-\rmi\om-1}$. Summing over all possible $\a$ and $\om$ and imposing the measure to be real by a suitable choice of the complex coefficients $c_\pm$, we obtain a multi-scale spacetime akin to deterministic multi-fractals, endowed with the measure weight \cite{frc2,frc6}
\bs\label{log}\ba
&&\text{log-oscillating measure:}\qquad \bar v(x):=\prod_\mu\left[\sum_{n,l} g_{\a_n,\om_l} v_{\a_n}(x^\mu)\,F_{\om_l}(\ln |x^\mu|)\right],\\
&&\qquad
F_{\om}(\ln |x|):= 1+A_{\a,\om}\cos\left[\om\ln\left(\frac{|x|}{\ell_\infty}\right)\right]+B_{\a,\om}\sin\left[\om\ln\left(\frac{|x|}{\ell_\infty}\right)\right],
\ea\es
where $g_{\a_n,\om_l}>0$, $A$ and $B$ are real, and $\ell_\infty$ is a fundamental length introduced for dimensional reasons and which can be identified with the Planck length \cite{ACOS}. This scale is smaller than the multi-scale hierarchy of the previous example, $\ell_\infty<\ell_n$ for all $n$. Geometric coordinates are $\bar q^\mu(x^\mu)=\int\rmd x^\mu\,\bar v_\mu(x^\mu)$. For each mode $\om$, the oscillatory part of the measure is invariant under the discrete dilation symmetry
\be
x\,\to\, \la_\om^m x\,,\qquad \la_\om=\exp\left(\frac{2\pi}{\om}\right)\,,
\ee
where $m$ is integer. This symmetry is called discrete scale invariance and appears also in chaotic systems and transport models \cite{Sor98,HOSS,StS}. All deterministic fractals possess a DSI. For example, smaller copies of the middle-third Cantor set $\cC$ have a fixed size with respect to larger ones, governed by the self-similarity ratio $\la_\om=1/3$. Only this predetermined zoom rescaling, and no one else, characterizes the set.


\subsection{Multi-scale theories}\label{must}

A spacetime equipped with a multi-scale measure with the above properties is not necessarily a fractal. That depends on the symmetries imposed on the Lagrangian. For any given measure weight $v(x)$, one can formulate inequivalent versions of multi-scale spacetimes. For instance, consider the action for a real scalar field:
\be\label{scafi}
S=\int\rmd^Dx\,v(x)\left[\frac{1}{2}\phi\,\cK\,\phi-W(\phi)\right],
\ee
where $W$ is the potential. The symmetries of $\cL$ determine the Laplace--Beltrami operator $\cK$. We consider four physically inequivalent theories classified according to the type of derivatives appearing in $\cK$. To this purpose, we first introduce the generic weighted derivative
\be\label{bD}
{}_\b\cD:=\frac{1}{v^\b}\p[v^\b\,\cdot\,]\,,
\ee
where $\b$ is an arbitrary constant whose value will change according to the situation. The cases $\b=1/2,1$ will be of special interest and, to avoid too many subscripts,  will be reserved the symbols $\cD:={}_{1/2}\cD$ and $\check{\cD}:={}_{1}\cD$, in agreement with the notation of \cite{frc6}.
\begin{enumerate}
\item \emph{Ordinary derivatives} \cite{fra2}:
\be\label{K1}
\cK^+:=\B=\eta^{\mu\nu}\p_\mu\p_\nu\,, \qquad (\cK^+)^\dagger =\B^\dagger=\check{\cD}^2=\frac{1}{v}\,\p_\mu\p^\mu\left[v\,\cdot\,\right]\,,
\ee
where $\eta={\rm diag}(-,+,\cdots,+)$ is the Minkowski constant diagonal metric. The Lagrangian $\cL$ is invariant under ordinary Poincar\'e transformations (since we will move to curved manifolds later, we replace spacetime indices $\mu,\nu$ with frame indices $I,J$ in the transformation law, thus stressing the fact that these are frame symmetries):
\be\label{xLax}
{x'}^I=\Lambda_J^{\ I}x^J+X^I\,,
\ee
where $\Lambda_J^{\ I}$ are the Lorentz matrices, acting in internal space, and $X^I$ is a constant vector. The differential-forms formalism used here is the usual one.

The operator $\cK^+$ is not self-adjoint with respect to the natural scalar product with measure weight $v$, but for a symmetric kinetic term it makes no difference, since $\vp\cK^+\vp\to\vp(\cK^+)^\dagger\vp$ when integrating by parts. However, the ordering of the operators in the Lagrangian is important and $\cK^+$ is not the square of a self-adjoint operator (quantum mechanics and quantum field theory are difficult to formulate here for this reason). Consequently, the kinetic term $-(1/2)\p_\mu\phi\p^\mu\phi$ is inequivalent to the one in eq.\ \Eq{scafi} and suggests to define another operator (similar to the covariant Laplacian)
\be\label{k-}
\cK^-:=\frac{1}{v}\,\p_\mu\left[v\p^\mu\,\cdot\,\right]=\check{\cD}_\mu\p^\mu\,.
\ee
This operator is self-adjoint but, again, is not a square. Notice that the model with \Eq{k-} is nothing but the covariant action on a background metric $\hat g_{\mu\nu}=v^{2/D}\eta_{\mu\nu}$. The conformal factor of the background encodes the different geometry of the effective spacetime. In the presence of gravity, a reinterpretation of the action via conformal rescalings of the metric will be more delicate.
\item \emph{Weighted derivatives} \cite{frc3,frc6}:
\be\label{Kv}
\cK_v=\cD^2=\frac{1}{\sqrt{v}}\,\p_\mu\p^\mu\left[\sqrt{v}\,\cdot\,\right]=\cK_v^\dagger\,.
\ee
Contrary to \Eq{k-}, this operator is both self-adjoint and the square of a first-order weighted derivative, a property necessary to define the momentum operator along each direction. In general, all the derivatives \Eq{bD} will appear in this theory, the weight depending on the field density it acts upon. Equation \Eq{Kv} is valid for the example \Eq{scafi} of a scalar-field density. For a vector density, again $\beta=1/2$ \cite{frc8}, but for bilinears such as the energy momentum tensor the weight is $\beta=1$ \cite{frc6}. As in the theory with ordinary derivatives, differentials are the usual ones, although a weighted version can be constructed. The Lagrangian $\cL$ is not invariant under ordinary Poincar\'e transformations, but in the free case symmetries are generated by the ordinary Poincar\'e algebra, where the momentum and Lorentz operators are weighted versions of the ordinary Poincar\'e generators \cite{frc6}. This model is dual to the previous one in the sense that they have exactly the same spectral dimension. Although they are multi-scale, none of them is a fractal, since $\dw\neq 2 \dh/\ds$ \cite{frc7}.
\item \emph{$q$-derivatives}:
\be\label{Kq}
\cK_q=\B_q:=\eta^{\mu\nu}\frac{\p}{\p q^\mu(x^\mu)}\frac{\p}{\p q^\nu(x^\nu)}=\eta^{\mu\nu}\frac{1}{v_\mu}\p_\mu\left[\frac{1}{v_\nu}\p_\nu\,\cdot\,\right]=\cK_q^\dagger\,.
\ee
The Lagrangian is formally identical to the usual one but the theory is non-trivial because physical momenta are conjugate to $x$, not $q$. This is, in fact, another way to state that geometric coordinates are anomalous under rescaling. The reason why one would consider this momentum choice rather than the one where $q$ is the canonical variable is simply the adoption of the multi-scale \emph{Ansatz}, implemented according to the symmetry imposed on the action. Both the measure and the Lagrangian separately possess $q$-Poincar\'e symmetries, i.e., Poincar\'e transformations on the geometric coordinates \cite{frc1,frc2}:
\be\label{qLaq}
q^I({x'}^I)=\Lambda_J^{\ I}q^J(x^J)+Q^J\,,
\ee
which are non-linear, non-invertible transformations of the coordinates $x$ \cite{frc1,frc2}. These spacetimes can be regarded as fractal, since their dimensions are related as in fractals by $\dw= 2 \dh/\ds$ \cite{frc7}. Despite the simplicity of the model in $q$ position space, it predicts new physical phenomena, as for instance the log-oscillating imprint in the observed cosmic microwave background spectrum \Eq{spec2}.
\item \emph{Multi-fractional derivatives} \cite{fra6,frc4}: 
\be
\cK_*=\sum_{n=1}^N g_n\cK_{\g_n}\,,\qquad \cK_\g:=-\frac{1}{2\cos(\pi\g)}\eta^{\mu\nu}({}_\infty\p^\g_\mu {}_\infty\p^\g_\nu+{}_\infty\bp^\g_\mu {}_\infty\bp^\g_\nu)=\cK_\g^\dagger\,.
\ee
Here the measure weight is ordinary ($v=1$), and $\dh=D$. All the anomalous scaling is transferred into the derivatives, which are fractional. The Laplace--Beltrami operator is split into a `left' part featuring the Liouville derivative $({}_\infty\p^\g f)(x) :=\Gamma^{-1}(m-\g)\int_{-\infty}^x\rmd x'\,(x'-x)^{m-1-\g}\p_{x'}^m f(x')$ and a `right' part with the Weyl derivative $({}_\infty\bp^\g f)(x) :=\Gamma^{-1}(m-\g)\int_x^{+\infty}\rmd x'\,(x'-x)^{m-1-\g}(-\p_{x'})^m f(x')$, where $m-1<\g\leq m$ and $m$ is a positive integer. Only a combination of the two sectors can give a self-adjoint operator \cite{frc1,frc4}. The differential structure clearly differs from the one of the other models, since here the fractional exterior derivative \cite{Tar12,CSN1} is employed \cite{frc1}. 
\end{enumerate}


\subsection{Status}

Table \ref{tab1} summarizes the status of each of the four models of multi-scale spacetime. In general, for all of them we have a rigorous definition of the space, the norm, and the differential structure \cite{frc1,frc2}, the geometry has a detailed stochastic characterization in terms of diffusive processes \cite{fra6,frc4,frc7}, we know both the Hausdorff and the spectral dimension and we have analytic control over dimensional flow \cite{frc4,frc2,frc7}. A discrete-to-continuum transition happens in log-oscillating geometries, where a hierarchy of scales emerges \cite{fra4,frc2}. Moreover, connections with independent models of quantum gravity have been establishes, especially with non-commutative spacetimes \cite{ACOS}, asymptotic safety \cite{fra7}, Ho\v{r}ava--Lifshitz gravity \cite{fra7}, and varying-$e$ and VSL models \cite{frc8}. However, many details remain to be explored, including renormalizability of some of the theories, the gravitational sector and cosmology. The focus of this paper is on these last two aspects. The model with fractional derivatives, which is also the least explored, will be left out from the discussion and dealt with elsewhere.
\begin{table}[ht]
\centering
\begin{tabular}{|l||c|c|c|c|}\hline
                           & $\cK^\pm$   & $\cD^2$          		 & $\B_q$ 					& $\cK_*$  \\\hline\hline
Momentum transform         & \ding{55}? \cite{frc3}  & \ding{51} \cite{frc3}      		 & \ding{51}? 			& \ding{51}? \\\hline
Non-relativistic mechanics & \ding{51} \cite{frc7}   & \ding{51} \cite{frc5}      		 & \ding{51}  			& ? \\\hline
Relativistic mechanics     & \ding{51}               & \ding{51} \cite{frc10}          & \ding{51} \cite{frc10}  			& ? \\\hline
Perturbative field theory & \ding{51}? \cite{fra1,fra2,fra3} & \ding{51} \cite{frc6,frc9} 		 & \ding{51}?   					& \ding{51}? \\\hline
Symmetries and dynamics of & 												 &								  		 &     	 					  &			 \\
scalar (Q)FT							 & ? \cite{fra2}             & \ding{51} \cite{frc6,frc9} 	   & \ding{51} \cite{frc6,frc9} & ?    \\\hline
Scalar QFT propagator      & ? \cite{fra2}				   & \ding{51} \cite{frc6} 				   & \ding{51}? 			&\ding{51}? \\\hline
Electrodynamics						 & ?  										 & \ding{51} \cite{frc8} 					 & \ding{51} 			  & ? \\\hline
Perturbative renormalizability          & ?  										 & \ding{55} \cite{frc9} & \ding{55} \cite{frc9}  & \ding{51}? \\\hline
Avoids Collins	{et al.}   & ? 											 & \ding{51} \cite{frc9} & \ding{51} \cite{frc9}  & ? \\\hline
Phenomenology (obs.\ constraints) & ?  & \ding{51} \cite{frc2,frc8} & ? & ? \\\hline
Gravity                    & \ding{51} \cite{fra2} 						  & ? & ? & ? \\\hline
Cosmology                  & \ding{51}? \cite{fra2}					  & ? & ? & ? \\\hline
\end{tabular}
\caption{\label{tab1} Status of multi-scale theories with ordinary ($\cK^\pm$), weighted ($\cD^2$), $q$- ($\B_q$) and multi-fractional ($\cK_*$) Beltrami--Laplace operators. For each entry, main references are given where an advanced analysis has been carried out. Trivial analyses with no particular reference are indicated only by their outcome (negative or problematic: \ding{55}; positive or neutral: \ding{51}). Aspects yet to be studied or incomplete are marked with a question mark, possibly with an indication of the expected outcome. The entry `Avoids Collins	{et al.}' refers to the enhancement of Lorentz-symmetry violation by quantum loop effects, which may happen in theories where the classical dispersion relation is modified \cite{CPSUV,CPS}. Both the weighted- and $q$-derivative models do not to suffer from this problem.}
\end{table}

In the present formulation of multi-fractional models, there is no guiding principle fixing the characteristic time-space scales in the measure. (The only exception is the length $\ell_\infty$ appearing in log-oscillating measures, identified with the Planck length \cite{ACOS}.) However, the latter can be constrained by observations. The most stringent constraints are for the case of a binomial measure, i.e., when the hierarchy of scales is only made by one characteristic time $t_*$ and spatial size $\ell_*$ \cite{frc2,frc8}. Various experiments of particle and atomic physics and astrophysics give independent upper bounds on the scales at which deviations from four spacetime dimensions may become appreciable. Some of these constraints have been worked out in the 1980s in toy models of dimensional regularization with fixed anomalous dimension, which we can recycle because the effect of small deviations from $\dh=4$ is about the same of multi-fractional geometries \cite{frc1}. Particle-physics experiment loosely suggest that no dimensionality effects occur at energies $M> 300\div 400~ {\rm GeV}$ \cite{She09}, roughly corresponding to an upper bound $\ell_*<10^{-18}\,\mbox{m}$. A stringent and more robust bound comes from the Lamb shift for the hydrogen atom, for which $|\dh-4|<10^{-11}$ at scales $\ell\sim 10^{-11}\,{\rm m}>\ell_*$ \cite{ScM,MuS}. The effectiveness of dimensional bounds at astrophysical and cosmological scales is weaker at larger and larger distances, up to the point where the cosmic microwave black-body spectrum only requires $|\dh-4|<10^{-5}$, corresponding to scales $\ell\sim 14.4\,\mbox{Gpc}$ \cite{CO}. Variation of the fine-structure constant during big-bang nucleosynthesis roughly limits the time scale in the binomial measure as $t_*<0.3$\,s \cite{frc8}. In general, the dimensionality of spacetime changes also with time, which requires a greater care in the interpretation of experiments performed at widely different cosmological scales. Incidentally, the present study aims also to introduce the tools to extract more precise phenomenology from gravitational multi-scale models (the above cosmological bounds do not include multi-scale and gravity effects properly).


\section{Multi-scaling gravity}\label{graco}


\subsection{Multi-scale paradigm versus scalar-tensor and unimodular theories}\label{coco0}

A theory of gravitation based on multi-scale spacetime should encode three main conceptual landmarks. First, we would like to obtain what is expected in a `covariant' description of a fractal or any other anomalous geometry with multi-scaling: namely, a non-trivial geometric \emph{and differential} structure at all points. Vielbeins, the frame vectors mapping a local inertial frame to another and curvilinear coordinate systems to local inertial frames, should move the measure around and maintain the anomalous scaling properties in all frames. Second, one should be able to describe a sensible phenomenology (including cosmology) in an economic and self-consistent way.

Third, it is clear that the geometry of multiscale manifolds is not going to be Riemannian: apart from the metric, which is determined dynamically, it possesses a measure structure which, in the absence of gravity, is given \emph{a priori}. Can we still regard this structure as non-dynamical when gravity is turned on? The answer is Yes, but delicate. We begin with general remarks which apply to all multi-scale theories, later specializing to specific cases.

\

\noindent {\bf Is the measure dynamical? Comparison with scalar-tensor models.} Just as in ordinary field theory, in multi-scale theories the equations of motions \cite{frc2,frc6,frc8} stem from the variation of the gravitational-matter action $S[g,v,\psi^i]$ with respect to all dynamical fields:
\be\label{eomge}
\Sigma_{\mu\nu}(g,v,\psi^i):=\frac{\de S[g,v,\psi^i]}{\de g^{\mu\nu}}=0\,,\qquad f_i(g,v,\psi^i):=\frac{\de S[g,v,\psi^i]}{\de \psi^i}=0\,,
\ee
which depend on the metric and measure structures as well as on the matter fields collectively called $\psi^i$. The multi-scale paradigm fixes once and for all the coordinate profile $v(x)$ as one of the distributions \Eq{mea2} and \Eq{log}. This profile does not change while the system evolves dynamically: it is simply fixed \emph{ab initio} by demanding that the integro-differential structure of the problem be determined by the lore of multi-fractal geometry briefly recalled in section \ref{mink}. The profile $v(x)$ is not a passive spectator either: through the equations of motion \Eq{eomge}, it determines the dynamical profiles $g_{\mu\nu}(x)$ and $\psi^i(x)$. (For the time being, we assume these equations can be consistently solved.) Thus, the function $v(x)$ is \emph{not} a scalar field, it affects the dynamics although it is not dynamical by itself. In particular, one does not consider the variation $\de S/\de v=0$, exactly as one does not treat friction terms in dissipative systems as dynamical \cite{frc2}.

We should also notice \cite{fra1} that a change in the measure is consistently accompanied by a new definition of functional variations, Dirac distribution, Poisson brackets, Lagrangian symmetries, momentum space, the line element and so on, in turn leading to an unfamiliar propagator and the deformation of the Poincar\'e algebra (see, e.g., \cite{frc2,frc6,frc10}). No such modifications occur in scalar-tensor theories where gravity is non-minimally coupled to a Lorentz scalar. The discussion on diffeo invariance of section \ref{graco2} should further convince the reader that there are heavy theoretical differences with respect to scalar-tensor theories: the latter are diffeo invariant, multi-scale theories in general are not, at least in the ordinary sense.

These considerations hold for all multi-scale theories. Next, we can make specific remarks for each multi-scale model separately. We begin with the theories with ordinary and weighted derivatives, which are those that most resemble scalar-tensor theories. In the absence of matter, we will see that, in the theories with ordinary and weighted derivatives, the equations of motion \Eq{eomge} can be written in the form
\be\label{gt}
\Sigma_{\mu\nu}=\cG_{\mu\nu}-\cT^v_{\mu\nu}=0\,,
\ee
where $\cT^v_{\mu\nu}$ is a contribution mostly dependent on the measure weight (what would correspond to the energy-momentum tensor of a scalar in a scalar-tensor theory) and $\cG_{\mu\nu}=\mathfrak{R}_{\mu\nu}-\tfrac12 g_{\mu\nu}\mathfrak{R}$ is the Einstein tensor or a model-dependent modification of it, where $\mathfrak{R}_{\mu\nu}$ and $\mathfrak{R}$ obey the same properties of the usual Ricci tensor $R_{\mu\nu}$ and scalar $R$. Equation \Eq{gt} is $D(D-1)/2$ equations determining the components of the metric. Thus, the profile $v(x)$ is dynamically found even without considering its would-be equation of motion $\de S[g,v]/\de v=0$. This redundancy happens because of the contracted Bianchi identities $2\N^\mu \mathfrak{R}_{\mu\nu}=\N_\nu \mathfrak{R}$ ($\N$ is the covariant derivative later to be defined). Taking the divergence of eq.\ \Eq{gt} with respect to $\N^\mu$, the Bianchi identities impose $\N^\mu \cG_{\mu\nu}=0$ and, consequently, $\N^\mu\cT^v_{\mu\nu}=0$. Since $v$ is a rank-0 tensor density and these $D$ equations are not independent, $\N^\mu\cT^v_{\mu\nu}=f(g,v,\p v) \N_\nu v=0$. One thus obtains the equation of motion $f(g,v,\p v)=0$ for the measure weight. Thus, there is no incompatibility with the notion that $v$ is not a dynamical field and the existence of a `Klein--Gordon' equation $\de S/\de v =0$. The main assumption, of course, is that the Bianchi identities hold, but such is the case in these theories. For ordinary derivatives, the action is \Eq{Sg2} and eq.\ \Eq{gt} is \Eq{ee2}, while for weighted derivatives the action and equations of motion are eqs.\ \Eq{eha} and \Eq{ee3}.

In order to get consistent solutions with viable properties (for instance, with ordinary matter content, or giving a certain cosmological evolution), in general it is necessary to include a non-vanishing `potential' term $U(v)$, and sometimes also what one would call a `kinetic' term $(\p v)^2$ for $v$. Both contributions are simply functional coordinates profiles introduced into the Lagrangian. This is the point where there is a major departure from scalar-tensor theories. If we regarded, as in ordinary scalar-tensor theories \cite{DeS}, $v(x)$ as a dynamical Lorentz scalar field
\be\label{vV}
v(x)\to V(x)\,,
\ee
all the models with non-trivial weight would be automatically Lorentz invariant in the usual sense, with dilaton-like actions where $V$ is non-minimally coupled with all the other fields. In particular, the generic profile $V(x)$ would no longer be factorizable and positive. These would no longer be multi-scale models in the sense of section \ref{mink} because the profile $V(x)$ would be solely determined by the dynamics for a given potential $U(V)$, and the scale hierarchy in $v(x)$ would be lost in the most general case. In fact, no such hierarchy would exist \emph{a priori} in solutions if one does not introduce it from first principles, and there would be no geometric and physical motivation why $V(x)$ should take a fractional polynomial or log-oscillating form as in eqs.\ \Eq{mea2} and \Eq{log}. If, on the other hand, we impose multi-scaling and one of the profiles of section \ref{meas}, the only way to accommodate the multi-scale \emph{Ansatz} within the gravitational dynamics is to fix the potential $U(v)$ to make them mutually consistent. In other words, we propose to change the dynamical problem into a problem of reconstruction.

Clearly, the technical difficulty of this problem is no different than having a genuine scalar field $V(x)$ with given potential and then getting $V(x)$ from the equations of motion. However, one should not overlook that the physical interpretation of multi-scale theories is radically different from a scalar-tensor theory, which results in a very characteristic cosmic evolution. In the multi-scale case, Lorentz invariance is broken because the texture of spacetime is assumed to be non-standard, following integration rules dictated by multi-fractal geometry. This bears an important consequence. By eliminating $U$ from the equations of motion, the choice of a fixed profile $v(x)$ will determine the gravitational dynamics (i.e., a solution $g_{\mu\nu}(x)$) univocally, in a way which scalar-tensor theories could not reproduce in general. This means that, if the equations of motion can be solved consistently, the resulting physics is completely determined by the multi-scaling and it can be tested experimentally.

The theory with $q$-derivatives is even more strikingly different from a scalar-tensor model: in that case, we will not even need `kinetic' or `potential' terms for $v$ to get consistent solutions. The equations of motion can be written in the form \Eq{gt} but with $\cT^v_{\mu\nu}=0$, and they are quite dissimilar from the scalar-tensor ones (see eqs.\ \Eq{Sgq} and \Eq{eeq}).

\

\noindent {\bf Comparison with unimodular gravity.} The situation created by a non-dynamical geometric structure may be remindful of theories of \emph{unimodular gravity}, where it is assumed that some of the metric components are non-dynamical. This can be implemented either as a partial gauge fixing of general coordinate transformations preserving the volume (so that $\sqrt{-g}$ transforms as a scalar rather than a density; in particular, $\sqrt{-g}=1$) \cite{AnF,Ray79,vvN,Zee85,BD88,BD89,Unr89,UW89,Ngv,Pet91,Alv05,ABGV,AFL,FiGa,AlV,BSZ,Eic13}, or preserving (in full or almost) general covariance but demanding $\sqrt{-g}$ to be non-dynamical \cite{HeT89,Wei89}, or replacing the volume weight $\sqrt{-g}$ with a scalar with some internal symmetries and no kinetic term \cite{GuK1,GuK2,Gue99,GuK3}. At the classical level, the dynamics of unimodular gravity is equivalent to general relativity or, when $\sqrt{-g}$ is not globally fixed to 1, to a scalar-tensor theory \cite{Wei89}. In these scenarios, the cosmological constant $\Lambda$ is then reinterpreted as an integration constant. Classically, this does not solve the cosmological constant problem, although in some models $\Lambda=0$ by the symmetries of the theory \cite{AlV} and dark energy is explained by a dynamical dilaton field \cite{BSZ}. Interesting physics emerges at the quantum level, where unimodular gravity deviates from the structure of general relativity. In Hamiltonian formalism, the wave-function of the Universe becomes a superposition of states with different values of the cosmological constant, which allows for a probabilistic reinterpretation of the $\Lambda$ problem. 

Multi-scale theories, especially that with ordinary derivatives, are somewhat akin to these models where the geometry of general relativity is modified from first principles. In this respect, our proposal has some antecedents and is not conceptually shocking. In particular, in the theory with weighted derivatives an extra consistency equation appears just like in these models. Still, there are notable differences, the main one being the fact that we are adding (or, more precisely, modifying) non-dynamical structure to the geometry rather than making some components of $g_{\mu\nu}$ non-dynamical. In fact, while one may regard the profile $v(x)$ in the action as a new non-dynamical structure, it is more economic to consider the non-dynamical structure as a multi-scale generalization of something which is already present for the start in standard general relativity, namely, the integro-differential structure of the action and, in particular, the coordinates charting \cite{frc1}. The metric structure is independent at the action level and all the $g_{\mu\nu}$ components are dynamical. This, together with a precise prescription for the non-dynamical part dictated by multi-fractal geometry, makes the dynamics of multi-scale theories quite dissimilar from previous ideas in this directions.\footnote{In particular, the point of view that the measure is dynamical because determined by the equations of motion of the other fields \cite{GuK1,GuK2,Gue99,GuK3} is not tenable in our case, as explain above.} 


\subsection{Acceleration and cosmological constant problem}\label{coco1}

To summarize, whether the measure is a Lorentz scalar or not splits theories with ordinary or weighted derivatives into two mutually exclusive formulations. The first case, where $v\to V$ is a dynamically determined Lorentz scalar and its potential $U(V)$ is fixed \emph{a priori}, is akin to traditional scalar-tensor theories. In the multi-scale formulation, on the other hand, one maintains both the scale hierarchy and the original motivation by multi-fractal geometry, the profile $v(x)$ is fixed \emph{a priori} and the potential $U(v)$ is tailored from one of the profiles discussed in section \ref{meas} and the resulting dynamics \Eq{gt}. This places the cosmological constant problem under a novel perspective.\footnote{On the other hand, this change of perspective where a geometric \emph{Ansatz} determined by some first principles governs the dynamics is no different from what attempted elsewhere. For instance, in asymptotic safety the cosmological and Newton constants are assumed to scale in a certain way compatible with the existence of a UV non-Gaussian fixed point \cite{BoR1}.} In fact, the quantity $U(v)$ is determined for self-consistency by the dynamics by keeping the multi-scale profile $v$ fixed. This is nothing but a spacetime-dependent cosmological constant 
\be\label{Lambda}
\Lambda(x,\{\ell_n\})\equiv \frac{1}{2\k^2}U[v(x,\{\ell_n\})]\,,
\ee
featuring the characteristic scales $\{\ell_n\}$ of the system. Thus, the energy scale of the effective $\Lambda$ is determined by the structure of the measure. At early times, we will see that there is the possibility to obtain an accelerating evolution without adding inflaton-like matter; this phase depends on the scales $\ell_n$. At late times, the measure weight tends to 1, meaning that the observed cosmological constant can be accounted for by this model only if $U(v=1)\sim H_0^2$, where $H_0$ is the Hubble parameter today. We will confirm this later (eq.\ \Eq{umin}). However, the metric structure is independent of the measure structure, which means that the constant $H_0$ is independent of the scales $\ell_n$. Thus, the cosmological constant problem can be reinterpreted but not solved. Still, the rigidity of the multi-scale \emph{Ansatz} allows one to constrain the scales $\ell_n$ against experiments and, through them, to test the prediction for the whole evolution of the universe, including possible early-time acceleration stages and the cosmological constant \Eq{Lambda}. The present paper also aims to begin such a study and check whether this reconstruction programme is feasible.

In a minimal formulation of the $q$-theory, the cosmological constant problem does not find a natural relaxation, since there is no immediate need to introduce a term $U(v)$ in the action to obtain consistent solutions. On the other hand, inflation can be replaced by alternative mechanisms where the hot-big-bang problems are tackled under a different perspective, as we shall see in section \ref{graco5}.


\subsection{Frames}\label{graco1}

The metric and measure structures are related to each other by their interplay under coordinate transformations. By that, we can already determine which of the multi-scale models will have a natural definition of local frames. Contrary to other proposals of non-Riemannian manifolds \cite{Pia12}, here we deal with metric spaces where a differential line element can be defined.

To find the metric, we follow the procedure of \cite{CSN1}, as in \cite{frc1}. Consider two coordinate systems $\{x^I\}$ and $\{{x'}^\mu\}$, the first (denoted with capital Roman indices) being the Cartesian system and the second a generic curvilinear one. The ordinary exterior derivative $\rmd$ must be coordinate invariant, giving $\rmd x^I \p_I=\rmd=\rmd {x'}^\mu\p_\mu$. Applying this relation to $x^J$, we get $\rmd x^J = \rmd {x'}^\mu \bar e_\mu^{\ J}$, where
\be\label{bare}
\bar e_\mu^{\ J}:=\frac{\p x^J}{\p {x'}^\mu}
\ee
is the usual $D\times D$ vielbein matrix (the inverse is such that $\bar e_{\ I}^\mu \bar e_\mu^{\ J} = \delta_I^J$). The multi-scale Jacobian $\cJ$ is simply $\bar\cJ=|\det \bar e|$ times the ratio of measure weight factors. In fact, the measure transforms as
\be\label{mes}
\rmd\vr(x)=\rmd\vr(x')\, \cJ(x')\,,\qquad \cJ(x')=\frac{v[x(x')]}{v(x')}\,\bar\cJ(x')\,.
\ee

If we defined the metric to coincide with the usual one, $\bar g_{\mu\nu}:= \eta_{IJ} \bar e_\mu^{\ I} \bar e_\nu^{\ J}$, we would soon meet a problem. In ordinary spacetime, the line element is
\be\label{ords}
\rmd \bar s^2:=\bar g_{\mu\nu}\rmd x^\mu \otimes \rmd x^\nu\,.
\ee
In finite form and space-like separation ($\Delta t=0$), this gives the square distance between two points. However, in multi-scale spacetimes the dimensional scaling of lengths, areas, and so on, is anomalous. For instance, in isotropically fractional spacetimes with measure \Eq{mea1} and $\a_\mu=\a$, the dimensional scaling is the usual one multiplied times $\a$ \cite{frc1}. Together with the measure structure \Eq{mes}, this points towards the line element
\be\label{line1}
\rmd s^2 \stackrel{?}{=} \bar g_{\mu\nu}\, \rmd q(x^\mu)\otimes \rmd q(x^\nu)=\bar g_{\mu\nu}\, v(x^\mu)\,v(x^\nu)\,\rmd x^\mu\otimes \rmd x^\nu\,,
\ee
where repeated indices are summed over and we omit indices in the symbols $q$ and $v$. However, it is easy to check that eq.\ \Eq{line1} is not invariant under coordinate transformations.

The crucial point is that multi-scale spacetimes have a non-trivial differential and geometric structure at all points. Like a self-affine fractal, one should see the same structure both when moving to different points and changing coordinate system. This is most naturally done in geometric coordinates. Equation \Eq{mes} can be rewritten for each direction as (repeated indices are not summed over in the second expression)
\be\label{fraq}
v(x^J)\,\rmd x^J = v({x'}^\mu)\,\rmd {x'}^\mu\, e_\mu^{\ J}\,,\qquad e_\mu^{\ J}:= \frac{v(x^J)}{v({x'}^\mu)}\bar e_\mu^{\ J}\,,
\ee
corresponding to
\be\label{newlor}
\rmd q^J = \rmd {q'}^\mu\, e_\mu^{\ J}\,,\qquad e_\mu^{\ J}:= \frac{\p q^J}{\p {q'}^\mu}\,.
\ee
In other words, the gravitational field in multi-scale spacetimes is most naturally defined \emph{with respect to geometric coordinates} and multi-scale frames map a curvilinear coordinate system to the Cartesian one \emph{with the same measure structure}.

The weights in the inverse $e^\mu_{\ I}$ of the weighted vielbein $e^\mu_{\ I}$ are swapped, $e^\mu_{\ I}:=\bar e^\mu_{\ I}$ $\times v({x'}^\mu)/v(x^I)$, so that $e_{\ I}^\mu e_\mu^{\ J} = \delta_I^J$. The \emph{multi-scale metric} is then defined as
\be\label{frm}
g_{\mu\nu}:= \eta_{IJ} e_\mu^{\ I} e_\nu^{\ J}\not\propto \bar g_{\mu\nu}\,,
\ee
which is not proportional to the ordinary metric due to the weight factors in the sum over $I$ and $J$. Thanks to the orthonormality of the multi-scale vielbeins, $g_{\mu\nu}g^{\mu\rho}=\de_\nu^\rho$ and indices can be raised and lowered as usual.

The multi-scale line element is formally identical to \Eq{line1}, except that the metric now is \Eq{frm}:
\be\label{fle}
\rmd s^2 := g_{\mu\nu}\, v(x^\mu)\,v(x^\nu)\,\rmd x^\mu\otimes \rmd x^\nu=g_{\mu\nu}\, \rmd q^\mu\otimes \rmd q^\nu\,.
\ee
Now it is possible to move from a curvilinear to a Cartesian multi-scale coordinate system:
\be
\rmd s^2 \ \stackrel{\text{\tiny \Eq{frm}}}{=}\ \eta_{IJ} e_\mu^{\ I} \rmd q^\mu\otimes e_\nu^{\ J}\rmd q^\nu \ \stackrel{\text{\tiny \Eq{newlor}}}{=}\  \eta_{IJ} \rmd {q'}^I\otimes \rmd {q'}^J\,.
\ee
Working in $q$ coordinates is time saving but one can repeat the same calculation in $x$ coordinates with explicit measure weight factors.

To summarize, trading local frames with multi-scale ones we are able to define the multi-scale version of a local inertial frame, where the metric is Minkowski. Because of the non-trivial measure, multi-scale inertial frames are not invariant under a Poincar\'e transformation of the coordinates, eq.\ \Eq{xLax}, but they are invariant under eq.\ \Eq{qLaq}. This is the transformation leading from a multi-scale local inertial frame charted by $\{x^J\}$ to another one charted by $\{{x'}^I\}$.  In multi-scale Minkowski spacetime, $g_{\mu\nu}=\eta_{\mu\nu}$, the vielbeins are $e_\mu^{\ I}=\de_\mu^{\ I}$, and there is no distinction between spacetime (Greek) and frame (capital Roman) indices. 

In Euclidean multi-scale space, the finite form of $\rmd s$ gives the distance between two points $x$ and $y$:
\be
\Delta_q(x,y):= \sqrt{\sum_{\mu=1}^D|q^\mu(x)-q^\mu(y)|^2}\,.
\ee
In the fractional isotropic case, $\Delta_q(x,y) \sim \text{`}\,\sqrt{|x^\a-y^\a|^2}\,\text{'}$, where quotation marks mean that we omitted details such as sums, indices, signs and constant prefactors. This is not a $2\a$-norm as in multi-scale spaces with fractional derivatives, where the distance in the same length units is $\Delta_\g(x,y) \sim \text{`}\,\sqrt{|x-y|^{2\g}}\,\text{'}$ \cite{frc1}. Therefore, in multi-scale spaces with $q$-derivatives there is no restriction of the range of $\g$ to the semi-interval $\g\geq 1/2$, as in \cite{frc1}.

All this discussion indicates that the $q$-theory is the only one realizing the curved, Lorentzian generalization of a genuine fractal behaviour, since in the other multi-scale models geometric coordinates do not appear in the dynamics and, therefore, the measure structure is not preserved by frame and coordinate transformations. This is in agreement with the findings on the relation between spectral and walk dimension \cite{frc7} recalled in section \ref{must}. Also, the line element in all the other cases is not eq.\ \Eq{fle}, and it does not correspond to the distance between two points: for the theory with ordinary derivatives it is eq.\ \Eq{ords}, for the one with weights it is a modification of it according to \cite{frc10}. In the theory with ordinary derivatives, the metric is actually $\bar g_{\mu\nu}$ and the frame matrices \Eq{bare} carry no information on the measure. This does not mean that the theories with ordinary and weighted derivatives are ill defined; simply, they do not realize multi-fractal geometries.


\subsection{Covariance and diffeomorphism invariance}\label{graco2}

We make a comment on covariance and diffeomorphism (in short, diffeo) invariance in multi-scale theories. They can be stated as follows \cite{Giu06}. Let $\Psi$ be some fields living on a manifold $\cM$ endowed with some non-dynamical structure $\Sigma$, and obeying the equations of motion $F[\Psi,\Sigma]=0$. Covariance determines that, under a diffeomorphism $f$, the transformed fields $f\cdot \Psi$ obey equations of motions with transformed non-dynamical structure: $F[\Psi,\Sigma]=0=F[f\cdot \Psi,f\cdot\Sigma]$. On the other hand, diffeo invariance limits the amount of non-dynamical structure: it requires that the same equation of motion be satisfied by the fields and their transforms, $F[\Psi,\Sigma]=0=F[f\cdot \Psi,\Sigma]$ (active diffeomorphism), or, equivalently, that any solution $\Psi$ of the equations of motion is also solution of a different set of equations parametrized by a transformed non-dynamical structure, $F[\Psi,\Sigma]=0=F[\Psi,f\cdot \Sigma]$ (passive diffeomorphism).

Any covariant theory with no non-dynamical structure is trivially diffeo invariant. Normally, one identifies covariance with diffeo invariance for this reason, but in multi-scale theories this is no longer true since the measure weight $v(x)$ is, strictly speaking, non-dynamical. Therefore, one should keep the concept of covariance and diffeo invariance distinct. In particular, multi-scale theories are covariant but not diffeo invariant in the usual sense. From the point of view of plain diffeomorphisms, the zeros or the singularities of $v$ are special points which do have an independent meaning, contrary to diffeo-invariant theories where a point acquires meaning only in relation to the happening of a physical event.

On the other hand, it would be desirable to describe multi-scale (and, in particular, fractal) geometry in a coordinate independent way. That is, there should be no meaning in statements such as `The measure singularity is located at this point, this much distant from the particle P.' Rather, the singularity should be `everywhere.' This is the significance of the multi-scale frames in eq.\ \Eq{newlor}. At each point on the multi-scale manifold, one can attach a local reference frame with a given distribution $\vr(x)=q(x)$. Thus, from the perspective of multi-scale geometry it is more natural to assume this modification of diffeo invariance which also includes the anomalous geometric distribution structure attached to each point. We saw this occurs only in the $q$-theory.

Obviously, the $q$-theory is invariant under active diffeomorphisms with respect to the geometric coordinates. The theory with ordinary derivatives does not possess such a property due to the overall non-trivial measure, while in the theory with weighted derivatives the problem stems from the interaction terms. All the theories are diffeo invariant when $v\to V$ is regarded as a dynamical scalar field. Take the example of a scalar field theory. In ordinary spacetime, consider the action
\be\label{bars}
\bar S_\vp=-\int\rmd^Dx\,\sqrt{-\bar g}\,\left(\frac12\bar g^{\mu\nu}\p_\mu\vp\p_\nu\vp+\frac{\la}{n}\vp^n\right),
\ee
where $\bar g$ is the determinant of the metric $\bar g_{\mu\nu}$. Since $\bar g'=\bar\cJ^2 \bar g$, the volume element $\rmd^Dx\,\sqrt{-\bar g}$ is invariant under a coordinate transformation. For the same reason, and due to invariance of the scalar field $\vp'(x')=\vp(x)$, both the kinetic and potential term are invariant. Thus, $\bar S_\vp$ is diffeo invariant.
\begin{enumerate}
\item The case of the theory with ordinary derivatives is the easiest and does not require many comments:
\bs\label{Sord1}\ba
S_{{\rm ord},\vp}&=&\int\rmd^Dx\,v\,\sqrt{-g}\,\left(\frac12 \vp\,\cK^\pm\vp-\frac{\la}{n}\vp^n\right),\\
&& \cK^-=\frac{1}{v}\,\N_\mu\left[v\N^\mu\,\cdot\,\right]\,,\qquad \cK^+= \B=g^{\mu\nu} \N_\mu\N_\nu\,.
\ea\es
Here and in what follows, we denote with $g_{\mu\nu}$ the metric of the chosen multi-scale model, reserving the symbol with a bar $\bar g_{\mu\nu}$ to the metric of ordinary spacetime (in this particular theory, however, $\bar g_{\mu\nu}=g_{\mu\nu}$). $\N$ is the covariant derivative with respect to $g$ ($\N_\mu=\p_\mu$ on a scalar). Notice that, after integration by parts, the kinetic term can be also written as
\be\label{spl}
\vp\,\cK^-\vp\to -g^{\mu\nu}\p_\mu\vp\,\p_\nu\vp\,,\qquad \vp\,\cK^+\vp\to -g^{\mu\nu}\check{\cD}_\mu\vp\,\p_\nu\vp\,,
\ee
or, in more compact form, $\vp\,\cK^\pm\vp\to -g^{\mu\nu}{}_{1/2\pm 1/2}\cD_\mu\vp\,\p_\nu\vp$, where we used eq.\ \Eq{bD}. The expression for the case $\cK^+$ can be easily symmetrized in the indices.

Since the measure is non-dynamical but does determine the dynamics of the fields, we can add kinetic and potential terms:
\be\label{Sord2}
S_{{\rm ord},\vp}=\int\rmd^Dx\,\sqrt{-g}\,v\left\{\frac12 \vp\cK^\pm\vp -\frac{\la}{n}\vp^n+\frac{1}{2\k^2}[\om v\cK^\pm v-U(v)]\right\},
\ee
where $\om$ is a function of $v$ (or just a constant) and $\k^2=8\pi G$ ($G$ is Newton's constant) has been introduced for later convenience. 
\item Consider now the multi-scale theory with weighted Laplacian:
\be\label{Sv1a}
S_{v,\phi}=-\int\rmd^Dx\,v\sqrt{-g}\,\left\{\frac12 g^{\mu\nu}\cD_\mu\phi\cD_\nu\phi+\frac{\la}{n}\phi^n+\frac{1}{2\k^2} [\om g^{\mu\nu}\cD_\mu v\cD_\nu v+U(v)]\right\}.
\ee
The field $\phi$ is a scalar density, which transforms as $\phi'(x')=\sqrt{v(x)/v(x')}\,\phi(x)$ \cite{frc6}. Thus, $\vp:=\sqrt{v}\,\phi$ is a scalar and the action \Eq{Sv1a} can be recast as
\be\label{Sv1b}
S_{v,\phi}=-\int\rmd^Dx\,\sqrt{-g}\,\left\{\frac12 g^{\mu\nu}\p_\mu\vp\p_\nu\vp+\frac{\la}{n}v^{1-\frac{n}2}(x)\vp^n+\frac{v}{2\k^2} [\om g^{\mu\nu}\cD_\mu v\cD_\nu v+U(v)]\right\}.
\ee
The kinetic term for $v$ could have been defined also with normal derivatives, in conformity with the $\vp$ sector. The difference is only a numerical factor, $(\cD v)^2=(9/4) (\p v)^2$. The interaction term breaks diffeo invariance even when $\om=0=U$, unless the measure itself be a dynamical scalar field, eq.\ \Eq{vV}.
\item In the $q$-theory, replacing everywhere in \Eq{bars} the coordinates $x^\mu$ with the geometric coordinates $q^\mu(x^\mu)$ and $\bar g_{\mu\nu}$ with the multi-scale metric \Eq{frm}, one has
\ba
S_{q,\vp} &=&-\int\rmd^Dq(x)\,\sqrt{-g(x)}\,\left\{\frac12 g^{\mu\nu}(x)\frac{\p \vp(x)}{\p q^\mu(x)}\frac{\p\vp(x)}{\p q^\nu(x)}+\frac{\la}{n}\vp^n(x)\right.\nonumber\\
&&\qquad\qquad\left.+\frac{1}{2\k^2} \left[\om g^{\mu\nu}
\frac{\p v(x)}{\p q^\mu(x)}\frac{\p v(x)}{\p q^\nu(x)}+U(v)\right]\right\}.\label{Sq1}
\ea
The field $\vp$ is a scalar and the action $S_{q,\vp}$ is diffeo invariant if eq.\ \Eq{vV} holds or if $\om=0=U$. The only change with respect to \Eq{bars} is that now coordinates are composite objects; accordingly, the metric is defined so that Jacobians always cancel with one another. For completeness, we have introduced a kinetic and potential term for the measure, but in this theory they are somewhat out of place, since eq.\ \Eq{Sq1} resembles a field theory coupled with a non-relativistic point particle. We will ignore them in section \ref{graco5}.
\end{enumerate}


\section{Theory with ordinary derivatives}\label{graco3}

In torsion-free (pseudo-)Riemannian manifolds, the requirement that the spin connection realizes parallel transport of angles and lengths translates into the compatibility equation $\bar\N_\s \bar g_{\mu\nu}=0$, stating that the metric is covariantly constant. Explicitly,
\be\label{ce1}
\bar\N_\s \bar g_{\mu\nu}=\p_\s\bar g_{\mu\nu}-\bar\Gamma_{\s\mu}^\tau \bar g_{\tau\nu}-\bar\Gamma_{\s\nu}^\tau \bar g_{\mu\tau}=0\,,
\ee
where
\be
\bar\G^\rho_{\mu\nu}:= \tfrac12\bar g^{\rho\s}\left(\p_{\mu} \bar g_{\nu\s}+\p_{\nu} \bar g_{\mu\s}-\p_\s \bar g_{\mu\nu}\right)\label{leci1}
\ee
is the Levi-Civita connection. The latter vanishes in a local inertial frame where $\bar g_{\mu\nu}=\eta_{\mu\nu}$. To have a Minkowski metric in local frames when the action measure is non-trivial, one possibility is to keep the same compatibility equation. This immediately leads to the model with ordinary derivatives. Its gravitational action, Einstein equations and some of the cosmology have been worked out in \cite{fra2} for the operator $\cK^-$. Here we report the main results, adding a potential for the measure. The reader can extend the discussion to $\cK^+$ by replacing $\p v\p v\to \check{\cD} v \p v$ in the equations of motion. 

The action for gravity reads (in the following, bars are omitted)
\be\label{Sg2}
S_g =\frac{1}{2\kappa^2}\int\rmd^Dx\,v\,\sqrt{-g}\,\left[R-\om\p_\mu v\p^\mu v-U(v)\right]\,,
\ee
where $U$ may include a cosmological constant term $\Lambda$ and the Ricci scalar $R:= g^{\mu\nu}R^\rho_{~\mu\rho\nu}$ is defined as usual, in terms of the Riemann tensor
\be\label{riem1}
R^\rho_{~\mu\s\nu} := \p_\s \G^\rho_{\mu\nu}-\p_\nu \G^\rho_{\mu\s}+\G^\tau_{\mu\nu}\G^\rho_{\s\tau}-\G^\tau_{\mu\s}\G^\rho_{\nu\tau}\,,
\ee
where $\Gamma$ is given by eq.\ \Eq{leci1}. Higher-derivative operators in the action are also possible, but we will limit the discussion to the Einstein--Hilbert term. 

In eq.\ \Eq{Sg2}, we added a `kinetic' and `potential' term for completeness in order to find non-trivial cosmological solutions accommodating usual matter, acceleration or other 
sensible features. Not doing so leads to the very restrictive scenarios of \cite{fra2} (matter with negative energy density, and so on). Including also the matter action $S_{\rm m}$, the Einstein equations are the usual ones but augmented by some terms in $v$:
\be
\kappa^2T_{\mu\nu} = R_{\mu\nu}-\frac12g_{\mu\nu}[R-U(v)]+g_{\mu\nu}\frac{\B v}{v} -\frac{\N_\mu\N_\nu v}{v}+\om\left(\frac12 g_{\mu\nu}\p_\s v\p^\s v-\p_\mu v \p_\nu v\right),\label{ee2}
\ee
where
\be\label{tmunu}
T_{\mu\nu}:=-\frac{2}{\sqrt{-g}}\frac{\de_v S_{\rm m}}{\de_v g^{\mu\nu}}=-\frac{2}{\sqrt{-g}\,v}\frac{\de S_{\rm m}}{\de g^{\mu\nu}}
\ee
is the energy momentum tensor of matter fields. The equation of motion for $v$ is
\be\label{veom}
R-U(v)=-2\k^2\cL_{\rm m}+v U_{,v}-\om(2v\B v+\p_\mu v\p^\mu v)\,,
\ee
where $\cL_{\rm m}$ is the Lagrangian density of matter fields and the subscript with comma denotes the derivative with respect to $v$. Taking the trace of eq.\ \Eq{ee2} gives
\be\label{tra}
R-\frac{D}2[R-U(v)]+(D-1)\frac{\Box v}{v}+\om\left(\frac{D}2-1\right)\p_\mu v\p^\mu v=\k^2T_\mu^{\ \mu}.
\ee
Plugging eq.\ \Eq{veom} into \Eq{tra}, we get 
\be
R-\frac{D}2v U_{,v}+(D-1)\frac{\Box v}{v}+\om[D v\B v+(D-1)\p_\mu v\p^\mu v] =\k^2\left(T_\mu^{\ \mu}-D\cL_{\rm m}\right)\label{tra2}\,.
\ee
Thus, $R$ in eq.\ \Eq{Sg2} can be replaced by the last expression and the dynamics of the metric is fully determined by the profile $v(x)$ (or by the dynamics of $V$ if the measure is interpreted as a Lorentz scalar).

Matter solutions on Minkowski background $g_{\mu\nu}=\eta_{\mu\nu}$ no longer obey the condition $\kappa^2T_{\mu\nu}=\Lambda\eta_{\mu\nu}$. One can also show that none of the profiles in section \ref{meas} admit Minkowski vacuum solutions.\footnote{The $0i$ component of eq.\ \Eq{ee2} with $R_{\mu\nu}=0=T_{\mu\nu}$ and $g_{\mu\nu}=\eta_{\mu\nu}$ gives $0=\p_i\dot v+\om v\dot v\p_i v$. If $v$ is factorizable, this equation admits solutions only if (i) $\p_i v=0$ or (ii) $\dot v=0$ or (iii) $\om=-1/v^2$. Then, the $00$ and $ii$ components of 
eq.\ \Eq{ee2} combine to give $\ddot v/v+\om \dot v^2=-\p_i^2v/v-\om (\p_iv)^2$. All three cases (i)--(iii) require that the left- and right-hand side of this equation vanish independently. The first case prescribes that $v=v(t)$, and $\ddot v/v+\om \dot v^2=0$ is solved by the inverse error function, completely different from the multi-scale profiles of section \ref{meas}. Case (ii) is identical. Case (iii) prescibes $\ddot v/v-(\dot v/v)^2=0=-\p_i^2v/v+(\p_iv/v)^2$, which is solved by an exponential $v=\exp(\sum_\mu c_\mu x_\mu)$, again not a multi-scale profile.}

The ordinary FLRW line element is
\bs\label{flrw}\be
\rmd s^2= g_{\mu\nu}\rmd x^\mu\rmd x^\nu=-\rmd t^2+a(t)^2\hat g_{ij}\rmd x^i\rmd x^j\,,
\ee
where $t=x^0$ is synchronous time, $a(t)$ is the scale factor,
\be
\hat g_{ij}\rmd x^i \rmd x^j=  \frac{\rmd r^2}{1-\textsc{k}\,r^2}+r^2\rmd O_{D-2}
\ee\es
is the line element of the maximally symmetric $(D-1)$-dimensional spatial slices of constant sectional curvature $\textsc{k}$ (equal to $-1$, 0 and $+1$ for, respectively, an open, flat, and closed universe with radius $a$) and $\rmd O_{D-2}$ is the angular part. The only non-zero components of the Ricci tensor and the expression for the Ricci scalar are
\ba
&& R_{00}   = -(D-1)(H^2+\dot H)\,,\qquad R_{ij} =\left[\frac{2\textsc{k}}{a^2}+(D-1)H^2+\dot H\right] g_{ij}\,,\\
&& R = (D-1)\left(\frac{2\textsc{k}}{a^2}+DH^2+2\dot{H}\right),
\ea
where 
\be
H := \frac{\dot{a}}{a}
\ee
is the Hubble parameter. In parallel, we assume that the measure $v(x)$ is non-trivial only along the time direction, $v(x)=v(t)=v_0(t)$, and that matter is a perfect fluid with energy-momentum tensor
\be\label{pefl}
T_{\mu\nu} =(\rho+P)u_\mu u_\nu+g_{\mu\nu}\,P\,,
\ee
where $\rho=T_{00}$ is the energy density, $P=T_i^{\ i}/(D-1)$ is the pressure and $u$ is the fluid relativistic velocity ($u_\mu u^\mu=-1$). The $00$ component of the Einstein equations \Eq{ee2} gives the first Friedmann equation
\be\label{00}
\left(\frac{D}2-1\right)H^2+H\frac{\dot v}{v}=\frac{\k^2}{D-1}\rho+\frac{1}{2}\frac{\om}{D-1}\dot v^2 +\frac{U(v)}{2(D-1)}-\frac{\textsc{k}}{a^2}\,,
\ee
while the second Friedmann equation is
\be\label{trm}
H^2+\dot H-H\frac{\dot v}{v}+\frac{\om}{D-1}v\B v=-\frac{\k^2}{D-1}(\rho+P)\,.
\ee
Early-universe cosmology was partially studied in \cite{fra2}, in the case where $v\sim |t_*/t|^{1-\a}$. For $U(v)=0$ and on a flat FLRW background, non-vacuum solutions exist only with exotic matter with negative energy density, which may be realized without violating unitarity by a fermionic condensate. A more detailed analysis of this model, taking into account the interpretation of section \ref{coco1}, will be presented elsewhere.


\section{Theory with weighted derivatives}\label{graco4}

In multi-scale spacetimes with weighted derivatives, we end up modifying the compatibility equation $\bar\N_\s \bar g_{\mu\nu}=0$ and, indirectly, the possibility to have $ g_{\mu\nu}=\eta_{\mu\nu}$ in local frames. Thus, either one keeps the notion of local inertial frame but abandons weighted derivatives (which leads back to the preceding theory) or modifies it in order not to discard field theory on multi-scale Minkowski background \cite{frc6,frc8,frc9} (which results in a Weyl integrable spacetime). In the second case, there is an interesting interplay between frames, the density behaviour of the gravitational field and the number of topological dimensions ($D=4$ plays a special role).

Below eq.\ \Eq{Kv}, we recalled that weighted derivatives \Eq{bD} act with different weights depending on the tensorial nature of the field densities. We saw the example of a scalar density, $\b=1/2$, and mentioned the case of vectors \cite{frc8}, with the same value of $\beta$. What about the metric? In principle, it should be a bilinear, made of two contracted tetrads, which makes $\beta=1$ the most natural case. However, it will be more instructive to leave the parameter $\beta$ free and let the theory itself choose the most natural value in due course.

We define various types of covariant derivatives and connections. One is eq.\ \Eq{ce1} under the replacement $\p\to {}_\b\cD$ everywhere:
\be\label{gGG}
{}^\b\N_\s g_{\mu\nu}:={}_\b\cD_\s g_{\mu\nu}-{}^\b\Gamma_{\s\mu}^\tau g_{\tau\nu}-{}^\b\Gamma_{\s\nu}^\tau g_{\mu\tau}\,,
\ee
where
\be\label{leci2}
{}^\b\G^\rho_{\mu\nu}[g]:= \tfrac12 g^{\rho\s}\left({}_\b\cD_{\mu} g_{\nu\s}+{}_\b\cD_{\nu} g_{\mu\s}-{}_\b\cD_\s g_{\mu\nu}\right)\,.
\ee
In particular, $\G[g]:={}^0\G[g]=\bar\G[g]$. Equations \Eq{gGG} and \Eq{leci2} also imply, after some simple algebra, that the ordinary covariant derivative $\N_\s={}^0\N_\s$ coincides with ${}^\b\N_\s$ when acting on the metric (and only then):
\be\label{gGG2}
\N_\s g_{\mu\nu}:=\p_\s g_{\mu\nu}-\Gamma_{\s\mu}^\tau\, g_{\tau\nu}-\Gamma_{\s\nu}^\tau\, g_{\mu\tau}={}^\b\N_\s g_{\mu\nu}\,.
\ee
The relation between $\G$ and ${}^\b\G$ is
\be\label{gaga}
{}^\b\G^\rho_{\mu\nu}=\G^\rho_{\mu\nu}+\tfrac12 \left(W_\mu \de_\nu^\rho+W_\nu \de_\mu^\rho- g_{\mu\nu}W^\rho\right)\,,\qquad W_\mu:=\p_\mu\ln v^{\b}.
\ee
A third and fourth type of derivative are a mixture of the above two, with weighted derivative and connection $\G$ or with ordinary derivative and connection ${}^\b\G$:
\ba
\N_\s^+ g_{\mu\nu} &:=& {}_\b\cD_\s g_{\mu\nu}-\Gamma_{\s\mu}^\tau g_{\tau\nu}-\Gamma_{\s\nu}^\tau g_{\mu\tau}=\N_\s g_{\mu\nu}+W_\s g_{\mu\nu}\,,\label{gGG3}\\
\N_\s^- g_{\mu\nu} &:=& \p_\s g_{\mu\nu}-{}^\b\Gamma_{\s\mu}^\tau g_{\tau\nu}-{}^\b\Gamma_{\s\nu}^\tau g_{\mu\tau}=\N_\s g_{\mu\nu}-W_\s g_{\mu\nu}\,.\label{gGG4}
\ea

In this section, we consider four geometries, each characterized by a given connection and a covariant conservation law for the metric:
\bs\ba
\text{Case $(a)$:}&\qquad {}^\b\Gamma\,,\qquad {}^\b\N_\s g_{\mu\nu}&=0\,,\\
\text{Case $(b)$:}&\qquad \Gamma\,,\qquad \N_\s g_{\mu\nu}&=0\,,\\
\text{Case $(c)$:}&\qquad \Gamma\,,\qquad \N^+_\s g_{\mu\nu}&=0\,,\\
\text{Case $(d)$:}&\qquad {}^\b\Gamma\,,\qquad \N^-_\s g_{\mu\nu}&=0\,.
\ea\es
We will discard cases $(a)$ and $(c)$, identify case $(b)$ with the previous multi-scale theory, and recognize case $(d)$ as a Weyl integrable spacetime. The reader not bothered by technical details may jump to section \ref{wist}.


\subsection{Model with standard frames}

Let us first consider case $(a)$ and write the metric 
\be\label{gte}
g_{\mu\nu}=\eta_{IJ}\tilde e_\mu^{\ I}\tilde e_\nu^{\ J}
\ee
in terms of the vielbeins $\tilde e_\mu^{\ I}$, which may differ from the previously introduced $\bar e$ and $e$. From eq.\ \Eq{leci2}, one gets $2g_{\rho\la}\,{}^\b\G^\rho_{\mu\nu}={}_\b\cD_\mu g_{\nu\la}+{}_\b\cD_\nu g_{\mu\la}-{}_\b\cD_\la g_{\mu\nu}$. Using the symmetry of ${}^\b\Gamma$ in the lower indices and adding and subtracting the same expression with, respectively, exchanged indices $\la\leftrightarrow\mu$ and $\la\leftrightarrow\nu$, we obtain ${}_\b\cD_\la g_{\mu\nu}=g_{\rho\nu}\,{}^\b\G^\rho_{\la\mu}+g_{\rho\mu}\,{}^\b\G^\rho_{\la\nu}$. On the other hand, applying ${}_\b\cD$ on $g_{\mu\nu}=\eta_{IJ}\tilde e_\mu^{\ I}\tilde e_\nu^{\ J}$ and comparing with the last expression, one arrives at
\be\label{Gee}
{}^\b\G^\rho_{\la\mu}=\tilde e^\rho_{\ I}\, {}_{\frac\b2}\cD_\la\tilde e_\mu^{\ I}\,.
\ee
As in the ordinary case \cite{wei72}, 
this derivation relies on the existence of a local inertial frame centered at some point P where the derivative of the metric vanishes, $\p_\s g_{\mu\nu}= 0$, meaning that $\eta_{\mu\nu}$ is the metric in the local inertial frame. 
For eq.\ \Eq{Gee} to be symmetric in $\la$ and $\mu$, the vielbein should be of the form $\tilde e_\mu^{\ I}=v^{-\b/2}\p_\mu (f^I x^I)$, where $f^I$ is some function. However, this expression leads to a metric $g_{\mu\nu}=v^{-\b}\eta_{IJ}\p_\mu (f^Ix^I)\, \p_\nu (f^J x^J)$ which, in turn, implies the following relation between curved and `Minkowski' line element: $v^{\b}g_{\mu\nu}\rmd x^\mu\otimes \rmd x^\mu=\eta_{IJ}\rmd (f^Ix^I)\otimes \rmd (f^J x^J)$. If $f^I=\sqrt{\om_I(s)}$ are line-element weights, the right-hand side is the line element found in \cite{frc10}, but the left-hand side does not have the same structure, unless $\b=0$ and $\om_I=1$ for all $I$. Another way to reach the same result is to express $\G[g]$ in terms of the vielbeins. Its definition gives
\be\label{deg}
\p_\la g_{\mu\nu}=g_{\rho\nu}\G^\rho_{\la\mu}+g_{\rho\mu}\G^\rho_{\la\nu}\,,
\ee
from which, via eq.\ \Eq{gte},
\be\label{Gee2}
\G^\rho_{\la\mu}=\tilde e^\rho_{\ I} \p_\la\tilde e_\mu^{\ I}\,.
\ee
Therefore, it must be $\tilde e_\mu^{\ I}=\p_\mu (f^I x^I)$, which, compared with the calculation above, requires $\b=0$, implying $g_{\mu\nu}=\bar g_{\mu\nu}$, $\tilde e_\mu^{\ I}=\bar e_\mu^{\ I}$, and $f^I=1$ for all $I$.

Ultimately, the source of the problem is the fact that ${}_\b\cD$ does not obey the usual Leibniz rule except when $\b=0$. Recall that the equation of motion of a free-falling relativistic particle worldline is $(\cD_s^2 x)^I:={}_I\cD_s^2 x^I=0$, where for each direction $I$ the derivative $\cD_s$ carries suitable (possibly all trivial) weights $\om_I(s)$ \cite{frc10}. This equation should coincide with the geodesic equation in a local frame where gravity is negligible. Using the ordinary frames \Eq{bare}, one can show that $0=(\cD_s^2 x)^I$ implies $0={}_I\cD_s^2 {x'}^\rho+\bar\Gamma_{\mu\nu}^\rho\p_s  {x'}^\mu\p_s  {x'}^\nu+f(x^I, {x'}^\rho){}_I\cD_s^2 1$. The last term vanishes if $\om_I(s)=1$ for all $I$, which corresponds to a spacetime with line element \Eq{ords}, i.e., $\b=0$ in eq.\ \Eq{ggbar}. Then, the metric in local frames is Minkowski. Attempts to construct a new geodesic equation ignoring the results of \cite{frc10} lead to somewhat complicated equations of motion for $x^I$, as well as to a non-flat metric in local frames.

All these arguments lead to eq.\ \Eq{ords}. The findings for the relativistic particle in weighted-derivative spacetimes \cite{frc10} suggest that, if $\om_I=1$ for all $I$, then the multi-scaling is only along spatial directions and $v_0(t)=1$. This information, if taken on board, would automatically lead to \emph{standard homogeneous cosmology}. If we ignore it but still account for the above results, we are forced to tick case $(a)$ off the list and adopt the second geometry, case $(b)$, where still  ${}^\b\N_\s g_{\mu\nu}=\N_\s g_{\mu\nu}=0$ but the connection is $\Gamma$. The ordinary Riemann tensor is the commutator of two covariant derivatives. Applying the same definition to multi-scale spacetimes with \Eq{gGG2}, $[\N_\mu,\N_\nu]u_\s=R^\tau_{\ \s\mu\nu}u_\tau$ for any vector $u$, and we obtain eq.\ \Eq{riem1}. This, in particular, guarantees than the Riemann curvature tensor vanishes in locally inertial frames, which are the usual ones where $\p_\s g_{\mu\nu}=0$.

Then, the natural gravitational action is eq.\ \Eq{Sg2} and we fall back into the theory with ordinary derivatives. 


\subsection{Weyl integrable spacetimes}\label{wist}

If one forfeits standard frames and allows for a general $\b\neq 0$, one obtains a model with weighted derivatives where $\p_\s g_{\mu\nu}\neq 0$ in local frames and geometry is non-Riemannian. One can still define a notion of local frame where the connection vanishes and the metric is not covariantly conserved in the usual sense. We thus fall in cases $(c)$ and $(d)$,
\be\label{tina}
\N^\pm_\s g_{\mu\nu}=0\qquad \Rightarrow\qquad \N_\s g_{\mu\nu}=\mp W_\s g_{\mu\nu}\,.
\ee
One can recognize these geometries as Weyl integrable spacetimes \cite{Wey52,NOSE,NoB08,PS}. The length $\ell_0$ of vectors changes under parallel transport by $\Delta \ell=\mp\ell_0 W_\mu\Delta x^\mu$, but this variation is zero in closed paths, provided the vector $W_\s$ is irrotational. Such is the present case, as
\be\label{Phi}
W_\mu=\p_\mu\Phi\,,\qquad \Phi:=\ln v^\b\,.
\ee
Since the non-metric part of the measure in the action is fixed to be $v=\rme^{\Phi/\b}$, cases $(c)$ and $(d)$ are physically inequivalent. Case $(c)$ is immediately ruled out because the commutator of two covariant derivatives yields the ordinary Riemann tensor, $[\N_\mu^+,\N_\nu^+]u_\s=R^\tau_{\ \s\mu\nu}u_\tau$, thus giving the usual Einstein--Hilbert action with plain derivatives. On the other hand, one could insist to include weighted derivatives in the action to get a tensor density with weighted derivatives and connection $\G$, i.e., of the form ${}^+R^\rho_{~\mu\s\nu}:= {}_\b\cD_\s \G^\rho_{\mu\nu}-{}_\b\cD_\nu \G^\rho_{\mu\s}+\G^\tau_{\mu\nu}\G^\rho_{\s\tau}-\G^\tau_{\mu\s}\G^\rho_{\nu\tau}$. However, this object is
not the commutator of two covariant derivatives and does not have the same index symmetries as $R^\rho_{~\mu\s\nu}$ (e.g., it is not symmetric in $\mu$ and $\nu$).

We are left with case $(d)$, which enjoys all the properties of a WIST. The connection ${}^\b\G_{\mu\nu}^\rho$ is the natural connection of these spacetimes \cite{NoB08} and is sometimes denoted as $C_{\mu\nu}^\rho$. The commutator of the covariant derivative \Eq{gGG4} on a vector gives $[\N_\mu^-,\N_\nu^-]u_\s=\cR^\tau_{\ \s\mu\nu}u_\tau$, where $\cR$ is the Riemann--Weyl curvature tensor with ordinary derivatives:
\be
\cR^\rho_{~\mu\s\nu}:= \p_\s {}^\b\G^\rho_{\mu\nu}-\p_\nu {}^\b\G^\rho_{\mu\s}+{}^\b\G^\tau_{\mu\nu}{}^\b\G^\rho_{\s\tau}-{}^\b\G^\tau_{\mu\s}{}^\b\G^\rho_{\nu\tau}\,.\label{rite2}
\ee
(Incidentally, also $[{}^\b\N,{}^\b\N]u=\cR u$, but we have already seen that this is not a WIST). In considering case $(d)$ in close conformity with our construction of multi-scale geometries, we will automatically establish a mapping of language between multi-scale spacetimes with weighted derivatives and WISTs.

The counterparts of the Ricci tensor and Ricci scalar are
\be\label{ricte2}
\cR_{\mu\nu}:= \cR^\rho_{~\mu\rho\nu}\,,\qquad \cR:= g^{\mu\nu}\cR_{\mu\nu}\,.
\ee
The simplest candidate action for multi-scale gravity is the generalization of the Einstein--Hilbert action:
\bs\label{eha}\ba
S_g &:=&\frac{1}{2\k^2}\int\rmd^Dx\,v\,\sqrt{-g}\left[\cR-\om\cD_\mu v\cD_\nu v-U(v)\right]\label{eha11}\\
 &=& \frac{1}{2\k^2}\int\rmd^Dx\,\rme^{\frac{1}{\b}\Phi}\,\sqrt{-g}\,\left(\cR-\Om\p_\mu\Phi\p^\mu\Phi-U\right)\,,\label{eha2}
\ea\es
where $\om$ is a function of $v$ and
\be\label{Om}
\Om:=\frac{9\om}{4\b^2}\rme^{\frac{2}{\b}\Phi}\,.
\ee
It is important to stress that, strictly speaking, WISTs are defined via a dynamical scalar field. To employ this naming rigorously, one should regard $v\to V$ as a dynamical Lorentz scalar. However, the following equations would be unchanged and we will stick with the interpretation of $v$ as a fixed profile throughout this section. The actual application of the paradigm established in section \ref{coco0}, and the differences with respect to scalar-tensor frameworks, will emerge only at the moment of finding explicit solutions to the equations of motion, as we shall do in section \ref{twdco}.

Actions in WISTs are characterized by an invariance under a double field redefinition which goes under the name of Weyl mapping (or simply gauge transformation):
\be\label{ggbar}
g_{\mu\nu}=:\rme^{-\chi}\tg_{\mu\nu}\,,\qquad g^{\mu\nu}=:\rme^{\chi}\tg^{\mu\nu}\,,\qquad \Phi=:\tilde\Phi+\chi\,,
\ee
so that $\tilde g_{\mu\tau} \tilde g^{\tau\nu}=\de_\mu^\nu=g_{\mu\tau} g^{\tau\nu}$. From eq.\ \Eq{gaga}, one sees that the connection is invariant,
\be
\widetilde{{}^\b\G}^\rho_{\mu\nu}={}^\b\G^\rho_{\mu\nu}\qquad \Rightarrow\qquad \tilde\cR^\rho_{~\mu\s\nu}=\cR^\rho_{~\mu\s\nu}\,,\qquad \tilde\cR_{\mu\nu}=\cR_{\mu\nu}\,,
\ee
while
\be\label{chir}
\cR =\rme^{\chi}\tilde\cR\,.
\ee
Applying the transformation \Eq{ggbar} to the action \Eq{eha}, we obtain
\ba
\hspace{-1cm}S_g &=& \frac{1}{2\k^2}\int\rmd^Dx\,\rme^{\frac{1}{\b}\tilde\Phi+\left(\frac{1}{\b}+1-\frac{D}{2}\right)\chi}\,\sqrt{-\tg}\,\left[\tilde\cR-\rme^{-\chi}U(\tilde\Phi+\chi)\vphantom{\frac11}\right.\nonumber\\
&&\qquad\qquad\qquad\qquad\qquad\qquad\qquad\qquad \left.-\Om(\tilde\Phi+\chi)\p_\mu(\tilde\Phi+\chi)\tilde\p^\mu(\tilde\Phi+\chi)\right],
\ea
where we used $\sqrt{-g}= \rme^{-D\chi/2} \sqrt{-\tg}$. Therefore, the action \Eq{eha} is a theory on a WIST with gauge invariance \Eq{ggbar} only when $\om=0$, $U=0$ and
\be\label{bstar}
\b=\b_*=\frac{2}{D-2}\,.
\ee
The kinetic term for the scalar field was dictated by natural considerations in the multi-scale framework. Relaxing the latter, another case of invariance is for $U=0$, $\om=\om(\Phi)= {\rm const}\times\rme^{-2\Phi/\b}\propto v^{-2}$ and the particular gauge transformation $\chi=-2\tilde\Phi$.

When $D=4$, one has $\b=1$ and ${}_\b\cD=\check{\cD}$, in which case $g_{\mu\nu}$ behaves as a field density with weight $-1$. 
Therefore, the metric can be regarded as a bilinear \Eq{gte} composed of density vectors $\tilde e$ with weight $-1/2$, even if $\tilde e$ does not behave as a standard frame vector.

Since the action \Eq{eha} is non-linear in the metric, one expects that it cannot be reduced to gravity in standard spacetime. In fact, only free multi-scale systems admit this mapping \cite{frc5,frc6}, while interacting multi-scale field theories typically show coordinate-dependent couplings \cite{frc6,frc8,frc9}. We check this expectation by first expressing the Riemann--Weyl and Ricci--Weyl curvature tensors in terms of the ordinary invariants, and then transforming to the Einstein frame defined below. From eqs.\ \Eq{gaga}, \Eq{rite2} and \Eq{ricte2},
\ba
\cR^\rho_{\ \mu\s\nu}&=& R^\rho_{\ \mu\s\nu}+r^\rho_{\ \mu\s\nu},\label{omnrs}\\
r^\rho_{\ \mu\s\nu} &:=&\de^\rho_{[\nu}\N_{\s]}\N_\mu\Phi+g_{\mu[\s}\N_{\nu]}\N^\rho\Phi+\frac12\left(\de^\rho_{[\s}\p_{\nu]}\Phi \p_\mu\Phi+g_{\mu[\nu}\p_{\s]}\Phi\p^\rho\Phi+\de^\rho_{[\nu}g_{\s]\mu}\p_\tau\Phi\p^\tau\Phi\right),\nonumber\\
\cR_{\mu\nu}&=& R_{\mu\nu}+r_{\mu\nu},\label{omn3}\\
r_{\mu\nu} &:=& r^\rho_{\ \mu\rho\nu}=-\left(\frac{D}2-1\right)\N_\mu\N_\nu\Phi-\frac12 g_{\mu\nu}\B\Phi
+\frac{D-2}{4}\left(\p_\mu\Phi\p_\nu\Phi-g_{\mu\nu}\p_\tau\Phi\p^\tau\Phi\right),\nonumber\\
\cR &=& R-(D-1)\left[\B\Phi+\frac{D-2}{4}\p_\mu\Phi\p^\mu\Phi\right],\label{crr}
\ea
where $A_{[\mu} B_{\nu]}=(A_\mu B_\nu-A_\nu B_\mu)/2$ and in the second line we used eq.\ \Eq{deg} (nowhere have we employed eq.\ \Eq{tina}). We can thus rewrite the action \Eq{eha}:
\ba
S_g &=& \frac{1}{2\k^2}\int\rmd^Dx\,\rme^{\frac{1}{\b}\Phi}\,\sqrt{-g}\,\left(R-\Om'\,\p_\mu\Phi\p^\mu\Phi-U\right),\label{sgn}\\
\Om' &:=& \Om+(D-1)\left(\frac{1}{2\b_*}-\frac{1}{\b}\right),
\ea
where we have integrated by parts the term in $\B\Phi$. 

The Weyl mapping \Eq{ggbar} is a frame transformation (in the sense of scalar-tensor theories \cite{DeS}) when 
\be\label{frat}
g_{\mu\nu}=\rme^{-\Phi} \bar g_{\mu\nu}\,,\qquad\chi=\Phi\,,\qquad \tilde\Phi=0\,,
\ee
moving from the Jordan frame with metric $g_{\mu\nu}$ to the Einstein frame with metric $\bar g_{\mu\nu}$. We call the new metric $\bar g$ because, contrary to the generic $\tg$, it is covariantly conserved, $\bar \N_\s \bar g_{\mu\nu}=0$. In multi-scale spacetimes, this transformation is called integer picture \cite{frc6} (where weighted derivatives acting on $g_{\mu\nu}=v^{-\b}\bar g_{\mu\nu}$ become ordinary derivatives acting on $\bar g_{\mu\nu}$): it tries to reduce the system to the standard one with usual measure and fields but possibly modified (i.e., measure-dependent) couplings. Combining eqs.\ \Eq{chir}, \Eq{sgn} and \Eq{frat}, we obtain $\overline{{}^\b\G}^\rho_{\mu\nu}=\G^\rho_{\mu\nu}$, $\cR =\rme^{\Phi}\bar R$ and
\be
S_g = \frac{1}{2\k^2}\int\rmd^Dx\,\rme^{\left(\frac{1}{\b}-\frac{1}{\b_*}\right)\Phi}\,\sqrt{-\bar g}\,\left(\bar R -\Om\p_\mu\Phi\bar\p^\mu\Phi-\rme^{-\Phi}U\right),\label{sgn2}
\ee
and a minimal coupling with gravity when eq.\ \Eq{bstar} holds. However, the terms in $\Phi$ persist, meaning that the dependence on the measure weight cannot be reabsorbed completely, except when
\be\label{sgn3}
\Om =0=\om\,,\qquad U=0\,.
\ee

Finally, we write the equations of motion in the Einstein frame, starting from eq.\ \Eq{sgn2}. This will help to compare with the dynamics of the other multi-scale models. The energy-momentum tensor \Eq{tmunu} transforms as
\be\label{tmunu2}
T_{\mu\nu}=-\rme^{-\frac{1}{\b}\Phi}\frac{2}{\sqrt{-g}}\frac{\de S_{\rm m}}{\de g^{\mu\nu}}=-\rme^{\left(\frac{1}{\b_*}-\frac{1}{\b}\right)\Phi}\frac{2}{\sqrt{-\bg}}\frac{\de S_{\rm m}}{\de \bg^{\mu\nu}}=\rme^{\left(\frac{1}{\b_*}-\frac{1}{\b}\right)\Phi}\bar T_{\mu\nu}\,.
\ee
The matter action is thus of the form
\be
S_{\rm m} =\int\rmd^Dx\,\rme^{\left(\frac{1}{\b}-\frac{1}{\b_*}\right)\Phi}\,\sqrt{-\bar g}\,\bar\cL_{\rm m}\,,\qquad \bar\cL_{\rm m}=\rme^{-\Phi}\cL_{\rm m}\,.
\ee
Taking into account the action for a scalar field, eq.\ \Eq{Sv1a} or \Eq{Sv1b}, we see that $\bar\cL_{\rm m}$ may in general depend on $\Phi$, unless $\bar\cL_{\rm m}$ coincided with the Lagrangian density in integer picture of a free system. Interestingly, this happens only if $\b=1$ and, if one wishes to have the Einstein frame, this constrains the number of topological dimensions to be $D=4$.

From the variations
\ba
\de\sqrt{-\bg} &=& -\tfrac12\,\bg_{\mu\nu}\sqrt{-\bg}\,\de \bg^{\mu\nu}\,,\\
\de\bar\G^\rho_{\mu\nu} &=& \tfrac12 \bg^{\rho\tau}\left(\bN_\mu\de \bg_{\nu\tau}+\bN_\nu\de \bg_{\mu\tau}-\bN_\tau\de \bg_{\mu\nu}\right),\\
\de\bR_{\mu\nu} &=& \bN_\s\de\bar\G^\s_{\mu\nu}-\bN_\nu\de\bar\G^\s_{\s\mu}\,,\\
\de\bR &=&\left(\bR_{\mu\nu}+\bg_{\mu\nu}\bB-\bN_\mu\bN_\nu\right)\de \bg^{\mu\nu}\,,
\ea
the Einstein equations in this model are
\ba
\hspace{-2cm}\k^2 \bar T_{\mu\nu}&=&\bR_{\mu\nu}-\frac12\bg_{\mu\nu}(\bR-\rme^{-\Phi}U)-\Om\left(\p_\mu\Phi\p_\nu\Phi+\frac12\bg_{\mu\nu}\p_\s\Phi\bar\p^\s\Phi\right)\nonumber\\
&&+\left(\frac{1}{\b}-\frac{1}{\b_*}\right)\left[\left(\frac{1}{\b}-\frac{1}{\b_*}\right)(\bg_{\mu\nu}\p_\s\Phi\bar\p^\s\Phi-\p_\mu\Phi\p_\nu\Phi)+\bg_{\mu\nu}\bB\Phi-\bN_\mu\bN_\nu\Phi\right],\label{ee3}
\ea
while the constitutive equation for the potential $U$ (i.e., the equation of motion for $v\to V$ if it was a scalar) is
\ba
0&=&\left(\frac{1}{\b}-\frac{1}{\b_*}\right)\left(\bR+\Om \p_\s\Phi\bar\p^\s\Phi-\rme^{-\Phi}U+\bar\cL_{\rm m}\right)-\Om_{,\Phi}\p_\s\Phi\bar\p^\s\Phi+2\Om\bB\Phi\nonumber\\
&&-\rme^{-\Phi}(U_{,\Phi}-U)+(\bar\cL_{\rm m})_{,\Phi}\,,\label{veom2}
\ea
where everything should be re-expressed in terms of $v$ as
\be
(\p\Phi)^2=\frac{\b^2}{v^2}(\p v)^2\,,\qquad \bB\Phi= \frac{\b}{v}\bB v- \frac{\b}{v^2}(\p v)^2,
\ee
and so on. As in the theory with ordinary derivatives, $\bar g_{\mu\nu}=\eta_{\mu\nu}$ is a vacuum solution only when the profile of the measure weight is tuned by the dynamics.

Equations \Eq{ee3} and \Eq{veom2} can be compared with their counterparts in the theory with ordinary derivatives, eqs.\ \Eq{ee2} and \Eq{veom}.\footnote{The choice between Einstein and Jordan frame can lead to two physically inequivalent models (with metric $\bar g$ and $g$, respectively) at the quantum level, although their classical cosmology yields the same physical predictions \cite{DeS10,ChY}. Here we work in the Einstein frame for simplicity.} The resemblance is stronger for $\b\neq\b_*$, but dynamics can be drastically simplified by setting $\b=\b_*$ or $\Om={\rm const}$. For instance, if both these conditions are met and $(\bar\cL_{\rm m})_{,\Phi}=0$, eq.\ \Eq{veom2} reduces to the ordinary equation of motion for a scalar field, $\bB\Phi-\rme^{-\Phi}(U_{,\Phi}-U)/(2\Om)=0$. If only $\b=\b_*$, eq.\ \Eq{ee3} becomes
\be\label{ee4}
\bR_{\mu\nu}-\frac12\bg_{\mu\nu}\bR=\k^2\,\bar T_{\mu\nu}+\frac{4\Om}{(D-2)^2}\frac{1}{v^2}\left(\p_\mu v\p_\nu v+\frac12\bg_{\mu\nu}\p_\s v\bar\p^\s v\right)-\bg_{\mu\nu}\frac1{2v^{\b_*}}U(v)\,.
\ee
The choice $\b=\b_*$ is attractive not only because it simplifies the Einstein equations, but also because, as said above, in $D=4$ it corresponds to having a bilinear metric ($\b=1$, ${}_\b\cD=\check{\cD}$). The trace of eq.\ \Eq{ee4} is
\be\label{ee5}
\left(\frac{D}{2}-1\right)\bR= -\k^2\,\bar T_{\mu}^{\ \mu}-\left(\frac{D}{2}+1\right)\frac{4\Om}{(D-2)^2}\frac{1}{v^2}\p_\s v\bar\p^\s v +\frac{D}{2v^{\b_*}} U(v)\,.
\ee


\subsection{Cosmology}\label{twdco}

In the scalar-tensor interpretation, the measure weight acts as a dynamical Lorentz scalar field in a Riemannian geometry, which is a particular form to recast a WIST. Depending on the choice of the parameters, the kinetic term of the scalar field can acquire a minus sign and become a phantom, thus leading to interesting phenomenology including bouncing solutions \cite{NOSE,NoB08}. On the other hand, spacetime is not a WIST in the standard sense in the multi-scale interpretation, where there exists a hierarchy of scales in the measure by default. We adopt this interpretation to illustrate the consequences of eq.\ \Eq{Lambda}.

Symmetry reduction of the dynamics is done by choosing the FLRW background \Eq{flrw} for the Einstein-frame metric $\bar g_{\mu\nu}$. Let $\b=\b_*$ and assume that $(\bar\cL_{\rm m})_{,\Phi}=0$. From eqs.\ \Eq{pefl} and \Eq{tmunu2}, $\bar T_{\mu\nu}=T_{\mu\nu}=(\rho+P)u_\mu u_\nu+g_{\mu\nu}\,P=v^{-\b_*}[(\rho+P)\bar u_\mu \bar u_\nu+\bg_{\mu\nu}\,P]=(\bar\rho+\bar P)\bar u_\mu \bar u_\nu+\bg_{\mu\nu}\,\bar P$, where $\bar\rho=v^{-\b_*}\rho$ and $\bar P=v^{-\b_*} P$ are the rescaled energy density and pressure. 

The 00 component of eq.\ \Eq{ee4} yields the first Friedmann equation
\be\label{cos1}
\left(\frac{D}{2}-1\right) H^2=\frac{\k^2}{D-1}\,\bar\rho+\frac{6\Om}{(D-1)(D-2)^2}\frac{\dot v^2}{v^2}+\frac{U(v)}{2(D-1) v^{\b_*}}-\frac{\textsc{k}}{a^2}\,,
\ee
while the trace equation is
\ba
\left(\frac{D}{2}-1\right) \left(\frac{2\textsc{k}}{a^2}+DH^2+2\dot{H}\right)&=&\frac{\k^2}{D-1}\,[\bar\rho-(D-1)\bar P]+\frac{2(D+2)\Om}{(D-1)(D-2)^2}\frac{\dot v^2}{v^2}\nonumber\\
&& +\frac{D\,U(v)}{2(D-1)v^{\b_*}} \,.\label{cos2}
\ea
These expressions can be compared with those of the previous section, eqs.\ \Eq{00} and \Eq{trm}; the structure is similar. Combining eqs.\ \Eq{cos1} and \Eq{cos2} to eliminate $U$, we find the master equation giving the cosmological evolution for a given measure profile:
\be\label{cos3}
(D-2)\dot{H}-\frac{2\textsc{k}}{a^2}+\k^2(\bar\rho+\bar P)=-\frac{4\Om}{(D-2)^2}\frac{\dot v^2}{v^2}\,.
\ee
Contrary to standard cosmology, the right-hand side is \emph{given}, dictated by multi-scale geometry. In particular, $v(x)$ is part of the differential structure of the theory and is not a dynamical scalar field, it does not possess a quantum propagator and it does not give rise to any instability if $\Om<0$ \cite{frc2,frc6}. Therefore, the right-hand side of, say, eq.\ \Eq{cos3} cannot be interpreted as (and does not share the problems of) the kinetic term of a ghost field when $\Om<0$. The other master equation of the system is obtained by plugging back eq.\ \Eq{cos3} in, say, \Eq{cos1}:
\be\label{cos4}
\frac{U(v)}{v^{\b_*}} = (D-2) [(D-1)H^2+3\dot H]+2(D-4)\frac{\textsc{k}}{a^2}+\k^2(\bar\rho+3\bar P)\,,
\ee
where the right-hand side, determined dynamically and geometrically by eq.\ \Eq{cos3} and the Raychaudhuri equation $\dot{\bar\rho}+(D-1)H(\bar\rho+\bar P)=0$, is a function of $t=t(v)$. Curiously, only in four dimensions does the curvature not contribute to the potential.

At this point, we abandon the general treatment and concentrate on an example that can be solved analytically: the flat case in four dimensions and in vacuum with
\be\label{fiso}
D=4\,,\qquad \textsc{k}=0\,,\qquad \bar\rho=0=\bar p\,,\qquad \Om=-\frac32.
\ee
Concerning the geometric profile $v$, we can take the multi-fractional measure \Eq{mea2} limited to the time direction and to two terms (binomial measure),
\be\label{binom}
v(t)=1+\left(\frac{t}{t_*}\right)^{\a-1}.
\ee
This is the simplest (but not toy) multi-scale model, where there is only one characteristic scale $t_*$  discriminating between `early' and `late' times. Then,
\be\label{tv}
t=t_* (v-1)^{\frac{1}{\a-1}}\,.
\ee
For the fractional charge $\a$, we choose the value $\a=1/2$, which is somewhat typical in these geometries \cite{frc1,frc2} (we checked that the solution is contracting for negative $\a$). Also, for simplicity we only consider positive times. We should notice immediately that choosing a different point $t_{\rm sing}\neq 0$ for the measure singularity and adopting the profile $\hat v(t)=1+[(t-t_{\rm sing})/t_*]^{\a-1}$ does not alter anything in the dynamics except the position of the origin of the time axis with respect to the features of the dynamical evolution (big bang, bounce, and so on). Therefore, neither the physics nor the background are changed by $v\to \hat v$ which amounts, rather, to a change of \emph{presentation} of the formul\ae\ \cite{frc1,fra7}.

Integrating twice eq.\ \Eq{cos3} yields the scale factor
\be\label{solu1}
a(t)=\left(1+\sqrt{\left|\frac{t_*}{t}\right|}\right)^{\frac38}\,\exp\left\{\frac98\left[H_0\left|\frac{t}{t_*}\right|+\sqrt{\left|\frac{t}{t_*}\right|}-\left|\frac{t}{t_*}\right|\ln\left(1+\sqrt{\left|\frac{t_*}{t}\right|}\right)\right]\right\},
\ee
where $H_0$ is an integration constant.
From this, one can extract the first slow-roll parameter
\be\label{srp}
\e(t):=-\frac{\dot H}{H^2}\,,
\ee
which determines whether the universe expands in deceleration ($\e>1$), accelerates ($0\leq \e <1$), or super-accelerates  ($\e<0$).

\begin{figure}
\centering
\includegraphics[width=7.4cm]{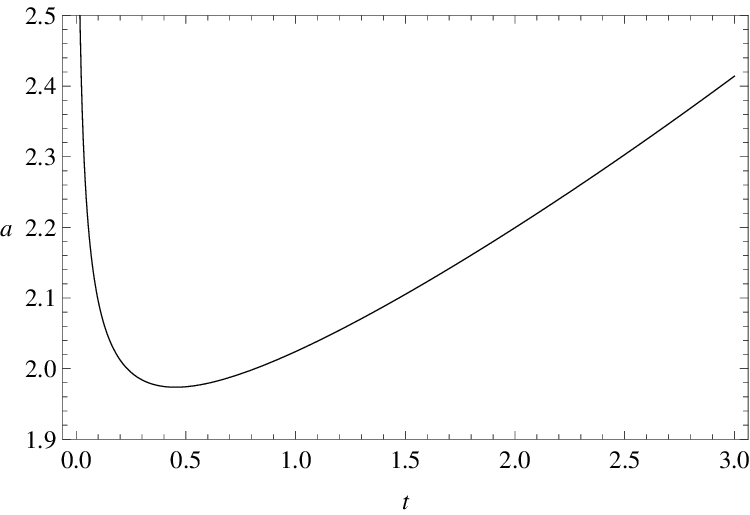}\hspace{.1cm}
\includegraphics[width=7.4cm]{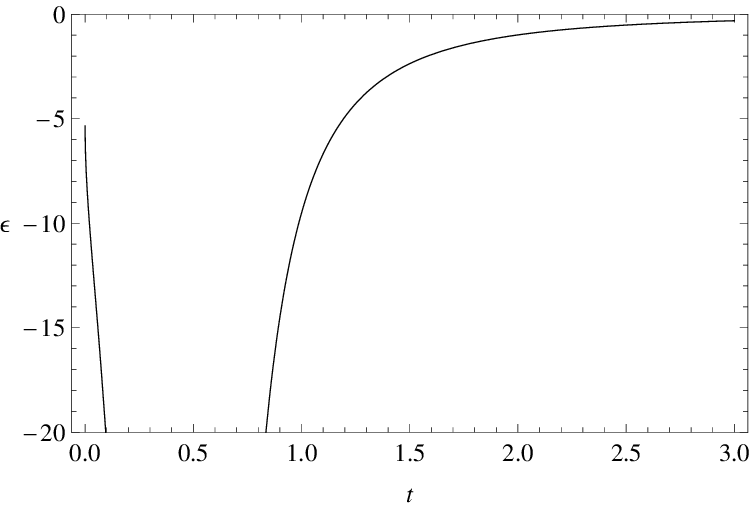}
\caption{\label{fig1} The solution \Eq{solu1} (left panel) and the slow-roll parameter \Eq{srp} (right panel) for $t_*=1$ and $H_0=4/45$.}
\end{figure}

At early times $|t- t_*|\ll 1$, the effects of the measure weight $v(t)$ are important and the Friedmann equations get non-trivial corrections. As the universe evolves, the weight in the time direction tends to unity and one obtains (in the homogeneous approximation) an ordinary cosmological evolution, since multi-scaling is confined to spatial slices. In the limit $t/t_*\ll 1$, $a\sim (t/t_*)^{-3/16}$ and the scale factor diverges, while at late times $t/t_*\gg 1$ there is an exponential behaviour $a\sim \rme^{H_0 t}$ and the solution expands in the late future only if $H_0>0$. This solution holds in the absence of matter but with a late-time cosmological constant $U(v)\sim U(v=1)$, hence the asymptotic de Sitter behaviour. Figure \ref{fig1} presents this evolution. Between $t=0$ and some critical time $t_{\rm bounce}$ the universe contracts, down to a bounce where $H(t_{\rm bounce})=0$ and the scale factor acquires a minimum non-zero value $a(t_{\rm bounce})\neq 0$. Depending on the value of the constant $H_0$, the bounce may happen before or after the characteristic time $t_*$. After that, the universe expands in super-acceleration, tending towards a de Sitter exponential law with late-time Hubble parameter $H_0$. 

\begin{figure}
\centering
\includegraphics[width=7.4cm]{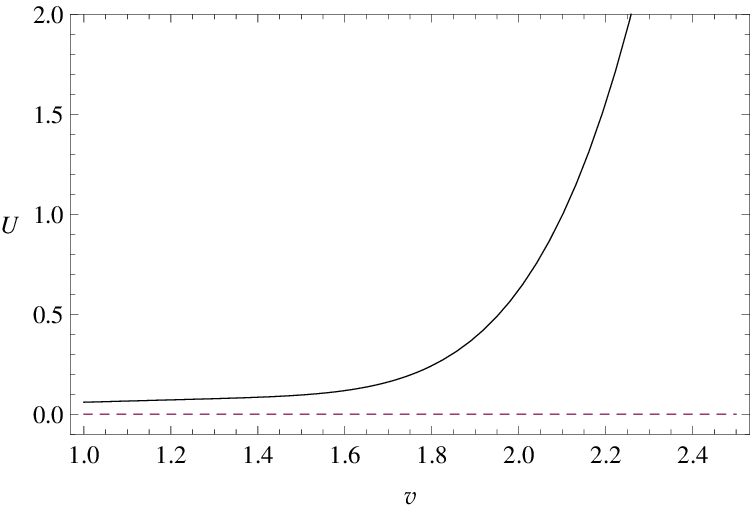}
\caption{\label{fig2} The potential $U(v)$ obtained from eqs.\ \Eq{cos4} and \Eq{solu1} for $t_*=1$ and $H_0=4/45$. The measure weight $v(t)$ rolls from $v(0)=+\infty$ down to the non-zero minimum at $v(\infty)=1$.}
\end{figure}

Realistic multi-scale spacetimes with weighted derivatives will always have a non-zero cosmological constant $\Lambda_0$ at late times. The evolution of the universe is governed by the natural rolling of the measure $v$ down its potential $U(v)$, which is shown in figure \ref{fig2} (the exact expression of $U$, which we do not write down because uninstructive, can be found by plugging eqs.\ \Eq{solu1} and \Eq{tv} in \Eq{cos4}). The minimum of the potential in this model is
\be\label{umin}
U_{\rm min}=U(v=1)=6H_0^2\qquad \Rightarrow\qquad \Lambda_0 =\frac{3 H_0^2}{\k^2}\,,
\ee
where we used eq.\ \Eq{Lambda}.

This example illustrates three characteristics of cosmological models in multi-scale spacetimes with weighted derivatives and multi-fractional measure:
\begin{enumerate}
\item At the classical level, the big-bang singularity can be replaced by a bounce.
\item The non-trivial measure weight induces an accelerating (actually, super-accelerating) phase and a late-time non-vanishing cosmological constant.
\item While the universe evolves, the Hausdorff dimension along the time direction changes with the time scale, from $\a$ to 1. The rolling of the measure weight towards the minimum of its potential $U(v)$ determines both the dynamics of the universe and the change of dimensionality in time (dimensional flow).
\end{enumerate}

\begin{figure}
\centering
\includegraphics[width=7.4cm]{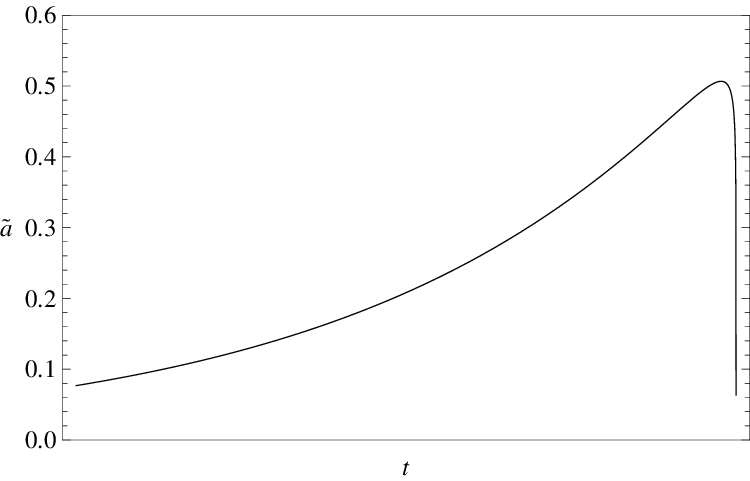}\hspace{.1cm}
\includegraphics[width=7.4cm]{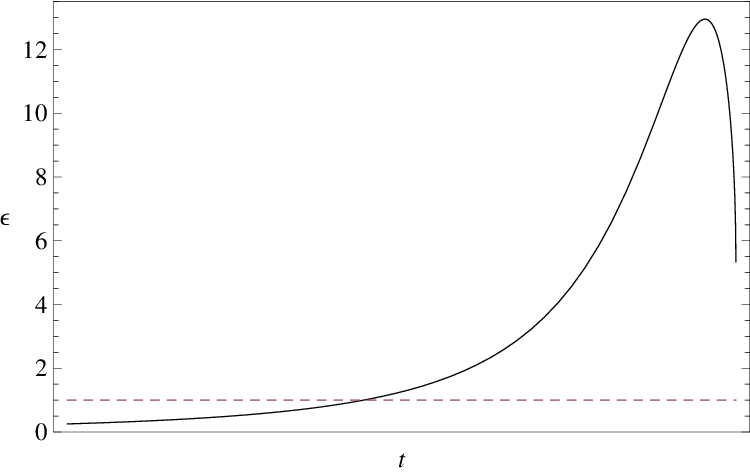}
\caption{\label{fig3} The solution $\tilde a$ (left panel) and the slow-roll parameter \Eq{srp} (right panel) for $t_*=1$ and $H_0=4/45$ (here $H_0$ does not correspond to the Hubble parameter today).}
\end{figure}

The bounce of the solution \Eq{solu1} occurs because $\Om<0$. 
 Vacuum solutions with $\Om>0$ also exist 
 and it may be interesting to show one such example, with the parameter choice \Eq{fiso} except $\Om=+3/2$ (the form of $\om$ is then obtained by inverting eq.\ \Eq{Om}). The scale factor $\tilde a(t)$ corresponding to this solution can be economically written as the inverse of eq.\ \Eq{solu1}, $\tilde a(t)=1/a(-t)$, and evolved backwards in time due to parity of the solution. The result, together with the first slow-roll parameter, is depicted in figure \ref{fig3}. The universe starts from a big bang at $\tilde a=0$ in the past infinity, expands in acceleration ($0<\e<1$), then in deceleration until it reaches a maximum `size', shortly after which it contracts into a big crunch. The potential $U(v)$ of this solution, which we do not show, is unbounded from below.


\section{Theory with \texorpdfstring{$q$}{}-derivatives}\label{graco5}

The action and equations of motion of this theory are straightforward in position space, but their solutions bear the full imprint of the hierarchy of scales in the geometry. Multi-scale frames have a clear interpretation and map a curvilinear coordinate system to the Cartesian one with the same measure structure. Gravity and cosmology are easy to work out by replacing $x\to q(x)$ in standard general relativity. For instance, the metric connection and the Riemann tensor are defined from the ordinary expressions \Eq{leci1} and \Eq{riem1}:
\ba
{}^q\G^\rho_{\mu\nu} &:=& \tfrac12 g^{\rho\s}\left(\frac{1}{v_\mu}\p_{\mu} g_{\nu\s}+\frac{1}{v_\nu}\p_{\nu} g_{\mu\s}-\frac{1}{v_\s}\p_\s g_{\mu\nu}\right)\,,\label{leciq}\\
{}^q R^\rho_{~\mu\s\nu} &:=& \frac{1}{v_\s}\p_\s {}^q\G^\rho_{\mu\nu}-\frac{1}{v_\nu}\p_\nu {}^q\G^\rho_{\mu\s}+{}^q\G^\tau_{\mu\nu}\,{}^q\G^\rho_{\s\tau}-{}^q\G^\tau_{\mu\s}\,{}^q\G^\rho_{\nu\tau}\,,\label{riemq}
\ea
and so on. From the action
\be\label{Sgq}
S =\frac{1}{2\kappa^2}\int\rmd^Dx\,v\,\sqrt{-g}\,({}^q R-2\Lambda)+S_{\rm m}\,,
\ee
we get the Einstein equations
\be\label{eeq}
{}^qR_{\mu\nu}-\frac12 g_{\mu\nu} ({}^q R-2\Lambda)=\k^2\, {}^qT_{\mu\nu}\,.
\ee
Unlike the models with ordinary and weighted derivatives, now the Bianchi identities impose $\p_\mu\Lambda=0$, i.e., a constant potential for $v(x)$. Although we could add a kinetic and potential term for the measure in eq.\ \Eq{Sgq} on general grounds, we prefer to restrict the dynamics for the purpose of highlighting certain features of multi-scale theories which could not be explored analytically in the other scenarios.

In the presence of a perfect fluid, the first Friedmann equation and the Raychaudhuri equation are
\ba
&&\left(\frac{D}{2}-1\right) \frac{H^2}{v^2}=\frac{\kappa^2}{D-1}\,\rho+\frac{\Lambda}{D-1}-\frac{\textsc{k}}{a^2}\,,\label{fri}\\
&& \dot\rho+(D-1)H(\rho+P)=0\,,\label{ra}
\ea
where $v=v_0(t)$. The integration of the measure weight gives the geometric time coordinate
\be\label{geom}
q(t)=\int^t\rmd t'\,v(t')\,.
\ee


\subsection{Power-law solutions and cosmological horizons}\label{cosol}

We now consider three types of solutions: with a barotropic fluid, with a scalar field, and with a cosmological constant.

If the perfect fluid obeys the equation of state $P=w\rho$ with $w$ constant, the solution of eqs.\ \Eq{fri} and \Eq{ra} with $\textsc{k}=0=\Lambda$ is simple:
\be\label{rgoa}
\rho=\rho_0\, a^{-\frac2p}\,,\qquad a(t)=\left[\frac{q(t)}{t_*}\right]^p\,,\qquad p:=\frac{2}{(D-1)(1+w)}\,,
\ee
where we chose a convenient normalization for the scale factor. The Hubble parameter in this class of cosmologies reads
\be\label{hup}
H=p\frac{\dot q(t)}{q(t)}=p\frac{v(t)}{q(t)}\,.
\ee
There are two horizons of interest in cosmology. One is the particle horizon, whose radius is the distance traveled by light (with speed $c=1$) since the big bang or, in the absence of a big bang, the initial time $t=0$. The comoving and proper particle-horizon radius are, respectively, 
\be\label{tau}
r_{\rm p} := \int_{0}^t \frac{\rmd t'\,v(t')}{a(t')}=t_*^p\frac{[q(t)]^{1-p}}{1-p}\,,\qquad R_{\rm p}:=a r_{\rm p} =\frac{q(t)}{1-p}\,,
\ee
where we assumed $q(0)=0$ and $p<1$.
Another cosmological scale is the Hubble horizon
\be\label{rh}
R_H = a\,r_H=\frac{q(t)}{p}\,,\qquad r_H:=\frac{v}{a H}= \ \frac{t_*^p}{p} [q(t)]^{1-p}.
\ee
In all power-law cosmologies, the particle and Hubble horizon are interchangeable since they differ only by an $O(1)$ factor, if $p$ is not too small:
\be
|r_{\rm p}|=\frac{p}{|1-p|}\,r_H\,.
\ee

If the fluid is a homogeneous scalar field,
\be
\rho_\vp=\frac1{2v^2}\dot\vp^2+W(\vp)\,,\qquad P_\vp=\frac1{2v^2}\dot\vp^2-W(\vp)\,,
\ee
the dynamical equations \Eq{fri} and \Eq{ra} in the absence of curvature and cosmological constant become
\ba
\left(\frac{D}{2}-1\right) H^2 &=&\frac{\kappa^2}{D-1}\left[\frac12\dot\vp^2+v^2\,W(\vp)\right],\label{frira2}\\
0 &=&\ddot\vp+\left[(D-1)H-\frac{\dot v}{v}\right]\dot\vp+v^2 W_{,\vp}\,.\label{frira3}
\ea
On the other hand, integrating eq.\ \Eq{fri} in the presence of only a $\Lambda$ term one gets
\be
a(t)=\exp\left[\sqrt{\frac{2\Lambda}{(D-1)(D-2)}}\, q(t)\right].\label{dS}
\ee

\

\noindent {\bf Evolution with multi-fractional measure.} If $0<p<1$, the comoving particle-horizon radius $r_{\rm p}$ and the proper particle horizon $R_{\rm p}$ increase in time provided $q(t)$ is monotonic. This is the case of the binomial measure \Eq{binom}, where the geometric coordinate \Eq{geom} is
\be\label{qt}
q(t) = t_*\left[\frac{t}{t_*}+\frac{1}{\a}\left(\frac{t}{t_*}\right)^\a\right].
\ee
The only difference with respect to standard cosmology is that the radius of the particle horizon increases faster in $t$ at times $t\lesssim t_*$ if $0<\a<1$. 

This implies that slow-roll inflation is still the natural mechanism solving the horizon and flatness problems. The reason is that we do not have a non-trivial cosmological constant in this theory, and we should fabricate it with matter so that $p> 1$, just like in standard cosmology. As one can see from eqs.\ \Eq{frira2} and \Eq{frira3}, ordinary slow-roll inflation ($p\gg 1$) is modified at times $t\lesssim t_*$: if the scalar field evolves slowly, neglecting its kinetic term does not lead to an almost constant Hubble parameter, since $H\sim v$ (eq.\ \Eq{dS}). In the binomial case \Eq{qt}, the rate of change of the scale factor is slightly larger than ordinary de Sitter expansion for $t\lesssim t_*$. In section \ref{ci}, however, it will become apparent that eq.\ \Eq{qt} can milden the slow-roll condition $p\gg 1$ and produce a nearly scale-invariant power spectrum just with $p\gtrsim 1$.

We conclude that multi-fractional geometry alone does \emph{not} lead to acceleration, unless it is sustained by a dynamical potential $U$ and a kinetic term for $v$. We will not switch on these contributions as before. In the $q$-theory, we are more interested in considering a case of multi-fractal measure which was analytically inaccessible in the theory with weighted derivatives: the measure with log-oscillations. While one cannot easily obtain analytic solutions when plugging a log-oscillating $v$ in the equations of motion for the weighted-derivative theory, the case of $q$-solutions such as \Eq{rgoa} and \Eq{dS} is most immediate. In the next sections, we will show that logarithmic oscillations, rather than multi-fractional scaling alone, can provide both a (partial) alternative to inflation and a cyclic inflationary scenario.

\

\noindent {\bf Evolution with log-oscillating measure.} Consider the log-oscillating measure \Eq{log} on a homogeneous background. We take a binomial multi-scaling with a combination of fractional charges $\a_n=1,\a$ and frequencies $\om_l=0,\om$ such that the measure weight is of the form $v(t)=1+v_\a(t)\,F_\om(\ln t)$. (An alternative measure where the whole multi-fractional part is multiplied times the log-oscillations is also possible, $v(t)=[1+v_\a(t)]F_\om(\ln t)$, but it leads to similar results.) The geometric time coordinate is
\bs\label{logos}\ba
q(t) &=& t+t_*\left(\frac{t}{t_*}\right)^\a F_\om(\ln t)\,,\\
F_\om(\ln t) &=& 1+A\cos\left[\om\ln\left(\frac{t}{\tp}\right)\right]+B\sin\left[\om\ln\left(\frac{t}{\tp}\right)\right]\,,
\ea\es
where $A$ and $B$ are some non-negative constants related to those in eq.\ \Eq{log} as follows: $A_{\a,\om}=A\a+B\om$, $B_{\a,\om}=B\a-A\om$. For definiteness, we chose the fundamental scale $t_\infty\ll t_*$ to be the Planck time $\tp$, as suggested by the relation between multi-scale log-oscillating geometries and non-commutative spacetimes \cite{ACOS}.

\begin{figure}
\centering
\includegraphics[width=7.4cm]{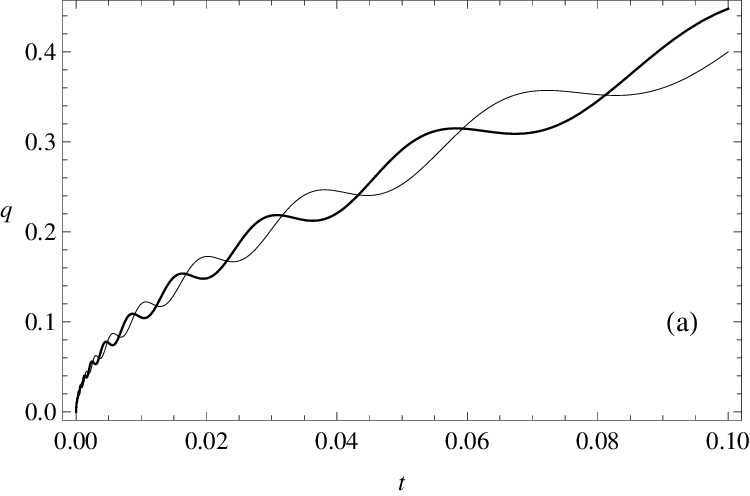}
\includegraphics[width=7.4cm]{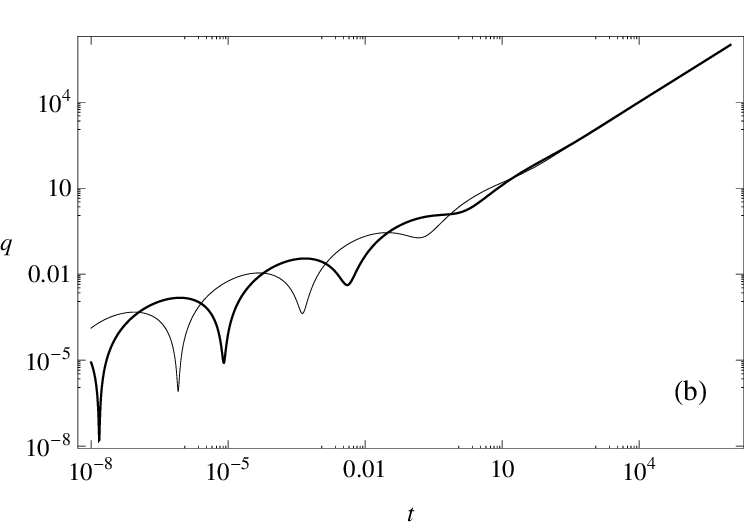}\\
\includegraphics[width=7.4cm]{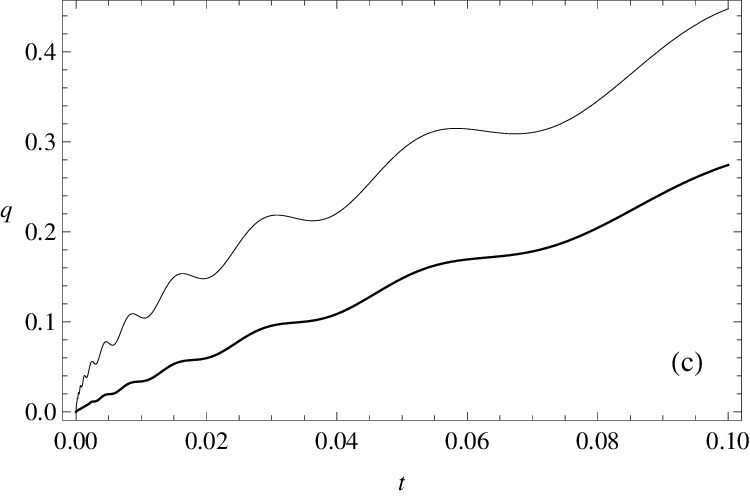}
\includegraphics[width=7.4cm]{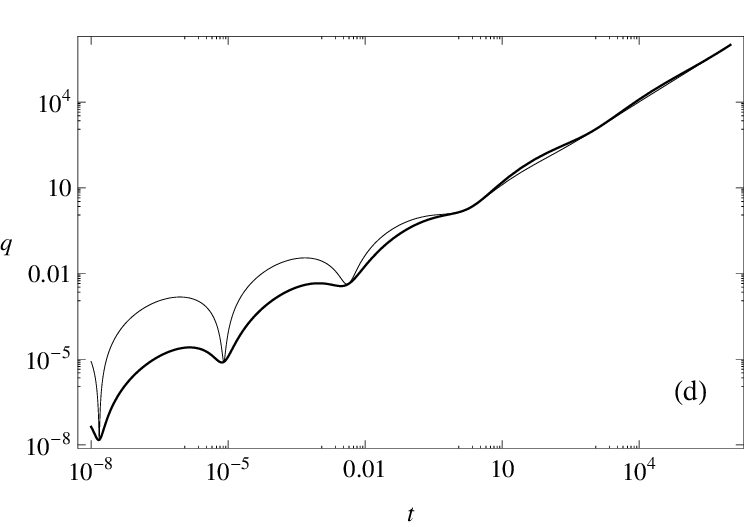}\\
\includegraphics[width=7.4cm]{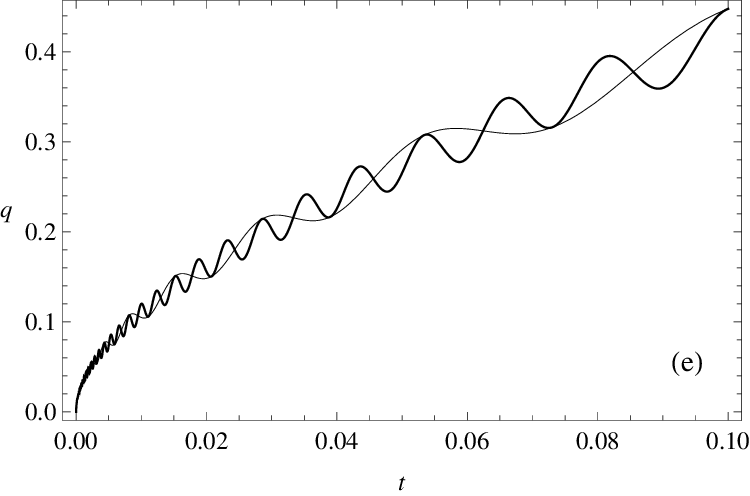}
\includegraphics[width=7.4cm]{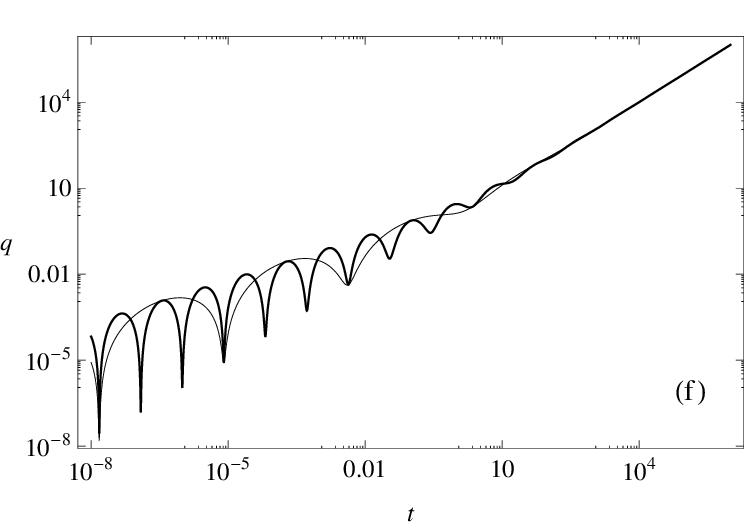}
\caption{\label{fig4} Linear-linear plots (left column) and log-log plots (right column) of the log-oscillating geometric coordinate \Eq{logos} with $t_*=1$, $B=0$, $A=10^{-1}$. In each log-log plot, the frequency $\om$ and the amplitudes $A$ and $B$ are, respectively, reduced and magnified 10 times with respect to the values of the corresponding linear-linear plot, which are: (a) $\tp=10^{-2},\,10^{-1}$ (increasing thickness), $\a=1/2$, $\om=10$; (c) $\tp=10^{-1}$, $\a=1/2,\,4/5$ (increasing thickness), $\om=10$; (e) $\tp=10^{-1}$, $\a=1/2$, $\om=10,\,30$ (increasing thickness).}
\end{figure}

\begin{figure}
\centering
\includegraphics[width=7.4cm]{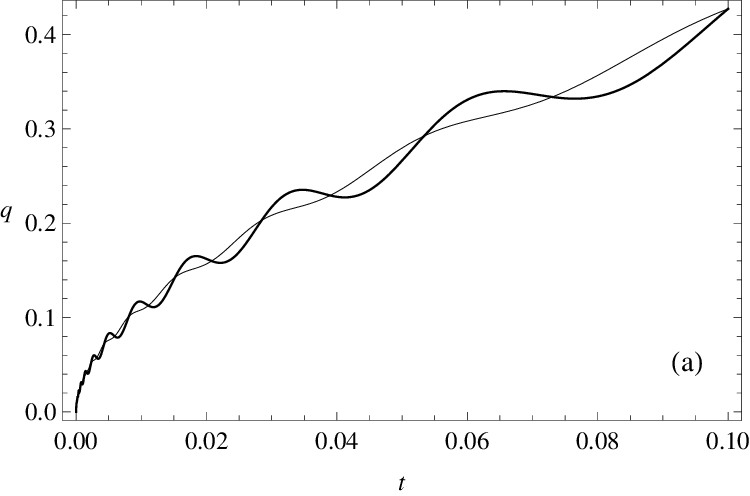}
\includegraphics[width=7.4cm]{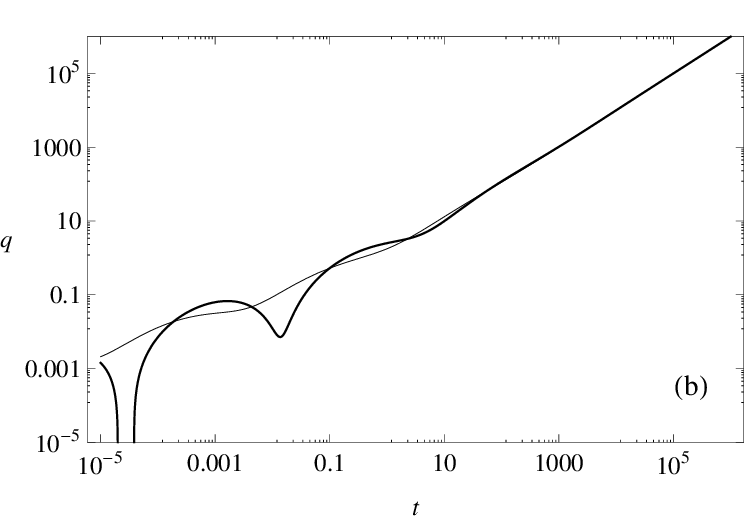}\\
\includegraphics[width=7.4cm]{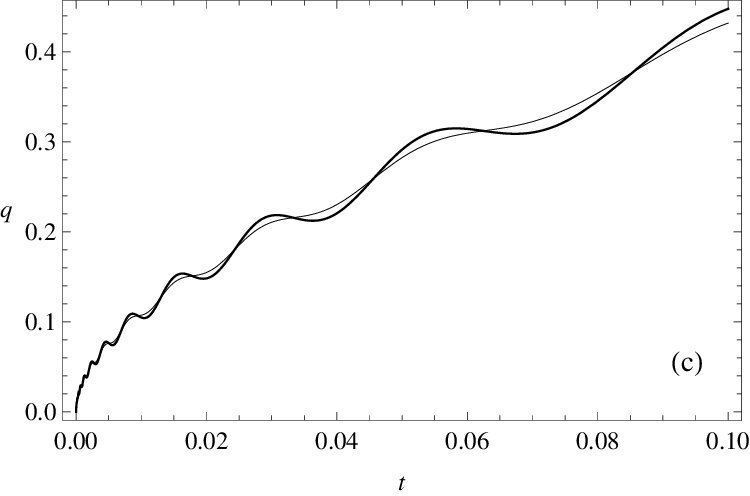}
\includegraphics[width=7.4cm]{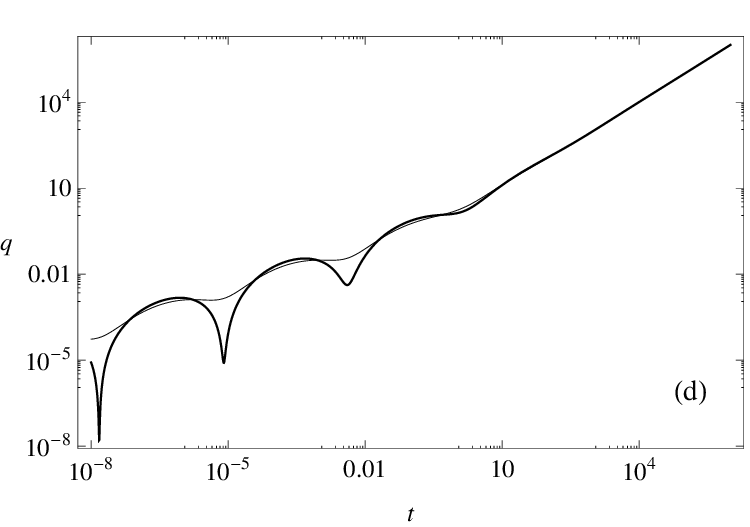}
\caption{\label{fig5} The log-oscillating geometric coordinate \Eq{logos} with $t_*=1$, $\tp=10^{-1}$, $\a=1/2$, and $\om=10$ (linear-linear plots) or $\om=1$ (log-log plots). In each log-log plot, the amplitudes $A$ and $B$ are magnified 10 times with respect to the values of the corresponding linear-linear plot, which are: (a) $B=0,\,10^{-1}$ (increasing thickness), $A=0.035$; (c) $B=0$, $A=0.05,\,0.1$ (increasing thickness).}
\end{figure}

\begin{figure}
\centering
\includegraphics[width=7.4cm]{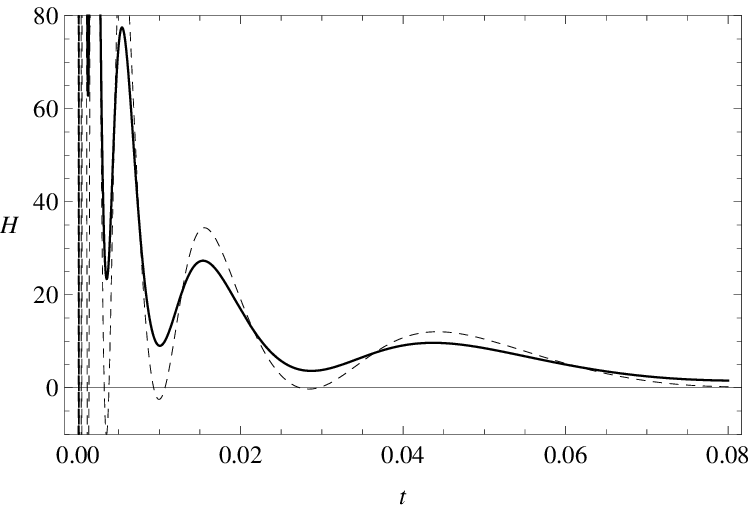}
\includegraphics[width=7.4cm]{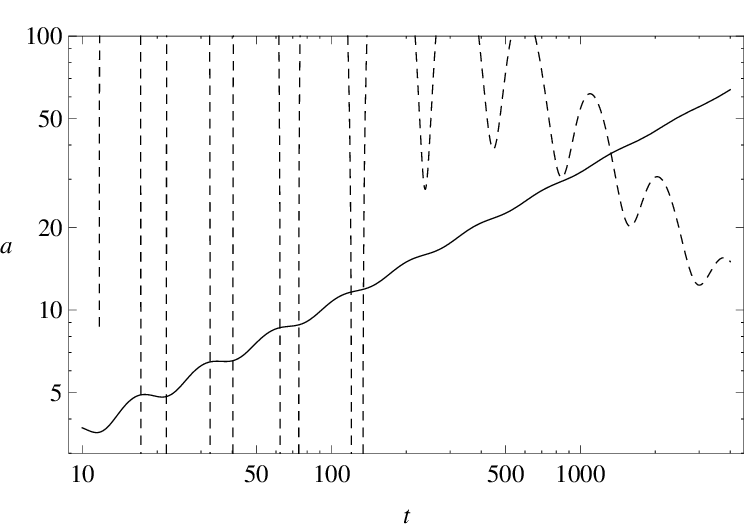}
\caption{\label{fig6} Left: linear-linear plot of the Hubble parameter \Eq{hup} with eq.\ \Eq{logos} for $t_*=1$, $\tp=0.5$, $\a=1/2$, $\om=6$, $B=0$, and $p=1/2$ (radiation in $D=4$). Solid curve: $A=0.06$, the universe is always expanding ($H>0$). Dashed curve: $A=0.1$, the universe undergoes cycles of contraction ($H<0$) and expansion ($H>0$). Right: log-log plot of the scale factor $a$ (solid curve) and of the (magnified) Hubble parameter $10^5H$ (dashed curve) for $t_*=1$, $\tp=10^{-1}$, $\a=1/2$, $\om=10$, $A=0.5=B$, and $p=1/2$. The critical time $t_{\rm crit}$ coincides with the last change of sign of $H$ (here, at $t_{\rm crit}\approx 150$), after which the evolution of the universe becomes monotonic.}
\end{figure}

We plot $q$ in figures \ref{fig4} and \ref{fig5} for some choices of the parameters. The scale factor \Eq{rgoa} for $p>0$ will follow the same pattern. As one can see,
\begin{itemize}
\item The fundamental time scale $\tp$ only affects the position of peaks (turn-arounds) and troughs (bounces).
\item The fractional charge $\a$ affects the slope of $q(t)$ averaged over the oscillations (i.e., the zero mode of the measure) as well as the amplitude of the oscillations. When $\a=0$, there is no modulation of the oscillations except due to the multi-fractional structure.
\item The parameter $\om$ modulates the frequency of the peaks and troughs as well as the steepness of the oscillations.
\item The parameter $B$ in front of the sine modulates the amplitude and position of the oscillations.
\item The parameter $A$ in front of the cosine modulates the amplitude of the oscillations. For sufficiently low values, the universe is always expanding. Otherwise, it undergoes logarithmic cycles of contractions and expansions up to some time $t_{\rm crit}$. The Hubble parameter \Eq{hup} vanishes, $H=0$ ($v=0$), at each peak and bounce (figure \ref{fig6}).
\item Log-oscillations disappear at time scales much larger than $t_*$, where evolution becomes monotonic and the Hubble parameter stops changing sign (figure \ref{fig6}). There is
a new critical scale $t_{\rm crit}\gg t_*\gg\tp$ which does not depend on $p$ and separates exotic from standard cosmological behaviour. At $t_{\rm crit}$, $H=0$ for the last time.
\end{itemize}

Up to a rescaling of $t_*$ and an overall rescaling, the geometric time coordinate \Eq{logos} is invariant under a discrete scale transformation:
\be\label{dsit}
t \to \lambda_\om^m t\,,\qquad \lambda_\om= \rme^{\frac{2\pi}{\om}}\,,\qquad m\in\mathbb{Z}\,.
\ee
Since $m$ is integer and can acquire also negative values, there are oscillations also between $t=0$ and $t=\tp$, and the total number of oscillations between $t=0$ and any given time $t$ is infinite. The position of two consecutive minima or maxima is governed by the constant ratio $t_{m+1}/t_m=\la_\om$:
\be\label{tnn}
\ln\frac{t_{m+1}}{t_m}=\frac{2\pi}{\om}\,.
\ee
The absolute position $t_{{\rm min},m}$ of, say, the minima (bounces) depends on the coefficients $A$ and $B$. For instance, if $A=1$ and $B=0$ (the measure has only the cosine), then $t_{{\rm min},m}= \tp \la_\om^{m+1/2}$, while if $A=0$ and $B=1$ (sine only), $t_{{\rm min},m}= \tp \la_\om^{3(m/2+1/4)}$.

From eq.\ \Eq{tnn}, we also get the expansion rate between two minima (or maxima) at early times, encoded in the net number of e-foldings per cycle
\be\label{ann}
\cN_{\uparrow\downarrow}:=\ln \frac{a_{m+1}}{a_m}\approx \frac{2\pi\a p}{\om}\,.
\ee
We can identify two regimes of the cosmic expansion. During the oscillatory era, $t\ll t_{\rm crit}$, the average slope $\de_{\uparrow\downarrow}$ of the trend of minima (or maxima) is
\be\label{mimi}
\de_{\uparrow\downarrow}=\frac{\ln (a_{m+1}/a_m)}{\ln (t_{m+1}/t_m)}=\cN_{\uparrow\downarrow}\,\frac{\om}{2\pi}\approx p\a\,,
\ee
while at times $t\gg t_{\rm crit}$ the slope is $p$. This result is easy to reach also by noting that these are the slopes of the zero mode of the measure (the power-law part), i.e., what remains after averaging $a$ over the log-oscillations. Notice that for $\a=0$ the net number of e-foldings per cycle is zero, $\cN_{\uparrow\downarrow}=0$: oscillations have constant zero mode.

\begin{figure}
\centering
\includegraphics[width=7.4cm]{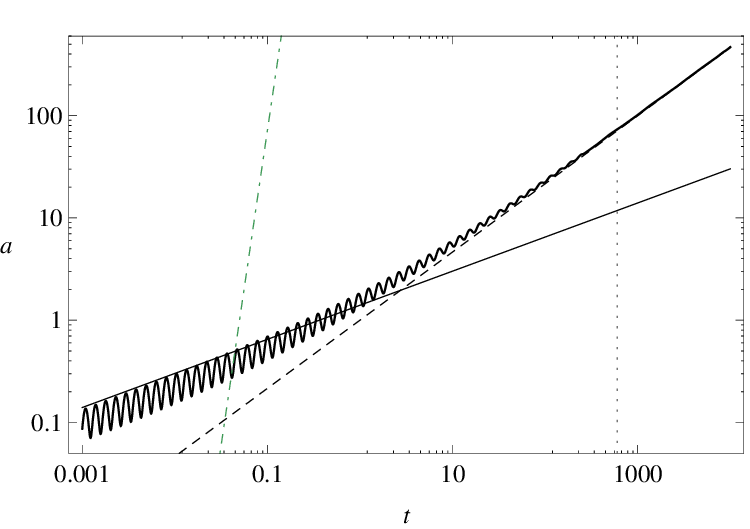}
\caption{\label{fig7} Log-oscillating scale factor (thick curve) for $t_*=1$, $\tp=10^{-1}$, $\a=1/2$, $\om=25$, $A=0.5$, $B=0$, and $p=2/3$ (dust if $D=4$). Thin line: evolution law $\sim t^{p\a}$ (eq.\ \Eq{mimi}) of the trend of minima (or maxima). Dashed line: late-time asymptotic evolution law $\sim t^p$ of the scale factor. Dot-dashed line: evolution law $\sim t^{\de_\uparrow}$ (eq.\ \Eq{mima}) of the expanding phase of an oscillation. The position of $t_{\rm crit}$ (end of log-periodic contraction epochs) is marked by the dotted vertical axis.}
\end{figure}

Equation \Eq{ann} can be found also by calculating the slopes of an expanding and a contracting phase. Assume for simplicity that the geometric coordinate has only the cosine ($B=0$; the case $A=0$, $B\neq 0$ is identical; the mixed case differs only in numbers). The relative position between any minimum and the next maximum is $t_{\rm max}/t_{\rm min}=\sqrt{\la_\om}=\rme^{\pi/\om}$. The ratio of two geometric coordinates at these points is
\be\label{mamiq}
\frac{q(t_{\rm max})}{q(t_{\rm min})}\approx \left(\frac{t_{\rm max}}{t_{\rm min}}\right)^\a\frac{1+A}{1-A}\,,
\ee
where $0<A<1$ and we have neglected the late-time contribution to the measure since it is subdominant in the oscillatory era. Therefore, the slope $\de_\uparrow$ of an expanding phase is
\be\label{mima}
\de_\uparrow=\frac{\ln (a_{\rm max}/a_{\rm min})}{\ln (t_{\rm max}/t_{\rm min})}\approx p\left(\a +\frac{\om}{\pi}\ln\frac{1+A}{1-A}\right)>p\,,
\ee
where the inequality holds because $A>0$. Notice that the larger the frequency $\om$, the steeper the oscillations. The number of e-foldings of an expansion phase is $\cN_\uparrow = \de_\uparrow \pi/\om$, which is partly undone in the next contraction. In fact, taking the inverse of eq.\ \Eq{mamiq} and noting that $t_{{\rm min}+1}/t_{\rm max}=t_{\rm max}/t_{\rm min}$, we get
\be\label{mami}
\de_\downarrow=\frac{\ln (a_{{\rm min}+1}/a_{\rm max})}{\ln (t_{{\rm min}+1}/t_{\rm max})}\approx p\left(\a -\frac{\om}{\pi}\ln\frac{1+A}{1-A}\right)\,,
\ee
so that the net number of e-folds per cycle is $\cN_{\uparrow\downarrow}=\cN_{\uparrow}+\cN_{\downarrow}=(\de_{\uparrow}+\de_{\downarrow})\pi/\om\approx 2p\a\pi/\om$, in agreement with eq.\ \Eq{ann}. Figure \ref{fig7} collects all these results.


\subsection{Alternative to inflation?}\label{cos}

\noindent {\bf Horizon problem.} To study the fate of cosmological perturbations and the shape of their spectrum, it is important to mark their evolution with a cosmological scale, distinguishing between thermalized modes in causal contact and modes outside the causal region. In standard cosmology, either the particle or Hubble horizon plays this role, and such is the case also in log-oscillating $q$-cosmology, since $r_{\rm p}\approx r_H$ never becomes singular except at the big bang. This happens because both $H$ and $v$ vanish at bounces and peaks, but $r_H\sim v/H$ remains finite. On the other hand, in usual cyclic cosmologies (e.g., \cite{NoB08} and references therein) the vanishing of the Hubble parameter is not compensated, the Hubble horizon diverges at bounces and turn-arounds, and one has to resort to a more involved cosmological time scale \cite{BiM,BMSh,BKM1}. This is one important difference with respect to ordinary cyclic inflation. The evolution of the Hubble horizon differs also with respect to the cyclic multiverse of \cite{Pia09,Pia10,ZLP,LP}.

That the particle or Hubble horizon is the correct milestone of the cosmological evolution can be also seen by considering the Mukhanov equation of a generic perturbation $u_k$ (for instance, a scalar fluctuation or one polarization mode of the graviton). In ordinary power-law cosmology, this equation is of the form $u_k''+(k^2-C/\tau^2)u_k=0$, where $k$ is the comoving wave-number, $C=p(2p-1)/(p-1)^2=O(1)$, and primes denote derivatives with respect to conformal time $\tau=r_{\rm p}/c=O(1) r_H/c$ ($c=1$ in our unit conventions). Up to an $O(1)$ constant, the effective mass term vanishes when the comoving perturbation $k=|{\bf k}|$ leaves the Hubble horizon. Horizon crossing is thus defined by the relation $k=1/r_H=aH$. We can apply the same discussion to the $q$-theory, modulo two differences. One is the form of the Hubble horizon, which is given by eq.\ \Eq{rh}. The other is the structure of momentum space. Instead of $k^i$, for each spatial direction $i$ one should replace the geometric coordinate of momentum space $p^i(k^i)$, whose functional form should be determined by the Fourier transform. For instance, it is easy to convince oneself that to $q^i(x^i)\sim x_i^\a$ there corresponds the same power, $p^i(k^i)\sim q^i(k^i)\sim k_i^\a$. Clearly, if spatial slices are ordinary, $q(x)=x$ and $p(k)=k$ for all $i$. We will discuss more complicated profiles $p(k)$ in section \ref{ci}. For the time being, we notice that the absolute value $k=|{\bf k}|$ in the Mukhanov equation is replaced by
\be\label{tik}
\tilde k:=\sqrt{\sum_i p_i^2(k^i)}=\sqrt{k^2+\dots}\,,
\ee
and the horizon-crossing condition reads
\be
\tilde k=\frac{1}{r_H}=\frac{a H}{v}.
\ee
Again, $r_H$ is interchangeable with the comoving particle horizon.

Let us now consider the horizon problem, which amounts to explain why even the largest perturbations we observe are in thermal equilibrium. In ordinary cosmology, the particle and Hubble horizons are monotonic in regimes where the expansion is described by a power law. For instance, assuming $\textsc{k}=0=\Lambda$ and ordinary matter, $0<p<1$, both $r_{\rm p}$ and $r_H$ increase with time, while for $p>1$ (inflation) they decrease and the horizon problem is solved. Here, the horizons oscillate without ever getting singular and originate a phenomenology completely different from the one of the previous sections.

\begin{figure}
\centering
\includegraphics[width=7.4cm]{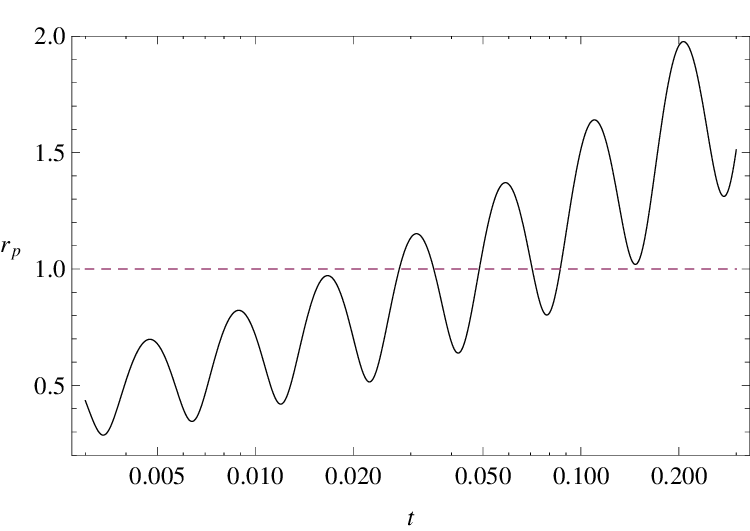}
\includegraphics[width=7.4cm]{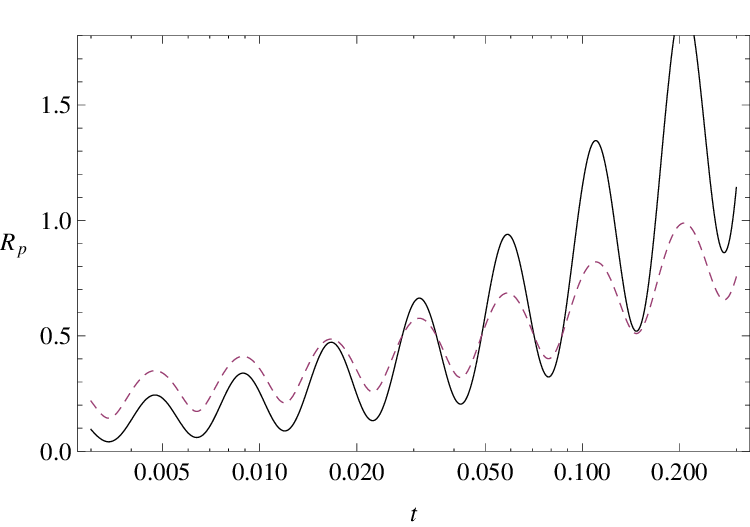}
\caption{\label{fig8} Horizon problem solved in log-oscillating spacetimes with $t_*=1$, $\tp=10^{-1}$, $\a=1/2$, $\om=10$, $A=0.5=B$, and $p=1/2$ (radiation if $D=4$). Left plot: comoving particle horizon $r_{\rm p}$ (thick curve) and a comoving perturbation of geometric wave-number $\tilde k=1$ (dashed curve), which exits and then re-enters in the causal horizon. Right plot: proper particle horizon $R_{\rm p}$ and a perturbation $\la =a/\tilde k=a$. Notice the logarithmic time scale.}
\end{figure}

A perturbations of wave-length $\la =a(t)/\tilde k$ can exit the particle horizon and then re-enter it; in comoving coordinates, the perturbation wave-number $k$ (and $\tilde k$) is constant, it exits the horizon at some point during the early evolution of the universe, and re-enters it as soon as the horizon reaches its scale (figure \ref{fig8}).

\begin{figure}
\centering
\includegraphics[width=7.4cm]{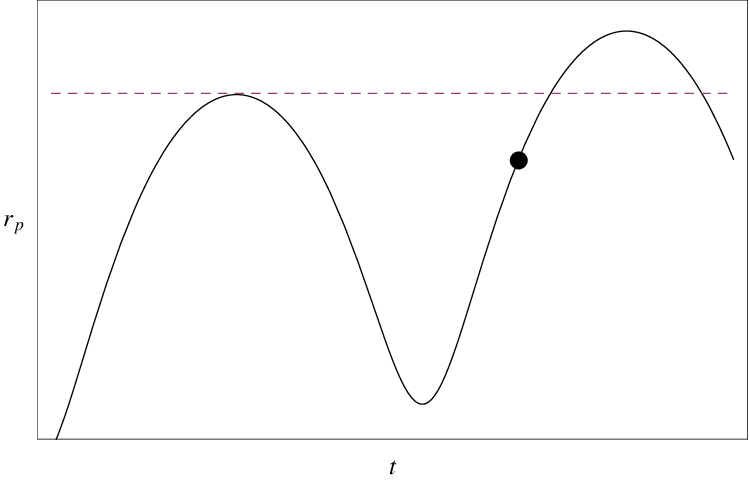}
\caption{\label{fig9} Alternative to inflation: we live in the expanding phase of one of the cycles, while thermalized fluctuations (all those below the dashed line) re-enter the horizon.}
\end{figure}

All perturbations which enter the horizon after $t_{\rm crit}$ have never thermalized. Therefore, in order to solve the horizon problem, we have two possibilities: either we assume $p>1$ (we will discuss this inflationary cyclic cosmology later), or we take $0<p<1$ but assume that $t_{\rm crit}$ is in our future, so that we are presently experiencing the expanding phase of one cycle. On one hand, this poses us at a special point during the evolution of the universe, since we have to be at a point (along the increasing slope of one cycle) which should not be higher than the peak of the preceding cycle (figure \ref{fig9}). On the other hand, this point is less special than deemed, since there are infinitely many cycles before us, all realizing about the same expansion rate \Eq{ann} (cycles close to $t_{\rm crit}$ will be $1/\a$ times longer than early ones, but this difference is negligible if $\a\lesssim O(1)$).

According to eq.\ \Eq{mima}, an observer on the ascending slope of a cycle will experience an effective cosmic expansion $a(t)\approx t^{\de_{\uparrow}}$. Since the universe did not always accelerate, a small exponent $\de_{\uparrow}$ would be more realistic. This would imply, in particular, that $\om=O(1)$. The present model is a simplified example of the possibilities of the theory and it cannot accommodate the variety of regimes of the history of the universe (radiation domination, matter domination, dark energy domination). We will not decide here whether more complex models of this scenario are viable.

\

\noindent {\bf Flatness problem.} Let $\Omega$ (not to be confused with the coefficient \Eq{Om} in the theory with weighted derivatives) be the density parameter
\be\label{omega}
\Omega := \frac{\rho}{\rho_{\rm crit}}\,,\qquad \rho_{\rm crit} := \frac{(D-1)(D-2)H^2}{2\k^2 v^2}\,.
\ee
Equation \Eq{fri} with $\Lambda=0$ becomes
\be\label{fr2bis}
\Omega-1 = \frac{2\textsc{k}}{D-2}\frac{v^2}{H^2 a^2}=\frac{2\textsc{k}}{D-2}\,r_H^2\,.
\ee
If the universe is spatially flat, $\Omega=1$. Otherwise, from eq.\ \Eq{rh},
\be\label{flapr}
|\Omega(t_0)-1|= |\Omega(t)-1| \left[\frac{q(t)}{q(t_0)}\right]^{-2(1-p)},
\ee
where $t<t_0$ and $t_0$ is today. The left-hand side is very close to zero, which may imply a fine tuning of the curvature of the universe at early times (flatness problem). In ordinary cosmology, the ratio $(t/t_0)^{-2(1-p)}$ would be a monotonic function of time, decreasing if $0<p<1$ (increasing $r_H$) and increasing if $p>1$ (decreasing $r_H$). The flatness problem would be resolved in the second case (inflation), since going backwards in time one can have $|\Omega(t)-1|\gg 1$ and still maintain the left-hand side small. 

\begin{figure}
\centering
\includegraphics[width=7.4cm]{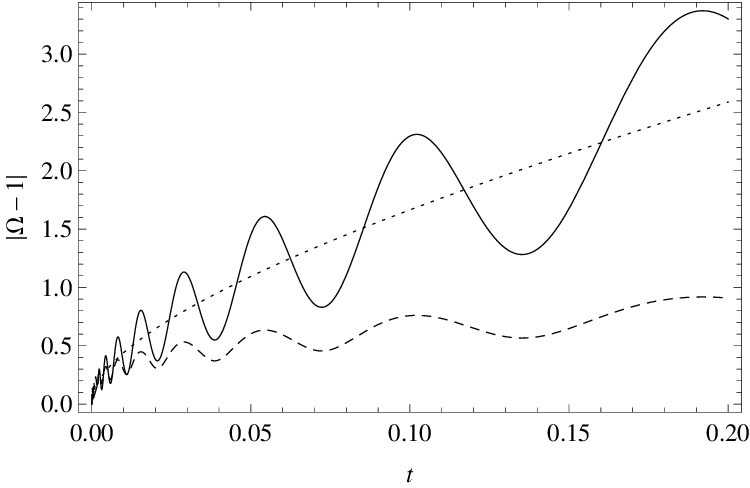}
\caption{\label{fig10} The density parameter $|\Om-1|=r_H^2$ (eq.\ \Eq{fr2bis}, solid curve) in log-oscillating spacetimes with $D=4$, $t_*=1$, $\tp=10^{-1}$, $\a=1/2$, $\om=10$, $A=0.5$, $B=0$, and $p=1/2$ (radiation). The zero mode (average evolution, dotted curve) and the scale factor (dashed curve) are shown for comparison.}
\end{figure}

As said above, in ordinary cyclic cosmologies the Hubble radius is ill-defined at the extrema and one has to gauge the flatness problem against a different cosmic scale \cite{BiM,BMSh,BKM1}. Here we can directly use eq.\ \Eq{rh}. As shown in figure \ref{fig10}, $\Omega$ undergoes an infinite number of oscillations in the past, whose maxima increase very slowly according to the law \Eq{ann} provided $\om\gtrsim 2\pi\a p=O(1)$. Thus, from a cycle to the next the density parameter $\Om$ does not deviate too much from unity, nor is there an appreciable increase in entropy (proportional to a positive power of the scale factor). However, in the past we had an infinite number of cycles and a steady growth of the zero mode of the measure. In the special case $\a=0$, the oscillations in the scale factor increase only because of the large-scale term $t$ in eq.\ \Eq{logos}. Going backwards in time, the offset of the oscillations of the Hubble radius decreases linearly, until it stops at some finite value at $t=0$, where $\langle q(0)\rangle=1$ in average and $|\Omega(0)-1|=(1-p)^2=O(1)$. Then, all the fine tuning in $|\Omega(t_0)-1|$ is transferred onto the amplitudes $A$ and $B$. Therefore, the flatness problem is not solved in these models with $0<p<1$.


\subsection{Cyclic mild inflation}\label{ci}

\begin{figure}
\centering
\includegraphics[width=7.4cm]{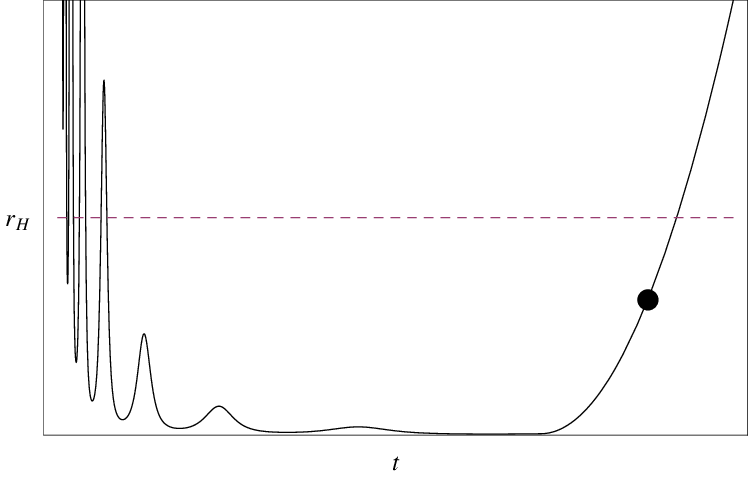}
\caption{\label{fig11} Qualitative pictorial of log-oscillating cyclic mild inflation. Our position in the evolution of the comoving Hubble horizon (solid line) is marked with a dot. The dashed line is a perturbation which exits the horizon during the cyclic era and then re-enters afterwards.}
\end{figure}

In the previous subsection, we studied log-oscillating models of the form \Eq{rgoa} with $0<p<1$ but we did not find a satisfactory solution to the flatness problem. If we take $p>1$, we enter into an inflationary scenario, since we need some matter with equation of state $P/\rho=w<-(D-3)/(D-1)$. Thus, instead of the evolution in figure \ref{fig9}, we experience another where we reside after a \emph{mildly} accelerating cyclic phase (figure \ref{fig11}). This strongly resembles models of emergent cyclic inflation \cite{Bis08,BiA,BiM,BMSh,BKM1,BKM2,DuB}, but with some important differences.

\emph{Origin of cyclic dynamics.} In CI models, the cyclic dynamics is driven by a negative cosmological constant via the standard Friedmann equation (in $D=4$)
\be
H^2=\frac{\kappa^2}{3}\,\rho-\frac{|\Lambda|}{3}\,,\label{frici}
\ee
while in log-oscillating cosmology the dynamics is modified by the non-trivial geometry (eq.~\Eq{fri}) which is also responsible for the cycles. Neither curvature nor $\Lambda$ plays any role.

\emph{Asymmetry and periodicity of cycles.} In CI the universe has no beginning and undergoes linear asymmetric cycles with constant period due to entropy exchange between different species (say, dust matter and radiation). Here, on the other hand, we have a flat cyclic universe with a finite past but, still, an infinite number of cycles. The cycles are \emph{log}-periodic and asymmetric by construction of the multi-fractal measure, even in the presence of only one matter species. We have not discussed whether the entropy of the universe is a meaningful concept in this multi-scale spacetime, but we expect it to be modified with respect to cosmological scenarios with standard differential structure.

\emph{Full analyticity.} We have seen that all the dynamical properties of log-oscillating cosmologies can be treated analytically with very few approximations, at least in the absence of curvature and of a cosmological constant. The amount of e-foldings of an expanding phase and of a single cycle is known analytically, too (eqs.\ \Eq{mimi}, \Eq{mima} and \Eq{mami}). The particle and Hubble horizons (eqs.\ \Eq{tau} and \Eq{rh}) are well defined throughout the whole evolution of the universe, and there is no need to construct another cosmological scale during the cycles. All this is possible thanks to the DSI \Eq{dsit} and the geometric structure of the theory, where the dynamical equations (derived from an action principle) are identical to the usual ones but coordinates are composite objects determined by multi-fractal geometry (eq.\ \Eq{logos}). General CI scenarios, on the other hand, require a number of semi-analytic approximations, and the evolution of the scale factor is sketched qualitatively.

\begin{figure}
\centering
\includegraphics[width=7.4cm]{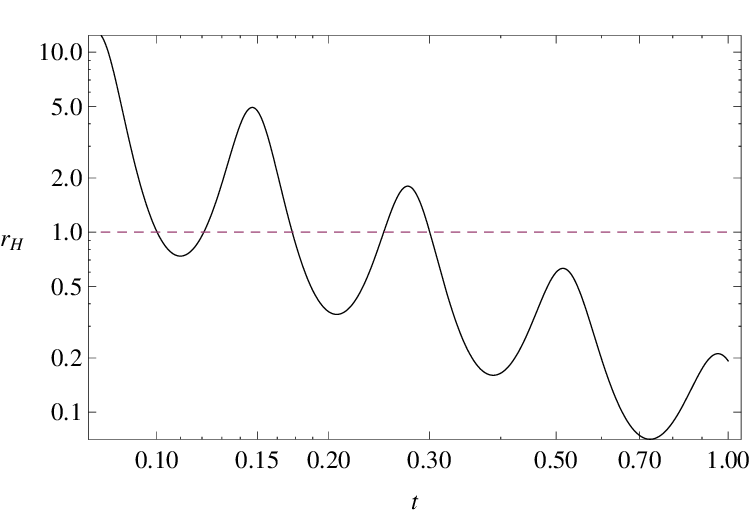}
\includegraphics[width=7.4cm]{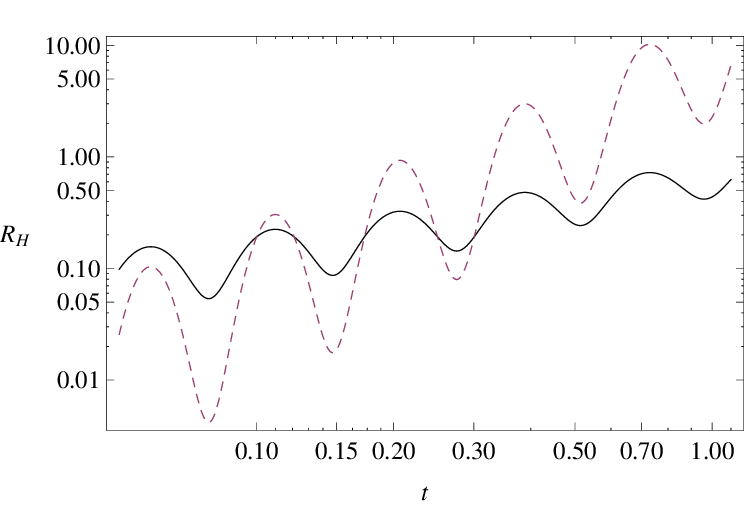}
\caption{\label{fig12} Comoving and proper Hubble horizon (solid curves) and a perturbation of comoving wave-number $\tilde k=1$ (dashed curves) for log-oscillating inflation. Parameters have the same values as in figure \ref{fig8}, except $p=3$ (mild inflation). The horizon re-entry is not shown.}
\end{figure}

\emph{Horizon problem.} The horizon problem is solved because perturbations, which begin inside the Hubble horizon, increase in amplitude so much that at some point they definitively exit the causal region and get frozen. This moment, called of \emph{last exit} in CI \cite{BiM}, is captured in figure \ref{fig12}. Later in the future, the average evolution of the Hubble horizon changes signature and perturbations can re-enter the thermalized patch. This is exactly the same mechanism of CI, the only difference being that here we do use the Hubble radius as a cosmological scale.

\begin{figure}
\centering
\includegraphics[width=7.4cm]{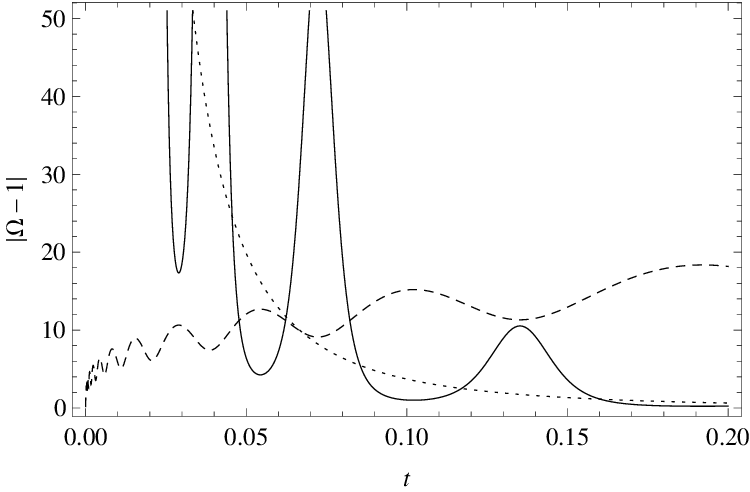}
\caption{\label{fig13} The density parameter $|\Om-1|=r_H^2$ (eq.\ \Eq{fr2bis}, solid curve) in log-oscillating spacetimes with the same parameters of figure \ref{fig10} except $p=3$ (mild inflation). The zero mode (average evolution, dotted curve) and the scale factor (dashed curve) are shown for comparison.}
\end{figure}

\emph{Flatness problem.} The flatness problem is also solved, since the density parameter can be arbitrarily large in the past (eq.\ \Eq{flapr} with $p>1$; see figure \ref{fig13}). Due to the lack of an analytic expression for a well-defined cosmological horizon, in CI scenarios the flatness problem is better stated in terms of the entropy of the universe. The latter increases monotonically little by little by the same factor in every cycle. The scale factor follows the same law, so that the overall growth during the cyclic era increases by a large number of e-foldings, thus mimicking an inflationary era. In log-oscillating cosmology, the expansion law per cycle is eq.\ \Eq{ann}, which is close to unity if $\om\sim 2\pi\a p$. This number is $O(1)$ for the model of section \ref{cos}, and large for large $p$. However, contrary to standard inflation we do not need $p$ to be very large, precisely because the cycles mimic the accelerated growth even when $p$ is smaller than or close to 1.

\emph{Graceful exit.} In emergent cyclic inflation, a scalar field is added to interrupt the era of cycles and thus recovering decelerating monotonic expansion. In log-oscillating scenarios, there is a natural exit from this period which does not require a by-hand mechanism. In fact, at some critical time $t_{\rm crit}$ the universe stops oscillating and standard expansion begins according to the matter content. However, we have seen that to solve the flatness problem we have to admit an inflationary matter, albeit of a very mild type, which thus reproposes the graceful-exit problem in the post-cyclic era just like in ordinary inflation. 

\emph{Perturbation spectrum.} Cyclic inflation can generate a scale-invariant spectrum by ascribing cosmological perturbations to statistical thermal fluctuations of a radiation bath \cite{BiA,BiM,BMSh,BKM1}. This mechanism is typically difficult to sustain in standard cosmology \cite{Pee93} but it finds an agile application both in CI and other scenarios \cite{AlBM,MaP,MaS,BBKM}. It should be possible to employ it also in log-oscillating cosmologies, but such a study goes beyond the scope of this paper. Here we only comment that the fluctuation spectrum will most likely be scale invariant and display logarithmic oscillations, both in the non-inflationary model of section \ref{cos} and in the mild inflationary one. The reason has been sketched in the context of CI \cite{BMSh} and we repeat the same argument here adapted to log-oscillating geometries. Consider two perturbations of geometric comoving wave-numbers $\tilde k$ and $\tilde k'$ exiting the horizon at consecutive cycles. From eq.\ \Eq{ann}, it follows that they represent the same perturbation with wave-length $\la$ if
\be
\frac{\tilde k'}{\tilde k}=\frac{a_{m+1}}{a_m}\approx \la_\om^{\a p}\,.
\ee
If $\om\gtrsim O(1)$, then the two wave-numbers are approximately the same. This leads to an almost scale-invariant spectrum with a log-oscillatory pattern.\footnote{Power spectra with log-oscillating features can be generated in ordinary cyclic inflation \cite{BMSh} but also in multi-field inflation \cite{Che11,BNV}, by generic initial-state effects of new physics at a fixed scale \cite{GSvS} and in string scenarios such as axion monodromy inflation \cite{FMPWX} and `unwinding' inflation \cite{DGKS}.} A computation of the power spectrum in $q$-theory can refine this argument. In ordinary inflationary cosmology, the spectrum of scalar perturbations is $P_{\rm s} \propto k^{n_{\rm s}-1}$, where $n_{\rm s}$ is the scalar spectral index, presently constrained to be about $0.96$ by Planck \cite{Ade13}. In the $q$-theory, we expect the same expression but with $k^\mu$ replaced by the geometric coordinate $p^\mu(k^\mu)$ in momentum space:
\be\label{spec1}
{}^q P_{\rm s} = \cA\, {\tilde k}^{{}^q n_{\rm s}-1},
\ee
where $\cA$ is a normalization constant, $\tilde k$ is given by eq.\ \Eq{tik} and the index ${}^q n_{\rm s}$ may be different from $n_{\rm s}$. The distribution $p(k)$ can be determined as follows. In Minkowski spacetime, the invertible unitary momentum transform of a function $\vp(x)$ expanded in a basis of the $q$-Laplace--Beltrami operator $\B_q$ is simply given by
\bs\label{motra}\ba
\tilde \vp(k) &:=&\int\frac{\rmd^D p(k)}{(2\pi)^\frac{D}{2}}\,\rme^{\rmi p_\mu(k^\mu) q^\mu(x^\mu)}\vp(x)\,,\\
\vp(x)        &=& \int\frac{\rmd^D q(x)}{(2\pi)^\frac{D}{2}}\,\rme^{-\rmi p_\mu(k^\mu) q^\mu(x^\mu)}\tilde\vp(k)\,,
\ea\es
where $p(k)=\prod_\mu p^\mu(k)\geq 0$ is factorizable and positive semi-definite. To determine the distribution $p(k)$ uniquely for a given geometric coordinate $q({\ell_n},x)$, we require that the ultraviolet and infrared limits of $p(k)$, as well as all intermediate regimes where $p$ has some characteristic asymptotic form, have the correct scaling matching the one of $q$. Consider, for instance, the one-dimensional multi-fractional measure \Eq{2}, so that $q(x)=x+{\rm sgn}(x)\,(\ell_*/\a)|x/\ell_*|^\a$. Naively, one might guess that $p(k)=q(\ell_*,k)=k+{\rm sgn}(k)(E_*/\a)|k/E_*|^\a$, where $E_*:=1/\ell_*$, but the infrared ($k\ll E_*$) and ultraviolet ($k\gg E_*$) limits would give the incorrect scaling. To get $p\sim k$ at low energies and $p\sim k^{\a}$ at high energies, one should have
\be
p(k)=\frac{1}{q\left(\frac{1}{E_*},\frac{1}{k}\right)}=\frac{E_*}{\frac{E_*}{k}+{\rm sgn}(k)\frac{E_*}{\a}\left|\frac{E_*}{k}\right|^\a},
\ee
so that $p\sim k$ and $pq\sim kx$ for $k\ll E_*$, while $p\sim \a E_* |k/E_*|^\a$ and $pq\sim |kx|^\a$ for $k\gg E_*$. Including also log-oscillations, the general form of the geometric momentum is
\be\label{pik}
p(k):=\frac{E_*}{\frac{E_*}{k}+{\rm sgn}(k)\frac{E_*}{\a}\left|\frac{E_*}{k}\right|^\a F_\om(\ln|k|)}\,.
\ee
In particular, the momentum transform \Eq{motra} is not an automorphism, since $p(k)\neq q(k)$.

Resuming the discussion of eq.\ \Eq{spec1} and calling $k_*$ the spatial characteristic momentum, if the perturbation spectrum is generated during the oscillatory phase, the term $1/k$ in the denominator of $p(k)$ can be neglected and $p(k)\sim (k/k_*)^\a F_\om^{-1}(\ln k)$, so that eq.\ \Eq{spec1} becomes, very roughly,
\be\label{spec2}
{}^q P_{\rm s} \sim \left(\frac{k}{k_*}\right)^{n_{\rm eff}-1} f_\om(\ln k)\,,\quad n_{\rm eff}-1=\a({}^q n_{\rm s}-1)\,,\quad f_\om(\ln k)\sim [F_\om(\ln k)]^{1-{}^q n_{\rm s}}\,.
\ee
A more rigorous expression in three spatial dimensions can be obtained by plugging \Eq{pik} for each direction into eq.\ \Eq{tik} and then $\tilde k$ into eq.\ \Eq{spec1}. The log-oscillatory pattern is directly inherited from the momentum measure structure. \emph{The effective spectral index $n_{\rm eff}$ can be very close to 1 even if $n_{\rm s}$ is not (which may be the case if $p=O(1)$), since there is a suppression by $\a$}.


\subsection{Big bang problem revisited}

\begin{figure}
\centering
\includegraphics[width=7.4cm]{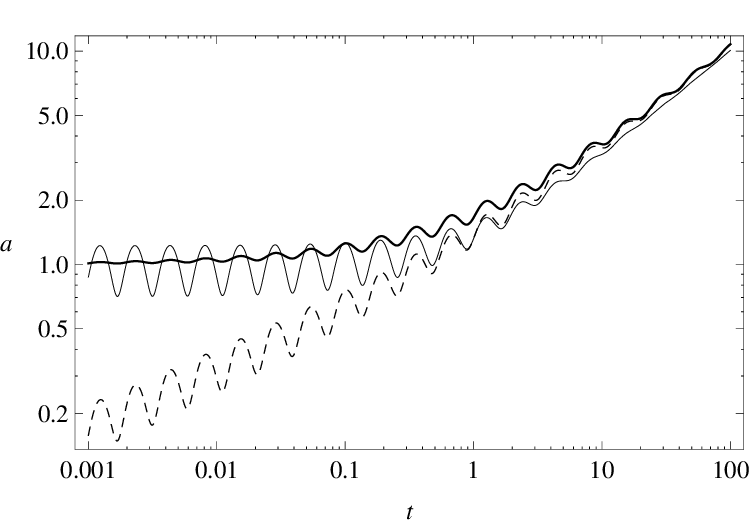}
\caption{\label{fig14}  Log-log plot of the scale factor \Eq{rgoa} with $t_*=1$, $\tp=10^{-1}$, $\om=10$, $A=0.5$, $B=0$, $p=1/2$, for $\a=1/2$ and $t_{\rm bb}=0$ (dashed curve), $\a=0$ and $t_{\rm bb}=0$ (thin curve), and $\a=1/2$ and $t_{\rm bb}=t_*$ (thick curve).}
\end{figure}

Equation \Eq{logos} is not the most general integration of the multi-scale log-oscillating measure weight. A shift of the distribution $q(x)$ by a constant does not change the Lebesgue measure, and one could consider cosmological profiles where the composite time coordinate is
\be\label{logos2}
q(t)\to t_{\rm bb}+q(t)\,.
\ee
Thus, if $t_{\rm bb}\neq 0$ the power-law solution \Eq{rgoa} becomes non-singular at $t=0$, where the scale factor acquires the value $a=(t_{\rm bb}/t_*)^p$. The history of the early universe is affected by the shift \Eq{logos2}, as one can see in the example of figure \ref{fig14}. In the $q$-theory, the big bang singularity can be removed simply by choosing a non-vanishing integration constant.\footnote{As a side remark, we note some similarity between the result \Eq{logos2} and the old idea that certain constants in the Lagrangian may be not fundamental but, rather, constants of integration. A well-known example is the cosmological constant $\Lambda$ when obtained as a constant of motion from a non-dynamical 3-form gauge field \cite{Wei89,DuV,ANT,HeT84} or by changing gravity as in unimodular theories (section \ref{coco0}). However, in our case the value of the scale factor at the big bounce is not a constant of motion but originates from a homogeneous contribution to the measure weight.} Another possibility is to stick with eq.\ \Eq{logos} but set $\a=0$, a rather extreme but still well-defined geometry. The zero mode becomes the constant $t_*$ which plays the same role as $t_{\rm bb}$. The oscillations are not damped going towards $t=0$, so this model is qualitatively distinguishable from the other.

In principle, if log-oscillating cosmologies turned out to be viable, independent experiments should be able to constrain not only the range of the scales $t_n,\ell_n$ of the theory (as shown in previous attempts \cite{frc2,frc8}) but also the value of the integration constant and of $\a$. This would open up the most intriguing possibility to have a cosmological model where the avoidance of the initial singularity can leave an observable imprint.


\section{Conclusions}\label{concl}

The main results of this paper have been summarized in the introduction and we will not repeat them here. There are many directions along which these models should deserve further attention. 

Much of the cosmology of the theory with ordinary derivatives is still to be explored. Numerical investigations can probe whether viable cosmological solutions with log-oscillating measures exist and modify the history of the early universe as in the $q$-theory.

In the theory with weighted derivatives, the cosmological constant problem is rephrased but not solved. In a vacuum flat model, the asymptotic value of the cosmological constant in the infinite future is determined by the arbitrary integration constant $H_0$, eq.\ \Eq{umin}. Adding matter and curvature, solutions can be found numerically. In that case, the value $\Lambda_0$ of the cosmological constant today will depend both on the curvature $\textsc{k}=\pm 1$ and on the details of the matter equation of state. This should help in linking the value $\Lambda_0$ with that of matter parameters which are constrained independently. Also in this theory it would be desirable to study log-oscillating solutions, possibly with semi-analytic or numerical methods.

Log-oscillating dynamics with matter admits an exact treatment in the $q$-theory. Without asking for it, we hit a concrete, fully analytic realization of cyclic inflation where a mild acceleration can produce an almost scale invariant spectrum of cosmological fluctuations modulated by logarithmic oscillations. Asymmetric cycles arise not because of an entropy exchange between different matter species as in normal cyclic inflation, but because of the DSI of spacetime. The enticing possibility to find evidence of deterministic multi-fractal geometry in the sky is subordinate to a careful study of cosmological perturbation theory in this framework and of the resulting cosmic microwave background spectrum, which has been only sketched here. Still within the $q$-theory, we have constructed also a partially successful alternative to inflation without acceleration but with a similar type of spectrum. This solves the problems of the hot big bang with the exception of the flatness problem. The latter may as well find a natural solution in multi-scale models. Reconsider eq.\ \Eq{fr2bis} in four dimensions, $|\Omega-1| = v^2/(H^2 a^2)$. If we could find a power-law solution $a\sim t^p$ at early times, then the only difference with respect to the standard flatness equation would be the measure factor $v^2$. The latter is very large at early times $t\ll t_*$ for the binomial case \Eq{binom}, and if $\a$ is close to zero and $t_*\sim 10^{32}\tp\approx 10^{-12}\,\mbox{s}$, then at the Planck scale $|\Omega-1|=O(1)$, as in varying-speed-of-light theories \cite{Mof92,AlM}. One can show that the horizon problem, too, can be solved in models with power-law solution $a\sim t^p$ for a certain range of the parameters $0<p,\a<1$. On the other hand, in our cyclic inflationary scenario a very mild average acceleration during the cyclic era solves the flatness problem, but then a graceful exit from inflation is required. Further studies of both scenarios will determine whether and how we can strike a balance between the flatness and graceful-exit problem.

Finally, we comment on the initial singularity. In the theory with weighted derivatives in vacuum, dynamical solutions exist where the big bang is replaced by a bounce. In the $q$-theory with matter, the resolution of the cosmic singularity is purely kinematical, i.e., determined solely by the structure of the measure. In one model, the big bounce replacing the initial bang is nothing but an integration constant $t_{\rm bb}$ (eq.\ \Eq{logos2}), which however can be constrained together with the dynamics by experiments. In another model where the fractional charge $\a$ in the measure is set to zero, the big bounce is not an integration constant but the zero mode of the log-oscillating scale factor. It is not clear whether this singularity resolution is related to dimensional flow of these scenarios \cite{frc7} in the same way as dimensional flow is related to the UV finiteness or renormalizability of quantum-gravity models, as mentioned in the introduction. Results on the renormalizability of field theories in multi-scale spacetimes with weighted and $q$-derivatives point towards a negative answer \cite{frc9}. Last, in scenarios where the big bang is not removed (for instance, in the $q$-theory when $\a\neq 0$ and the integration constant $t_{\rm bb}$ is set to zero) log-oscillations determine the fate of the initial singularity. The latter is reached only by the zero mode, since the universe approaches the big bang at $t=0$ through infinitely many oscillations, packed with logarithmic progression around the origin. This is evocative \cite{frc2} of BKL scenarios, where the oscillations are due to anisotropy \cite{BKL70,BKL82}. The present analysis is a promising starting point from which to investigate this possible relation with the BKL singularity, via the simple but powerful analytic tools of log-oscillating geometries.



\begin{acknowledgments}
The author thanks M.\ Scalisi for useful comments. This work is under a Ram\'on y Cajal contract.
\end{acknowledgments}



\begin{thebibliography}{99}

\bibitem{bH}    D.\ ben-Avraham and S.\ Havlin, \book{Diffusion and Reactions in Fractals and Disordered Systems}{Cambridge University Press}{Cambridge}{U.K.}{2000}.
\bibitem{MeK00} R.\ Metzler and J.\ Klafter, \tia{The random walk's guide to anomalous diffusion: a fractional dynamics approach} \doinn{10.1016/S0370-1573(00)00070-3}{Phys.\ Rep.}{339}{1}{2000}.
\bibitem{Zas02} G.M.\ Zaslavsky, \tia{Chaos, fractional kinetics, and anomalous transport} \doinn{10.1016/S0370-1573(02)00331-9}{Phys.\ Rep.}{371}{461}{2002}.
\bibitem{Sok12} I.M.\ Sokolov, \tia{Models of anomalous diffusion in crowded environments} \doinn{10.1039/C2SM25701G}{Soft Matter}{8}{9043}{2012}.

\bibitem{Har01} D.\ Harte, \book{Multifractals: Theory and Applications}{Chapman \& Hall/CRC}{Boca Raton}{U.S.A.}{2001}
\bibitem{Fal03} K.\ Falconer, \book{Fractal Geometry}{Wiley}{New York}{U.S.A.}{2003}.

\bibitem{tH93}  G.\ 't Hooft, \tia{Dimensional reduction in quantum gravity} in \emph{Salamfestschrift}, A.\ Ali, J.\ Ellis and S.\ Randjbar-Daemi eds., World Scientific, Singapore (1993) [\oarX{gr-qc/9310026}].
\bibitem{Car09} S.\ Carlip, \tia{Spontaneous dimensional reduction in short-distance quantum gravity?} \doinn{10.1063/1.3284402}{AIP Conf.\ Proc.}{1196}{72}{2009} [\arX{0909.3329}].
\bibitem{fra1}  G.\ Calcagni, \tia{Fractal universe and quantum gravity} \doinn{10.1103/PhysRevLett.104.251301}{Phys.\ Rev.\ Lett.}{104}{251301}{2010} [\arX{0912.3142}].
\bibitem{Car10} S.\ Carlip, \tia{The small scale structure of spacetime} in \emph{Foundations of Space and Time}, G.\ Ellis, J.\ Murugan and A.\ Weltman eds., Cambridge University Press, Cambridge U.K.\ (2012) [\arX{1009.1136}].
\bibitem{AJL4}  J.\ Ambj{\o}rn, J.\ Jurkiewicz and R.\ Loll, \tia{Spectral dimension of the universe} \doinn{10.1103/PhysRevLett.95.171301}{Phys.\ Rev.\ Lett.}{95}{171301}{2005} [\oarX{hep-th/0505113}].
\bibitem{BeH}   D.\ Benedetti and J.\ Henson, \tia{Spectral geometry as a probe of quantum spacetime} \doin{10.1103/PhysRevD.80.124036}{Phys.\ Rev.}{D}{80}{124036}{2009} [\arX{0911.0401}].
\bibitem{SVW1}  T.P.\ Sotiriou, M.\ Visser and S.\ Weinfurtner, \tia{Spectral dimension as a probe of the ultraviolet continuum regime of causal dynamical triangulations} \doinn{10.1103/PhysRevLett.107.131303}{Phys.\ Rev.\ Lett.}{107}{131303}{2011} [\arX{1105.5646}].
\bibitem{LaR5}  O.\ Lauscher and M.\ Reuter, \tia{Fractal spacetime structure in asymptotically safe gravity} \doij{10.1088/1126-6708/2005/10/050}{JHEP}{10}{050}{2005} [\oarX{hep-th/0508202}].
\bibitem{CES}   G.\ Calcagni, A.\ Eichhorn and F.\ Saueressig, \tia{Probing the quantum nature of spacetime by diffusion} \doin{10.1103/PhysRevD.87.124028}{Phys.\ Rev.}{D}{87}{124028}{2013} [\arX{1304.7247}].
\bibitem{Mod08} L.\ Modesto, \textit{Fractal structure of loop quantum gravity} \doinn{10.1088/0264-9381/26/24/242002}{Classical Quant.\ Grav.}{26}{242002}{2009} [\arX{0812.2214}].
\bibitem{CaM}   F.\ Caravelli and L.\ Modesto, \tia{Fractal dimension in 3d spin-foams} \arX{0905.2170}.
\bibitem{MPM}   E.\ Magliaro, C.\ Perini and L.\ Modesto, \tia{Fractal space-time from spin-foams} \arX{0911.0437}.
\bibitem{Hor3}  P.\ Ho\v{r}ava, \tia{Spectral dimension of the universe in quantum gravity at a Lifshitz point} \doinn{10.1103/PhysRevLett.102.161301}{Phys.\ Rev.\ Lett.}{102}{161301}{2009} [\arX{0902.3657}].
\bibitem{Con06} A.\ Connes, \tia{Noncommutative geometry and the standard model with neutrino mixing} \doij{10.1088/1126-6708/2006/11/081}{JHEP}{11}{081}{2006} [\oarX{hep-th/0608226}].
\bibitem{CCM}   A.H.\ Chamseddine, A.\ Connes and M.\ Marcolli, \tia{Gravity and the standard model with neutrino mixing} \doinn{10.4310/ATMP.2007.v11.n6.a3}{Adv.\ Theor.\ Math.\ Phys.}{11}{991}{2007} [\oarX{hep-th/0610241}].
\bibitem{AA}    E.\ Alesci and M.\ Arzano, \tia{Anomalous dimension in semiclassical gravity} \doin{10.1016/j.physletb.2011.12.026}{Phys.\ Lett.}{B}{707}{272}{2012} [\arX{1108.1507}].
\bibitem{Ben08} D.\ Benedetti, \tia{Fractal properties of quantum spacetime} \doinn{10.1103/PhysRevLett.102.111303}{Phys.\ Rev.\ Lett.}{102}{111303}{2009} [\arX{0811.1396}].
\bibitem{ACOS}  M.\ Arzano, G.\ Calcagni, D.\ Oriti and M.\ Scalisi, \tia{Fractional and noncommutative spacetimes} \doin{10.1103/PhysRevD.84.125002}{Phys.\ Rev.}{D}{84}{125002}{2011} [\arX{1107.5308}].
\bibitem{Mod11} L.\ Modesto, \tia{Super-renormalizable quantum gravity} \doin{10.1103/PhysRevD.86.044005}{Phys.\ Rev.}{D}{86}{044005}{2012} [\arX{1107.2403}].
\bibitem{CaG}   S.\ Carlip and D.\ Grumiller, \tia{Lower bound on the spectral dimension near a black hole} \doin{10.1103/PhysRevD.84.084029}{Phys.\ Rev.}{D}{84}{084029}{2011} [\arX{1108.4686}].
\bibitem{Mur12} J.R.\ Mureika, \tia{Primordial black hole evaporation and spontaneous dimensional reduction} \doin{10.1016/j.physletb.2012.08.029}{Phys.\ Lett.}{B}{716}{171}{2012} [\arX{1204.3619}].
\bibitem{AC1}   M.\ Arzano and G.\ Calcagni, \tia{Black-hole entropy and minimal diffusion} \doin{10.1103/PhysRevD.88.084017}{Phys.\ Rev.}{D}{88}{084017}{2013} [\arX{1307.6122}].
\bibitem{MoN}   L.\ Modesto and P.\ Nicolini, \tia{Spectral dimension of a quantum universe} \doin{10.1103/PhysRevD.81.104040}{Phys.\ Rev.}{D}{81}{104040}{2010} [\arX{0912.0220}].
\bibitem{DJW1}  B.\ Durhuus, T.\ Jonsson and J.F.\ Wheater, \tia{Random walks on combs} \doin{10.1088/0305-4470/39/5/002}{J.\ Phys.}{A}{39}{1009}{2006} [\oarX{hep-th/0509191}].
\bibitem{AGW}   M.R.\ Atkin, G.\ Giasemidis and J.F.\ Wheater, \tia{Continuum random combs and scale dependent spectral dimension} \doin{10.1088/1751-8113/44/26/265001}{J.\ Phys.}{A}{44}{265001}{2011} [\arX{1101.4174}].
\bibitem{GWZ1}  G.\ Giasemidis, J.F.\ Wheater and S.\ Zohren, \tia{Dynamical dimensional reduction in toy models of $4D$ causal quantum gravity} \doin{10.1103/PhysRevD.86.081503}{Phys.\ Rev.}{D}{86}{081503(R)}{2012}
[\arX{1202.2710}].
\bibitem{GWZ2}  G.\  Giasemidis, J.F.\ Wheater and S.\ Zohren, \tia{Multigraph models for causal quantum gravity and scale dependent spectral dimension} \doin{10.1088/1751-8113/45/35/355001}{J.\ Phys.}{A}{45}{355001}{2012} [\arX{1202.6322}].

\bibitem{frc4}  G.\ Calcagni, \tia{Diffusion in multiscale spacetimes} \doin{10.1103/PhysRevE.87.012123}{Phys.\ Rev.}{E}{87}{012123}{2013} [\arX{1205.5046}].
\bibitem{fra2}  G.\ Calcagni, \tia{Quantum field theory, gravity and cosmology in a fractal universe} \doij{10.1007/JHEP03(2010)120}{JHEP}{03}{120}{2010} [\arX{1001.0571}].
\bibitem{fra3}  G.\ Calcagni, \tia{Gravity on a multifractal} \doin{10.1016/j.physletb.2011.01.063}{Phys.\ Lett.}{B}{697}{251}{2011} [\arX{1012.1244}].
\bibitem{HX1}   H.-J.\ He and Z.-Z.\ Xianyu, \tia{Unitary standard model from spontaneous dimensional reduction and weak boson scattering at the LHC} \doinn{10.1140/epjp/i2013-13040-2}{Eur.\ Phys.\ J.\ Plus}{128}{40}{2013} [\arX{1112.1028}].
\bibitem{HX2}   H.-J.\ He and Z.-Z.\ Xianyu, \tia{Spontaneous spacetime reduction and unitary weak boson scattering at the LHC} \doin{10.1016/j.physletb.2013.01.06}{Phys.\ Lett.}{B}{720}{142}{2013} [\arX{1301.4570}].
\bibitem{STW}   A.\ Sheykhi, Z.\ Teimoori and B.\ Wang, \tia{Thermodynamics of fractal universe} \doin{10.1016/j.physletb.2012.12.072}{Phys.\ Lett.}{B}{718}{1203}{2013} [\arX{1212.2137}].
\bibitem{frc6}  G.\ Calcagni and G.\ Nardelli, \tia{Symmetries and propagator in multi-fractional scalar field theory} \doin{10.1103/PhysRevD.87.085008}{Phys.\ Rev.}{D}{87}{085008}{2013} [\arX{1210.2754}].
\bibitem{frc7}  G.\ Calcagni and G.\ Nardelli, \tia{Spectral dimension and diffusion in multiscale spacetimes} \doin{10.1103/PhysRevD.88.124025}{Phys.\ Rev.}{D}{88}{124025}{2013} [\arX{1304.2709}].
\bibitem{frc8}  G.\ Calcagni, J.\ Magueijo and D.\ Rodr\'iguez Fern\'andez, \tia{Varying electric charge in multiscale spacetimes} {\it Phys.\ Rev.} {\bf D} (in press) [\arX{1305.3497}].
\bibitem{frc10} G.\ Calcagni, \tia{Relativistic particle in multiscale spacetimes} \doin{10.1103/PhysRevD.88.065005}{Phys.\ Rev.}{D}{88}{065005}{2013} [\arX{1306.5965}].
\bibitem{fra4}  G.\ Calcagni, \tia{Discrete to continuum transition in multifractal spacetimes} \doin{10.1103/PhysRevD.84.061501}{Phys.\ Rev.}{D}{84}{061501(R)}{2011} [\arX{1106.0295}].
\bibitem{frc1}  G\ Calcagni, \tia{Geometry of fractional spaces} \doinn{10.4310/ATMP.2012.v16.n2.a5}{Adv.\ Theor.\ Math.\ Phys.}{16}{549}{2012} [\arX{1106.5787}].
\bibitem{frc2}  G.\ Calcagni, \tia{Geometry and field theory in multi-fractional spacetime} \doij{10.1007/JHEP01(2012)065}{JHEP}{01}{065}{2012} [\arX{1107.5041}].
\bibitem{frc3}  G.\ Calcagni and G.\ Nardelli, \tia{Momentum transforms and Laplacians in fractional spaces} \doinn{10.4310/ATMP.2012.v16.n4.a5}{Adv.\ Theor.\ Math.\ Phys.}{16}{1315}{2012} [\arX{1202.5383}].
\bibitem{fra6}  G.\ Calcagni, \tia{Diffusion in quantum geometry} \doin{10.1103/PhysRevD.86.044021}{Phys.\ Rev.}{D}{86}{044021}{2012} [\arX{1204.2550}].
\bibitem{frc5}  G.\ Calcagni, G.\ Nardelli and M.\ Scalisi, \tia{Quantum mechanics in fractional and other anomalous spacetimes} \doinn{10.1063/1.4757647}{J.\ Math.\ Phys.}{53}{102110}{2012} [\arX{1207.4473}].
\bibitem{fra7}  G.\ Calcagni, \tia{Multifractional spacetimes, asymptotic safety and Ho\v{r}ava--Lifshitz gravity} \doin{10.1142/S0217751X13500929}{Int.\ J.\ Mod.\ Phys.}{A}{28}{1350092}{2013} [\arX{1209.4376}].
\bibitem{frc9}  G.\ Calcagni and G.\ Nardelli, \tia{Quantum field theory with varying couplings} \arX{1306.0629}.

%
\bibitem{Pad98} T.\ Padmanabhan, \tia{Quantum structure of spacetime and black hole entropy} \doinn{10.1103/PhysRevLett.81.4297}{Phys.\ Rev.\ Lett.}{81}{4297}{1998} [\oarX{hep-th/9801015}].
\bibitem{Pad99} T.\ Padmanabhan, \tia{Event horizon: Magnifying glass for Planck length physics} \doin{10.1103/PhysRevD.59.124012}{Phys.\ Rev.}{D}{59}{124012}{1999} [\oarX{hep-th/9801138}].

\bibitem{Sor98} D.\ Sornette, \tia{Discrete scale invariance and complex dimensions} \doinn{10.1016/S0370-1573(97)00076-8}{Phys.\ Rep.}{297}{239}{1998} [\oarX{cond-mat/9707012}].

\bibitem{Bis08} T.\ Biswas, \tia{Emergence of a cyclic universe from the Hagedorn soup} \arX{0801.1315}.
\bibitem{BiA}   T.\ Biswas and S.\ Alexander, \tia{Cyclic inflation} \doin{10.1103/PhysRevD.80.043511}{Phys.\ Rev.}{D}{80}{043511}{2009} [\arX{0812.3182}].
\bibitem{BiM}   T.\ Biswas and A.\ Mazumdar, \tia{Inflation with a negative cosmological constant} \doin{10.1103/PhysRevD.80.023519}{Phys.\ Rev.}{D}{80}{023519}{2009} [\arX{0901.4930}].
\bibitem{BMSh}  T.\ Biswas, A.\ Mazumdar and A.\ Shafieloo, \tia{Wiggles in the cosmic microwave background radiation: Echoes from nonsingular cyclic inflation} \doin{10.1103/PhysRevD.82.123517}{Phys.\ Rev.}{D}{82}{123517}{2010} [\arX{1003.3206}].
\bibitem{BKM1}  T.\ Biswas, T.\ Koivisto and A.\ Mazumdar, \tia{Could our universe have begun with $-\Lambda$?} \arX{1105.2636}.
\bibitem{BKM2}  T.\ Biswas, T.\ Koivisto and A.\ Mazumdar, \tia{Phase transitions during cyclic inflation and non-Gaussianity} \doin{10.1103/PhysRevD.88.083526}{Phys.\ Rev.}{D}{88}{083526}{2013} [\arX{1302.6415}].
\bibitem{DuB}   W.\ Duhe and T.\ Biswas, \tia{Emergent cyclic inflation, a numerical investigation} \arX{1306.6927}.

\bibitem{Wey52} H.\ Weyl, \book{Space, Time, and Matter}{Dover}{Mineola}{U.S.A}{1952}.
\bibitem{NOSE}  M.\ Novello, L.A.R.\ Oliveira, J.M.\ Salim and E.\ Elbaz, \tia{Geometrized instantons and the creation of the universe} \doin{10.1142/S021827189200032X}{Int.\ J.\ Mod.\ Phys.}{D}{1}{641}{1992}.
\bibitem{NoB08} M.\ Novello and S.E.\ Perez Bergliaffa, \tia{Bouncing cosmologies} \doinn{10.1016/j.physrep.2008.04.006}{Phys.\ Rep.}{463}{127}{2008} [\arX{0802.1634}].
\bibitem{PS}    F.P.\ Poulis and J.M.\ Salim, \tia{Weyl geometry and gauge-invariant gravitation} \arX{1305.6830}.

\bibitem{El07a} R.A.\ El-Nabulsi, \tia{Differential geometry and modern cosmology with fractionally differentiated Lagrangian function and fractional decaying force term} \ndoinn{http://www.nipne.ro/rjp/2007_52_3-4.html}{Rom.\ J.\ Phys.}{52}{467}{2007}. 
\bibitem{El07b} R.A.\ El-Nabulsi, \tia{Some fractional geometrical aspects of weak field approximation and Schwarzschild spacetime} \ndoinn{http://www.nipne.ro/rjp/2007_52_5-7.html}{Rom.\ J.\ Phys.}{52}{705}{2007}. 
\bibitem{El07c} R.A.\ El-Nabulsi, \tia{Cosmology with a fractional action principle} \ndoinn{http://www.rrp.infim.ro/2007_59_3.html}{Rom.\ Rep.\ Phys.}{59}{763}{2007}. 
\bibitem{El08}  R.A.\ El-Nabulsi, \tia{Increasing effective gravitational constant in fractional ADD brane cosmology} \ndoinn{http://www.ejtp.com/ejtpv5i17}{Electronic J.\ Theor.\ Phys.}{5 No.\ 17}{103}{2008}. 
\bibitem{El10a} R.A.\ El-Nabulsi, \tia{Fractional action-like variational approach, perturbed Einstein's gravity and new cosmology} \ndoin{http://fizika.hfd.hr/fizika_b/bv10/b19p103.htm}{Fizika}{B}{19}{103}{2010}. 
\bibitem{El10b} R.A.\ El-Nabulsi, \tia{Oscillating flat FRW dark energy dominated cosmology from periodic functional approach} \doinn{10.1088/0253-6102/54/1/03}{Commun.\ Theor.\ Phys.}{54}{16}{2010}. 
\bibitem{Shc10} V.K.\ Shchigolev, \tia{Cosmological models with fractional derivatives and fractional action functional} \doinn{10.1088/0253-6102/56/2/34}{Commun.\ Theor.\ Phys.}{56}{389}{2011} [\arX{1011.3304}].
\bibitem{JRMRE} M.\ Jamil, M.A.\ Rashid, D.\ Momeni, O.\ Razina and K.\ Esmakhanova, \tia{Fractional action cosmology with power law weight function} \doinn{10.1088/1742-6596/354/1/012008}{J.\ Phys.: Conf.\ Ser.}{354}{012008}{2012} [\arX{1106.2974}].
\bibitem{DCJ}   U.\ Debnath, S.\ Chattopadhyay and M.\ Jamil, \tia{Fractional action cosmology: some dark energy models in emergent, logamediate, and intermediate scenarios of the universe} \doinn{10.1186/2251-7235-7-25}{J.\ Theor.\ Appl.\ Phys.}{7}{25}{2013} [\arX{1107.0541}].
\bibitem{DJC}   U.\ Debnath, M.\ Jamil and S.\ Chattopadhyay, \tia{Fractional action cosmology: emergent, logamediate, intermediate, power law scenarios of the universe and generalized second law of thermodynamics}, \doinn{10.1007/s10773-011-0961-1}{Int.\ J.\ Theor.\ Phys.}{51}{812}{2012} [\arX{1109.1506}].
\bibitem{El12a} R.A.\ El-Nabulsi, \tia{Gravitons in fractional action cosmology} \doinn{10.1007/s10773-012-1290-8}{Int.\ J.\ Theor.\ Phys.}{51}{3978}{2012}. 
\bibitem{Shc12} V.K.\ Shchigolev, \tia{Cosmic evolution in fractional action cosmology} \doinn{10.5890/DNC.2013.04.002}{Discont.\ Nonlin.\ Complexity}{2}{115}{2013} [\arX{1208.3454}]. 

\bibitem{El05}  R.A.\ El-Nabulsi, \tia{Fractional unstable Euclidean universe} \ndoinn{http://www.ejtp.com/ejtpv2i8}{Electronic J.\ Theor.\ Phys.}{2 No.\ 8}{1}{2005} 
\bibitem{Rob09} M.D.\ Roberts, \tia{Fractional derivative cosmology} \arX{0909.1171}.
\bibitem{Mun10} J.\ Munkhammar, \tia{Riemann--Liouville fractional Einstein field equations} \arX{1003.4981}.
\bibitem{Vac10} S.I.\ Vacaru, \tia{Fractional dynamics from Einstein gravity, general solutions, and black holes} \doinn{10.1007/s10773-011-1010-9}{Int.\ J.\ Theor.\ Phys.}{51}{338}{2012} [\arX{1004.0628}].

%
\bibitem{MN}    R.\ Mansouri and F.\ Nasseri, \tia{Model universe with variable space dimension: Its dynamics and wave function} \doin{10.1103/PhysRevD.60.123512}{Phys.\ Rev.}{D}{60}{123512}{1999} [\oarX{gr-qc/9902043}].

\bibitem{Mag03} J.\ Magueijo, \tia{New varying speed of light theories} \doinn{10.1088/0034-4885/66/11/R04}{Rep.\ Prog.\ Phys.}{66}{2025}{2003} [\oarX{astro-ph/0305457}].
\bibitem{Mof92} J.W.\ Moffat, \tia{Superluminary universe: a possible solution to the initial value problem in cosmology} \doin{10.1142/S0218271893000246}{Int.\ J.\ Mod.\ Phys.}{D}{2}{351}{1993} [\oarX{gr-qc/9211020}].
\bibitem{AlM}   A.\ Albrecht and J.\ Magueijo, \tia{Time varying speed of light as a solution to cosmological puzzles} \doin{10.1103/PhysRevD.59.043516}{Phys.\ Rev.}{D}{59}{043516}{1999} [\oarX{astro-ph/9811018}].
\bibitem{BaM1}  J.D.\ Barrow and J.\ Magueijo, \tia{Varying alpha theories and solutions to the cosmological problems} \doin{10.1016/S0370-2693(98)01294-5}{Phys.\ Lett.}{B}{443}{104}{1998} [\oarX{astro-ph/9811072}].
\bibitem{BaM2}  J.D.\ Barrow and J.\ Magueijo, \tia{Solutions to the quasi-flatness and quasilambda problems} \doin{10.1016/S0370-2693(99)00008-8}{Phys.\ Lett.}{B}{447}{246}{1999} [\oarX{astro-ph/9811073}].
\bibitem{BaM3}  J.D.\ Barrow and J.\ Magueijo, \tia{Solving the flatness and quasiflatness problems in Brans--Dicke cosmologies with a varying light speed} \doinn{10.1088/0264-9381/16/4/030}{Classical Quant.\ Grav.}{16}{1435}{1999} [\oarX{astro-ph/9901049}].
\bibitem{Bar99} J.D.\ Barrow, \tia{Cosmologies with varying light speed} \doin{10.1103/PhysRevD.59.043515}{Phys.\ Rev.}{D}{59}{043515}{1999} [\oarX{astro-ph/9811022}].
\bibitem{LSV}   S.\ Landau, P.D.\ Sisterna and H.\ Vucetich, \tia{Charge conservation and time varying speed of light} \doin{10.1103/PhysRevD.63.081303}{Phys.\ Rev.}{D}{63}{081303(R)}{2001} [\oarX{astro-ph/0007108}].
\bibitem{ElU}   G.F.R.\ Ellis and J.-P.\ Uzan, \tia{$c$ is the speed of light, isn't it?} \doinn{10.1119/1.1819929}{Am.\ J.\ Phys.}{73}{240}{2005} [\oarX{gr-qc/0305099}].
\bibitem{Mag00} J.\ Magueijo, \tia{Covariant and locally Lorentz invariant varying speed of light theories} \doin{10.1103/PhysRevD.62.103521}{Phys.\ Rev.}{D}{62}{103521}{2000} [\oarX{gr-qc/0007036}].
\bibitem{MBS}   J.\ Magueijo, J.D.\ Barrow and H.B.\ Sandvik, \tia{Is it $e$ or is it $c$? Experimental tests of varying alpha} \doin{10.1016/S0370-2693(02)02928-3}{Phys.\ Lett.}{B}{549}{284}{2002} [\oarX{astro-ph/0202374}].
\bibitem{Ma08a} J.\ Magueijo, \tia{Speedy sound and cosmic structure} \doinn{10.1103/PhysRevLett.100.231302}{Phys.\ Rev.\ Lett.}{100}{231302}{2008} [\arX{0803.0859}].
\bibitem{Ma08b} J.\ Magueijo, \tia{Bimetric varying speed of light theories and primordial fluctuations} \doin{10.1103/PhysRevD.79.043525}{Phys.\ Rev.}{D}{79}{043525}{2008} [\arX{0807.1689}].
\bibitem{Ma08c} J.\ Magueijo, \tia{DSR as an explanation of cosmological structure} \doinn{10.1088/0264-9381/25/20/202002}{Classical Quant.\ Grav.}{25}{202002}{2008} [\arX{0807.1854}].
\bibitem{Mag10} J.\ Magueijo, J.\ Noller and F.\ Piazza, \tia{Bimetric structure formation: non-Gaussian predictions} \doin{10.1103/PhysRevD.82.043521}{Phys.\ Rev.}{D}{82}{0435251}{2010} [\arX{1006.3216}].

%
\bibitem{DeS}   A.\ De Felice and S.\ Tsujikawa, \tia{$f(R)$ theories} \doinn{10.12942/lrr-2010-3}{Living Rev.\ Rel.}{13}{3}{2010} [\arX{1002.4928}].

\bibitem{Wei79} S.\ Weinberg, \tia{Ultraviolet divergences in quantum gravity} in \emph{General Relativity, an Einstein Centenary Survey}, S.W.\ Hawking and W.\ Israel eds., Cambridge University Press, Cambridge U.K.\ (1979), pg.\ 790--831.
\bibitem{Reu1}  M.\ Reuter, \tia{Nonperturbative evolution equation for quantum gravity}  \doin{10.1103/PhysRevD.57.971}{Phys.\ Rev.}{D}{57}{971}{1998} [\oarX{hep-th/9605030}].
\bibitem{Nie06} M.\ Niedermaier, \tia{The asymptotic safety scenario in quantum gravity: an introduction} \doinn{10.1088/0264-9381/24/18/R01}{Classical Quant.\ Grav.}{24}{R171}{2007} [\oarX{gr-qc/0610018}].
\bibitem{RSnax} M.\ Reuter and F.\ Saueressig, \tia{Asymptotic safety, fractals, and cosmology} \doinn{10.1007/978-3-642-33036-0_8}{Lect.\ Notes Phys.}{863}{185}{2013} [\arX{1205.5431}].
\bibitem{BoR1}  A.\ Bonanno and M.\ Reuter, \tia{Cosmology of the Planck era from a renormalization group for quantum gravity} \doin{10.1103/PhysRevD.65.043508}{Phys.\ Rev.}{D}{65}{043508}{2002} [\oarX{hep-th/0106133}].
\bibitem{ReS3}  M.\ Reuter and F.\ Saueressig, \tia{From big bang to asymptotic de Sitter: complete cosmologies in a quantum gravity framework} \doij{10.1088/1475-7516/2005/09/012}{JCAP}{09}{012}{2005} [\oarX{hep-th/0507167}].
\bibitem{BoR3}  A.\ Bonanno and M.\ Reuter, \tia{Entropy signature of the running cosmological constant} \doij{10.1088/1475-7516/2007/08/024}{JCAP}{08}{024}{2007} [\arX{0706.0174}].
\bibitem{Bon12} A.\ Bonanno, \tia{An effective action for asymptotically safe gravity} \doin{10.1103/PhysRevD.85.081503}{Phys.\ Rev.}{D}{85}{081503}{2012} [\arX{1203.1962}].

\bibitem{RYS}   F.-Y.\ Ren, Z.-G.\ Yu and F.\ Su, \tia{Fractional integral associated to the self-similar set or the generalized self-similar set and its physical interpretation} \doin{10.1016/0375-9601(96)00418-5}{Phys.\ Lett.}{A}{219}{59}{1996}.
\bibitem{YRZ}   Z.-G.\ Yu, F.-Y.\ Ren and J.\ Zhou, \tia{Fractional integral associated to generalized cookie-cutter set and its physical interpretation} \doin{10.1088/0305-4470/30/15/036}{J.\ Phys.}{A}{30}{5569}{1997}.
\bibitem{RYZLM} F.-Y.\ Ren, Z.-G.\ Yu, J.\ Zhou, A.\ Le M\'ehaut\'e and R.R.\ Nigmatullin, \tia{The relationship between the fractional integral and the fractal structure of a memory set} \doin{10.1016/S0378-4371(97)00353-1}{Physica}{A}{246}{419}{1997}.
\bibitem{Yu99}  Z.-G.\ Yu, \tia{Flux and memory measure on net fractals} \doin{10.1016/S0375-9601(99)00316-3}{Phys.\ Lett.}{A}{257}{221}{1999}.
\bibitem{QL} W.-Y.\ Qiu and J.\ L\"u, \tia{Fractional integrals and fractal structure of memory sets} \doin{10.1016/S0375-9601(00)00448-5}{Phys.\ Lett.}{A}{272}{353}{2000}.
\bibitem{RQLW} F.-Y.\ Ren, W.-Y.\ Qiu, J.-R.\ Liang and X.-T.\ Wang, \tia{Determination of memory function and flux on fractals} \doin{10.1016/S0375-9601(01)00544-8}{Phys.\ Lett.}{A}{288}{79}{2001}.
\bibitem{RLWQ} F.-Y.\ Ren, J.-R.\ Liang, X.-T.\ Wang and W.-Y.\ Qiu, \tia{Integrals and derivatives on net fractals} \doinn{10.1016/S0960-0779(02)00211-4}{Chaos Solitons Fract.}{16}{107}{2003}.
\bibitem{LMNN}  A.\ Le M\'ehaut\'e, R.R.\ Nigmatullin and L.\ Nivanen, \book{Fl\`eches du Temps et G\'eom\'etrie Fractale}{Hermes}{Paris}{France}{1998}.
\bibitem{NLM}   R.R.\ Nigmatullin and A.\ Le M\'ehaut\'e, \tia{Is there geometrical/physical meaning of the fractional integral with complex exponent?} \doinn{10.1016/j.jnoncrysol.2005.05.035}{J.\ Non-Cryst.\ Solids}{351}{2888}{2005}.

\bibitem{DIL}   B.\ Derrida, C.\ Itzykson and J.M.\ Luck, \tia{Oscillatory critical amplitudes in hierarchical models} \doinn{10.1007/BF01212352}{Commun.\ Math.\ Phys.}{94}{115}{1984}.
\bibitem{KiL}   J.\ Kigami and M.L.\ Lapidus, \tia{Weyl's problem for the spectral distribution of Laplacians on P.C.F.\ self-similar fractals} \doinn{10.1007/BF02097233}{Commun.\ Math.\ Phys.}{158}{93}{1993}.
\bibitem{Tep05} A.\ Teplyaev, \tia{Spectral zeta functions of fractals and the complex dynamics of polynomials} \doinn{10.1090/S0002-9947-07-04150-5}{T.\ Am.\ Math.\ Soc.}{359}{4339}{2007} [\oarX{math.SP/0505546}].
\bibitem{LvF}   M.L.\ Lapidus and M.\ van Frankenhuysen, \book{Fractal Geometry, Complex Dimensions and Zeta Functions}{Springer}{New York}{U.S.A.}{2006}.
\bibitem{Akk1}  E.\ Akkermans, G.V.\ Dunne and A.\ Teplyaev, \tia{Physical consequences of complex dimensions of fractals} \doinn{10.1209/0295-5075/88/40007}{Europhys.\ Lett.}{88}{40007}{2009} [\arX{0903.3681}].
\bibitem{ABS}   A.\ Allan, M.\ Barany and R.S.\ Strichartz, \tia{Spectral operators on the Sierpinski gasket I} \doinn{10.1080/17476930802272978}{Complex Var.\ Elliptic Equ.}{54}{521}{2009}.
\bibitem{Kaj10} N.\ Kajino, \tia{Spectral asymptotics for Laplacians on self-similar sets} \doinn{10.1016/j.jfa.2009.11.001}{J.\ Funct.\ Anal.}{258}{1310}{2010}.

\bibitem{HOSS}  Y.\ Huang, G.\ Ouillon, H.\ Saleur and D.\ Sornette, \tia{Spontaneous generation of discrete scale invariance in growth models} \doin{10.1103/PhysRevE.55.6433}{Phys.\ Rev.}{E}{55}{6433}{1997}.
\bibitem{StS}   D.\ Stauffer and D.\ Sornette, \tia{Log-periodic oscillations for biased diffusion in 3D random lattices} \doin{10.1016/S0378-4371(97)00680-8}{Physica}{A}{252}{271}{1998} [\oarX{cond-mat/9712085}].

%
\bibitem{Tar12} V.E.\ Tarasov, \tia{Fractional vector calculus and fractional Maxwell's equations} \doinn{10.1016/j.aop.2008.04.005}{Annals Phys.}{323}{2756}{2008} [\arX{0907.2363}].
\bibitem{CSN1}  K.\ Cottrill-Shepherd and M.\ Naber, \tia{Fractional differential forms} \doinn{10.1063/1.1364688}{J.\ Math.\ Phys.}{42}{2203}{2001} [\oarX{math-ph/0301013}].

%
\bibitem{CPSUV} J.\ Collins, A.\ Perez, D.\ Sudarsky, L.\ Urrutia and H.\ Vucetich, \tia{Lorentz invariance and quantum gravity: an additional fine-tuning problem?} \doinn{10.1103/PhysRevLett.93.191301}{Phys.\ Rev.\ Lett.}{93}{191301}{2004} [\oarX{gr-qc/0403053}].
\bibitem{CPS}   J.\ Collins, A.\ Perez and D.\ Sudarsky, \tia{Lorentz invariance violation and its role in quantum gravity phenomenology} \oarX{hep-th/0603002}.
\bibitem{She09} V.I.\ Shevchenko, \tia{Phenomenology of scale-dependent space-time dimension} \arX{0903.0565}.
\bibitem{ScM}   A.\ Sch\"afer and B.\ M\"uller, \tia{Bounds for the fractal dimension of space} \doin{10.1088/0305-4470/19/18/034}{J.\ Phys.}{A}{19}{3891}{1986}.
\bibitem{MuS}   B.\ M\"uller and A.\ Sch\"afer, \tia{Improved bounds on the dimension of space-time} \doinn{10.1103/PhysRevLett.56.1215}{Phys.\ Rev.\ Lett.}{56}{1215}{1986}.
\bibitem{CO}    F.\ Caruso and V.\ Oguri, \tia{The cosmic microwave background spectrum and a determination of fractal space dimensionality}
 \doinn{10.1088/0004-637X/694/1/151}{Astrophys.\ J.}{694}{151}{2009} [\arX{0806.2675}].

\bibitem{AnF}   J.L.\ Anderson and D.\ Finkelstein, \tia{Cosmological constant and fundamental length} \doinn{10.1119/1.1986321}{Am.\ J.\ Phys.}{39}{901}{1971}.
\bibitem{Ray79} J.\ Rayski, \tia{The problems of quantum gravity} \doinn{10.1007/BF00756668}{Gen.\ Rel.\ Grav.}{11}{19}{1979}.
\bibitem{vvN} J.J.\ van der Bij, H.\ van Dam and Y.J.\ Ng, \tia{The exchange of massless spin-two particles} \doin{10.1016/0378-4371(82)90247-3}{Physica}{A}{116}{307}{1982}.
\bibitem{Zee85} A.\ Zee, \tia{Remarks on the cosmological constant paradox} in \emph{High-Energy Physics: Proceedings of the 20th Orbis Scientiae, 1983}, S.L.\ Mintz and A.\ Perlmutter eds., Plenum Press, New York U.S.A.\ (1985), pg.\ 211.
\bibitem{BD88}  W.\ Buchm\"uller and N.\ Dragon, \tia{Einstein gravity from restricted coordinate invariance} \doin{10.1016/0370-2693(88)90577-1}{Phys.\ Lett.}{B}{207}{292}{1988}. 
\bibitem{BD89}  W.\ Buchm\"uller and N.\ Dragon, \tia{Gauge fixing and the cosmological constant} \doin{10.1016/0370-2693(89)91608-0}{Phys.\ Lett.}{B}{223}{313}{1989}. 
\bibitem{Unr89} W.G.\ Unruh, \tia{Unimodular theory of canonical quantum gravity} \doin{10.1103/PhysRevD.40.1048}{Phys.\ Rev.}{D}{40}{1048}{1989}. 
\bibitem{UW89}  W.G.\ Unruh and R.M.\ Wald, \tia{Time and the interpretation of canonical quantum gravity} \doin{10.1103/PhysRevD.40.2598}{Phys.\ Rev.}{D}{40}{2598}{1989}. 
\bibitem{Ngv}   Y.J.\ Ng and H.\ van Dam, \tia{Unimodular theory of gravity and the cosmological constant} \doinn{10.1063/1.529283}{J.\ Math.\ Phys.}{32}{1337}{1991}. 
\bibitem{Pet91} A.N.\ Petrov, \tia{On the cosmological constant as a constant of integration} \doin{10.1142/S0217732391002281}{Mod.\ Phys.\ Lett.}{A}{6}{2107}{1991}. 
\bibitem{Alv05} E.\ \'Alvarez, \tia{Can one tell Einstein's unimodular theory from Einstein's general relativity?} \doij{10.1088/1126-6708/2005/03/002}{JHEP}{03}{002}{2005} [\oarX{hep-th/0501146}].
\bibitem{ABGV}  E.\ \'Alvarez, D.\ Blas, J.\ Garriga and E.\ Verdaguer, \tia{Transverse Fierz--Pauli symmetry} \doin{10.1016/j.nuclphysb.2006.08.003}{Nucl.\ Phys.}{B}{756}{148}{2006} [\oarX{hep-th/0606019}].
\bibitem{AFL}  E.\ \'Alvarez, A.F.\ Faedo and J.J.\ L\'opez-Villarejo, \tia{Ultraviolet behavior of transverse gravity} \doij{10.1088/1126-6708/2008/10/023}{JHEP}{10}{023}{2008} [\arX{0807.1293}].
\bibitem{FiGa}  B.\ Fiol and J.\ Garriga, \tia{Semiclassical unimodular gravity} \doij{10.1088/1475-7516/2010/08/015}{JCAP}{08}{015}{2010} [\arX{0809.1371}].
\bibitem{AlV}   E.\ \'Alvarez and R.\ Vidal, \tia{Weyl transverse gravity (WTDiff) and the cosmological constant} \doin{10.1103/PhysRevD.81.084057}{Phys.\ Rev.}{D}{81}{084057}{2010} [\arX{1001.4458}].
\bibitem{BSZ}   D.\ Blas, M.\ Shaposhnikov and D.\ Zenh\"ausern, \tia{Scale-invariant alternatives to general relativity} \doin{10.1103/PhysRevD.84.044001}{Phys.\ Rev.}{D}{84}{044001}{2011} [\arX{1104.1392}].
\bibitem{Eic13} A.\ Eichhorn, \tia{On unimodular quantum gravity} \doinn{10.1088/0264-9381/30/11/115016}{Classical Quant.\ Grav.}{30}{115016}{2013} [\arX{1301.0879}].
\bibitem{HeT89} M.\ Henneaux and C.\ Teitelboim, \tia{The cosmological constant and general covariance} \doin{10.1016/0370-2693(89)91251-3}{Phys.\ Lett.}{B}{222}{195}{1989}. 
\bibitem{Wei89} S.\ Weinberg, \tia{The cosmological constant problem} \doinn{10.1103/RevModPhys.61.1}{Rev.\ Mod.\ Phys.}{61}{1}{1989}.
\bibitem{GuK1}  E.I.\ Guendelman and A.B.\ Kaganovich, \tia{Principle of nongravitating vacuum energy and some of its consequences} \doin{10.1103/PhysRevD.53.7020}{Phys.\ Rev.}{D}{53}{7020}{1996}.
\bibitem{GuK2}  E.I.\ Guendelman and A.B.\ Kaganovich, \tia{Gravitational theory without the cosmological constant problem} \doin{10.1103/PhysRevD.55.5970}{Phys.\ Rev.}{D}{55}{5970}{1997} [\oarX{gr-qc/9611046}].
\bibitem{Gue99} E.I.\ Guendelman, \tia{Scale invariance, new inflation and decaying lambda terms} \doin{10.1142/S0217732399001103}{Mod.\ Phys.\ Lett.}{A}{14}{1043}{1999} [\oarX{gr-qc/9901017}].
\bibitem{GuK3}  E.I.\ Guendelman and A.B.\ Kaganovich, \tia{Dynamical measure and field theory models free of the cosmological constant problem} \doin{10.1103/PhysRevD.60.065004}{Phys.\ Rev.}{D}{60}{065004}{1999} [\oarX{gr-qc/9905029}].

%
\bibitem{Pia12} F.\ Piazza, \tia{Infrared-modified Universe} \arX{1204.4099}.
\bibitem{Giu06} D.\ Giulini, \tia{Some remarks on the notions of general covariance and background independence}  \doinn{10.1007/978-3-540-71117-9_6}{Lect.\ Notes Phys.}{721}{105}{2007} [\oarX{gr-qc/0603087}].
\bibitem{wei72} S.\ Weinberg, \book{Gravitation and Cosmology}{Wiley}{New York}{U.S.A.}{1972}.
\bibitem{DeS10} N.\ Deruelle and M.\ Sasaki, \tia{Conformal equivalence in classical gravity: the example of ``veiled'' general relativity}
 \doinn{10.1007/978-3-642-19760-4_23}{Springer Proc.\ Phys.}{137}{247}{2011} [\arX{1007.3563}]. 
\bibitem{ChY}   T.\ Chiba and M.\ Yamaguchi, \tia{Conformal-frame (in)dependence of cosmological observations in scalar-tensor theory} \doij{10.1088/1475-7516/2013/10/040}{JCAP}{10}{040}{2013} [\arX{1308.1142}].
\bibitem{Pia09} Y.-S.\ Piao, \tia{Proliferation in cycle} \doin{10.1016/j.physletb.2009.05.009}{Phys.\ Lett.}{B}{677}{1}{2009} [\arX{0901.2644}].
\bibitem{Pia10} Y.-S.\ Piao, \tia{Design of a cyclic multiverse} \doin{10.1016/j.physletb.2010.06.039}{Phys.\ Lett.}{B}{691}{225}{2010} [\arX{1001.0631}].
\bibitem{ZLP}   J.\ Zhang, Z.-G.\ Liu and Y.-S.\ Piao,\tia{Amplification of curvature perturbations in cyclic cosmology} \doin{10.1103/PhysRevD.82.123505}{Phys.\ Rev.}{D}{82}{123505}{2010} [\arX{1007.2498}].
\bibitem{LP}    Z.-G.\ Liu and Y.-S.\ Piao, \tia{Scalar perturbations through cycles} \doin{10.1103/PhysRevD.86.083510}{Phys.\ Rev.}{D}{86}{083510}{2012} [\arX{1201.1371}].

\bibitem{Pee93} P.J.E.\ Peebles, \book{Principles of Physical Cosmology}{Princeton University Press}{Princeton}{U.S.A.}{1993}.
\bibitem{AlBM}  S.\ Alexander, R.\ Brandenberger and J.\ Magueijo, \tia{Noncommutative inflation} \doin{10.1103/PhysRevD.67.081301}{Phys.\ Rev.}{D}{67}{081301}{2003} [\oarX{hep-th/0108190}].
\bibitem{MaP}   J.\ Magueijo and L.\ Pogosian, \tia{Could thermal fluctuations seed cosmic structure?} \doin{10.1103/PhysRevD.67.043518}{Phys.\ Rev.}{D}{67}{043518}{2003} [\oarX{astro-ph/0211337}].
\bibitem{MaS}   J.\ Magueijo and P.\ Singh, \tia{Thermal fluctuations in loop cosmology} \doin{10.1103/PhysRevD.76.023510}{Phys.\ Rev.}{D}{76}{023510}{2007} [\oarX{astro-ph/0703566}].
\bibitem{BBKM}  T.\ Biswas, R.\ Brandenberger, T.\ Koivisto and A.\ Mazumdar, \tia{Cosmological perturbations from statistical thermal fluctuations} \doin{10.1103/PhysRevD.88.023517}{Phys.\ Rev.}{D}{88}{023517}{2013} [\arX{1302.6463}].

%
\bibitem{Che11} X.\ Chen, \tia{Primordial features as evidence for inflation} \doij{10.1088/1475-7516/2012/01/038}{JCAP}{01}{038}{2012} [\arX{1104.1323}].
\bibitem{BNV}   T.\ Battefeld, J.C.\ Niemeyer and D.\ Vlaykov, \tia{Probing two-field open inflation by resonant signals in correlation functions}
 \doij{10.1088/1475-7516/2013/05/006}{JCAP}{05}{006}{2013} [\arX{1302.3877}].
\bibitem{GSvS}  B.\ Greene, K.\ Schalm, J.P.\ van der Schaar and G.\ Shiu, \tia{Extracting new physics from the CMB}  eConf C {\bf 041213} (2004) 0001  [\oarX{astro-ph/0503458}].
\bibitem{FMPWX} R.\ Flauger, L.\ McAllister, E.\ Pajer, A.\ Westphal and G.\ Xu, \tia{Oscillations in the CMB from axion monodromy inflation}
 \doij{10.1088/1475-7516/2010/06/009}{JCAP}{06}{009}{2010} [\arX{0907.2916}].
\bibitem{DGKS}  G.\ D'Amico, R.\ Gobbetti, M.\ Kleban and M.\ Schillo, \tia{Unwinding Inflation} \doij{10.1088/1475-7516/2013/03/004}{JCAP}{03}{004}{2013}
  [\arX{1211.4589}].
\bibitem{Ade13} \textsc{Planck} Collaboration, P.A.R.\ Ade {et al.}, \tia{Planck 2013 results. XXII. Constraints on inflation} \arX{1303.5082}.

\bibitem{DuV}   M.J.\ Duff and P.\ van Nieuwenhuizen, \tia{Quantum inequivalence of different field representations} \doin{10.1016/0370-2693(80)90852-7}{Phys.\ Lett.}{B}{94}{179}{1980}.
\bibitem{ANT}   A.\ Aurilia, H.\ Nicolai, and P.K.\ Townsend, \tia{Hidden constants: The $\theta$ parameter of QCD and the cosmological constant of $N=8$ supergravity} \doin{10.1016/0550-3213(80)90466-6}{Nucl.\ Phys.}{B}{176}{509}{1980}.
\bibitem{HeT84} M.\ Henneaux and C.\ Teitelboim, \tia{The cosmological constant as a canonical variable} \doin{10.1016/0370-2693(84)91493-X}{Phys.\ Lett.}{B}{143}{415}{1984}.

\bibitem{BKL70} V.A.\ Belinsky, I.M.\ Khalatnikov and E.M.\ Lifshitz, \tia{Oscillatory approach to a singular point in the relativistic cosmology} \doinn{10.1080/00018737000101171}{Adv.\ Phys.}{19}{525}{1970}.
\bibitem{BKL82} V.A.\ Belinsky, I.M.\ Khalatnikov and E.M.\ Lifshitz, \tia{A general solution of the Einstein equations with a time singularity} \doinn{10.1080/00018738200101428}{Adv.\ Phys.}{31}{639}{1982}.

\end{thebibliography}
\end{document}